\documentclass{emulateapj}
\usepackage{color}
\usepackage{booktabs}

\newcommand{\Y}{Y_{lm}(\theta , \phi)}
\newcommand{\diff}{\mathrm{d}}

\newcommand{\Laplace}{\mathcal{L}}

\newcommand{\gtsim}{\protect\raisebox{-0.5ex}{$\:\stackrel{\textstyle >}{\sim}\:$}} 
\newcommand{\jump}[1]{\ensuremath{[\![#1]\!]} }

\newcommand{\up}{^\mathrm{(u)}}
\newcommand{\down}{^\mathrm{(d)}}
\newcommand{\Res}{\mathrm{Res}}

\slugcomment{}

\shorttitle{Links between the shock instabilities and asymmetric accretions}
\shortauthors{Takahashi, \textcolor{black}{Iwakami, Yamamoto} \& Yamada}

\begin{document}


\title{Links between the shock instability in core-collapse supernovae and asymmetric accretions of envelopes}


\author{Kazuya Takahashi\altaffilmark{1,\textcolor{black}{2}}, Wakana Iwakami\altaffilmark{3,4}, Yu Yamamoto\altaffilmark{3} and Shoichi Yamada\altaffilmark{5}}
\affil{$^1$Frontier Research Institute for Interdisciplinary Sciences, Tohoku University, Sendai, 980-8578, Japan}


\altaffiltext{2}{\textcolor{black}{Astronomical Institute, Tohoku University, Sendai, 980-8578, Japan}}
\altaffiltext{3}{Advanced Research Institute for Science and Engineering, Waseda University, 3-4-1 Okubo, Shinjuku, 169-8555, Japan}
\altaffiltext{4}{Yukawa Institute for Theoretical Physics, Kyoto University, Oiwake-cho, Kitashirakawa, Sakyo-ku, Kyoto, 606-8502, Japan}
\altaffiltext{5}{Science and Engineering, Waseda University, 3-4-1 Okubo, Shinjuku, 169-8555, Japan}


\begin{abstract}
The explosion mechanism of core-collapse supernovae has not been fully understood yet but multi-dimensional fluid instabilities such as standing accretion shock instability (SASI) and convection are now believed to be crucial for shock revival. Another multi-dimensional effect that has been recently argued is the asymmetric structures in progenitors, which are induced by violent convections in silicon/oxygen layers that occur before the onset of collapse, as revealed by recent numerical simulations of the last stage of massive star evolutions. Furthermore, it has been also demonstrated numerically that accretions of such non-spherical envelopes could facilitate shock revival. These two multi-dimensional may hence hold a key to successful explosions. In this paper, we performed a linear stability analysis of the standing accretion shock in core-collapse supernovae, taking into account non-spherical, unsteady accretion flows onto the shock to clarify the possible links between the two effects. We found that such pre-shock perturbations can excite the fluid instabilities efficiently and hence help the shock revive in core-collapse supernovae. 
\end{abstract}





\section{Introduction}
The explosion mechanism of core-collapse supernovae (CCSNe) is one of the biggest interests in high-energy astrophysics. The challenge is how the stagnant shock wave revives and goes through an iron core. The most anticipated scenario is based on the so-called neutrino-heating mechanism, in which fluid behind the shock gains energy from neutrinos that diffuse out of  the proto-neutron star (PNS) and as a result the shock is pushed outward. In this scenario, multi-dimensional fluid instabilities such as the standing accretion shock instability (SASI) and/or neutrino-driven convections are expected to play an important role to enhance the efficiency of the neutrino heating and produce shock revival. In fact, it is almost a consensus that one-dimensional, spherically symmetric simulations do not lead shock revival even with the most detailed physics implemented.

The multi-dimensional fluid instabilities in CCSNe have been systematically investigated by numerical simulations \citep{Burrows12, Bruenn13, Hanke13,Iwakami, IwakamiB, Takiwaki14,Abd,Fernandez15} and also by linear analysis \citep{Foglizzo07,Yamasaki07,Yamasaki08,Foglizzo09,Sato,Guilet12,Foglizzo15}. The neutrino-driven convection is characterized by small-scale multiple buoyant bubbles and is driven by negative entropy gradient caused by neutrino heating. SASI, on the other hand, is characterized by large-scale deformations of the shock surface induced by sloshing and spiral motions. It is a global instability caused by the so-called advective-acoustic cycle, in which the spherical shock surface is deformed by repetitious round trips of upcoming sonic waves and  down-going vorticity waves between the shock and PNS surface. Which of these instabilities, convection or SASI, dominates the shock dynamics is dependent on the mass accretion rate, neutrino luminosity, and hence the progenitors.

Multi-dimensionality is recently discussed not only in fluid instabilities but also in structures of progenitors, which was revealed by a series of papers by the groups of Arnett \citep{Arnett, Asida, Bazan, Meakin06, Meakin07} and \citet{Chat}. They showed numerically that the silicon and oxygen shells outside the iron core in progenitors substantially deviate from spherically symmetric configurations at the onset of collapse due to nuclear burnings and associated violent convections. They have found that the power spectrum of the turbulence peaks at $l = 4$ or $5$. Another \textcolor{black}{group} reported, however, the development of a larger-scale convection with $l=2$ in a different progenitor just before the collapse \citep{Muller16}. Their results imply that the patterns of the fluctuation may be dependent on the progenitor.

Such fluctuations in the outer shells will accrete onto the stagnant shock and may affect the subsequent shock dynamics. In fact, it is demonstrated by linear analysis \citep{Lai,KT} that the fluctuations can grow rather than damp during the super-sonic infall. Furthermore, \citet{Couch, Couch2,MJ14} and \citet{Couch3} showed that the interaction of such non-spherical outer layers with the stagnant shock can produce shock revival for the progenitors that failed to explode without the fluctuations. The systematic study by \citet{MJ14} reported that large scale velocity perturbations efficiently contribute to the shock revival. \citet{Fernandez14} also found that upstream perturbations with $l=1,2$ do not turn the SASI-dominant flows into convectively dominated flows .

Thanks to the above studies, the multi-dimensional structures of progenitors have been focused as a new key ingredient to successful explosions of CCSNe in addition to the fluid instabilities and other multi-dimensional players such as stellar rotation and magnetic fields \citep{Guilet10, IwakamiB, Sawai,Moesta,Takiwaki}. However, it has not been fully understood yet how the asymmetry in the envelopes interacts with the shock instabilities and eventually helps the stalled shock revive.

In this paper, we focus on the links between the shock instabilities and the asymmetric accretion flows that occur as consequences of the multi-dimensional structures of progenitors. We perform a linear stability analysis of the standing accretion shock by taking into account such asymmetric, unsteady accretion flows in front of the shock as the outer boundary condition, which has never been considered in previous liner analyses. We systematically study the dependence of the results on the typical frequency of the accreting fluctuations for several types of background flows, which are designed to crudely mimic the collapse of fluctuating progenitors. We also obtain some general relations analytically.

This paper is organized as follows. We explain the method and setups in the next section, where basic equations and Laplace transform are introduced together with our models. In Sec.~\ref{NumResults}, we present the results of the global linear stability analysis of the standing accretion shock. Discussions are given in Sec.~\ref{sec.discussion}. We summarize our investigation in Sec.~\ref{sec.summary2}. Following the main body, some appendices are added, which include analytic treatments that give the general relations between shock dynamics and perturbations.

\section{Method}\label{sec.2.method}
As stated in the introduction, we investigate the linear stability of the standing accretion shock against the multi-dimensional upstream perturbations.
We ignore the time variation of the (spherically symmetric, unperturbed) flows, since the typical timescales of the changes in the neutrino luminosity and/or mass accretion rate are much longer than the timescale of the instability.

In this section, we introduce the system of basic equations and the method of Laplace transform, which is the principal tool for the investigation of the linear analysis in this paper.

\subsection{Basic equations} \label{basic eq2}
The basic equations that govern accretion flows in the supernovae core are given as follows:
\begin{eqnarray}
\frac{\partial \rho}{\partial t} +{\bf \nabla }\cdot (\rho {\bf v}) = 0, \\
\frac{\partial }{\partial t}(\rho {\bf v}) + {\bf \nabla} \cdot (\rho {\bf v v} + p{\bf I}) = -\rho \frac{GM}{r^2}\frac{{\bf r}}{r}, \\
\frac{\diff \varepsilon}{\diff t} +p\frac{\diff }{\diff t}\left( \frac{1}{\rho} \right) = q, \\
\frac{\partial }{\partial t}(n Y_e) +{\bf \nabla}\cdot (nY_e {\bf v}) = \lambda, 
\end{eqnarray}
in addition to an equation of state (EoS). In the above expression, $\rho $, $p$, $n$, $Y_e$, $\varepsilon $ and ${\bf v}$ are density, pressure, number density, electron fraction, specific internal energy and velocity, respectively. 
The mass of the PNS is denoted by $M$, which is assumed to be constant, and $G$ is the gravitational constant. The self-gravity is neglected.
We incorporate the reactions between the electron-type neutrinos and matter, which are symbolically denoted by $q$ and $\lambda$: the former is the net heating rate and the latter is the net reaction rate of electrons and positrons.

We use the light-bulb approximation instead of solving the neutrino transport \citep{Ohnishi, Scheck}. In this prescription, neutrino luminosities ($L_{\nu _e}$ and $L_{\bar{\nu}_e}$) and temperatures ($T_{{\nu}_e}$ and $T_{\bar{{\nu}}_e}$) are arbitrary parameters. 
Then the radius of the neutrino sphere, $r_{\nu _e}$, is given by the following relation.
\begin{equation}
L_{\nu_e} = \frac{7}{16}4\pi r_{\nu_e}^2 \sigma T_{\nu_e}^4 ,
\end{equation}
where $\sigma $ is the Stefan-Boltzmann constant. The radius of the anti-electron neutrino sphere is also given in the same manner.
The unperturbed flows are given as spherically symmetric, time-independent solutions to these equations with appropriate boundary conditions.

Following \citet{Lai} and our previous paper, we linearize these equations for perturbed quantities written as
\begin{eqnarray}
\delta X ({\bf r},t) &=& \sum _{l,m} \delta X^{(l, m)} (r,t) Y_{lm}(\theta , \phi),  \\
\delta {\bf v} ({\bf r},t) &=& \sum _{l,m} \delta v_r^{(l, m)} (r,t) \Y \hat{{\bf r}} \nonumber \\
 &&+\delta v _\perp ^{(l, m)} (r,t) \left[\hat{{\bf \theta}} \frac{\partial Y_{lm}}{\partial \theta} + \frac{\hat{{\bf \phi}}}{\sin \theta} \frac{\partial Y_{lm}}{\partial \phi} \right] \nonumber \\
&&+\delta v_{rot} ^{(l, m)} (r,t)\left[ -\hat{{\bf \phi}} \frac{\partial Y_{lm}}{\partial \theta} + \frac{\hat{{\bf \theta}}}{\sin \theta} \frac{\partial Y_{lm}}{\partial \phi} \right], 
\end{eqnarray}
where $X$ denotes scalar variables. $\Y$ is the spherical harmonics with the indices $l, m$. $\hat{\bf r}, \hat{\bf \theta}$, and $\hat{\bf \phi}$ are the unit vectors in the spherical coordinates. 
For spherically symmetric background flows, the system of linearized equations for each combination of indices, $(l,m)$, is decoupled with each other and is given symbolically as follows:
\begin{equation}
\label{eq.linearized}
\frac{\partial \bf{y}}{\partial r} = A\frac{\partial \bf{y}}{\partial t} +B\bf{y} ,
\end{equation}
where ${\bf y} = {\bf y}(r,t)$ denotes the vector of perturbed quantities given as
\begin{equation}
\label{eq.perturbedstate}
{\bf y}(r,t) = \left(\frac{\delta \rho}{\rho _0}, \frac{\delta v_r}{v_{r0} }, \frac{\delta v_\perp}{v_{r0} }, \frac{\delta \varepsilon}{\varepsilon_0}, \frac{\delta Y_e}{Y_{e0} }, \frac{\delta v_{rot}}{v_{r0} } \right)^T,
\end{equation}
where $(\cdots)^T$ means a transposed vector and the subscript, $0$, denotes unperturbed states. Here and hereafter the superscript of indices is omitted for notational simplicity. The coefficient matrices, $A = A(r)$ and $B = B(r)$, whose components are made of the background quantities, are given in Appendix \ref{app.matrix}.
While the matrices are dependent on $l$, they are independent of $m$ thanks to the assumption of spherically symmetric  background.

We solve the linearized equations (\ref{eq.linearized}) in the region between the standing shock and the PNS surface which is assumed to coincide with the electron-type-neutrino sphere. The initial perturbed state at $t=0$ in the shocked region is given as
\begin{equation}
\label{eq.initial}
{\bf y}(r,t=0) = {\bf y}_0(r) \ \ (r_{\nu _e} < r < r_\mathrm{sh}),
\end{equation}
where $r_\mathrm{sh}$ is the unperturbed shock radius. 

The outer boundary condition imposed at the shock radius is given by the perturbed quantities in front of the shock through the linearized Rankine-Hugoniot relations, which are schematically given by
\begin{equation}
\label{eq.outerboundary}
{\bf y}(r_\mathrm{sh},t) = R{\bf z}(t) +\frac{\partial }{\partial t}\frac{\delta r_\mathrm{sh}}{r_\mathrm{sh}} {\bf c} +\frac{\delta r_\mathrm{sh}}{r_\mathrm{sh}} {\bf d},
\end{equation}
where $\delta r_\mathrm{sh}$ is the time-dependent perturbed shock radius; $R$ is another matrix, whose components are described only by background quantities; The components of the vectors, ${\bf c}$ and ${\bf d}$, are also given by the unperturbed states; Their explicit forms are found in Appendix \ref{app.matrix}; A given fluctuation in front of the shock is denoted by ${\bf z}(t)$.

The inner boundary is set at the PNS surface. We note here that we can impose only one condition, which determines a remaining degree of freedom at the outer boundary, i.e., the perturbed shock radius, once the upstream flow is given. Otherwise there is generally no solution that satisfies the inner and outer boundary conditions at the same time. Hence the inner boundary condition is symbolically represented as
\begin{equation}
\label{eq.innerboundary}
f({\bf y}(r_{\nu _e},t),t) = 0.
\end{equation}

To summarize, we solve the initial-boundary value problem that is described by Eqs.~(\ref{eq.linearized}), (\ref{eq.initial}), (\ref{eq.outerboundary}) and (\ref{eq.innerboundary}) to find the time-dependent shock radius, $\delta r_\mathrm{sh}/r_\mathrm{sh}$.

\subsection{Mode analysis in Laplace transform} \label{sec.modes}
To solve this initial-boundary value problem, we use the Laplace transform with respect to time (See \citet{Text} or \citet{KT} for properties of Laplace transform). 
Laplace transformed equations are given as
\begin{eqnarray}
\label{eq.L-linearized}
\frac{\diff {\bf y}^* }{\diff r} (r,s) &=& (sA+B){\bf y}^* -A{\bf y}_0(r), \\
\label{OB}
{\bf y}^*(r_\mathrm{sh},s) &=& (s{\bf c} + {\bf d})\frac{\delta r_\mathrm{sh}^* (s)}{r_\mathrm{sh}} +R{\bf z}^*(s), \\
\label{IB}
f^*({\bf y}^*(r_{\nu _e},s),s) &=& 0,
\end{eqnarray}
where $s$ is the conjugate variable of $t$ and is a complex number. The superscript, $*$, means Laplace-transformed variables, which are complex functions of $s$ in general.

We here emphasize that this is a system of ordinary differential equations with $s$ being a parameter. We find $\delta r_\mathrm{sh}^*/r_\mathrm{sh}$ for each $s$ so that both the inner and outer boundary conditions would be satisfied. We note that such a value of $\delta r_\mathrm{sh}^*/r_\mathrm{sh}$ is easily found for a given $s$ by integrating Eq.~(\ref{eq.L-linearized}) twice as explained in Appendix \ref{app.solvetheeigenvalueproblem}.

\begin{figure*}
 \begin{tabular}{cc}
  \begin{minipage}{0.5\hsize}
   \begin{center}
    \includegraphics[bb = 0 0 574 361, width = 80mm]{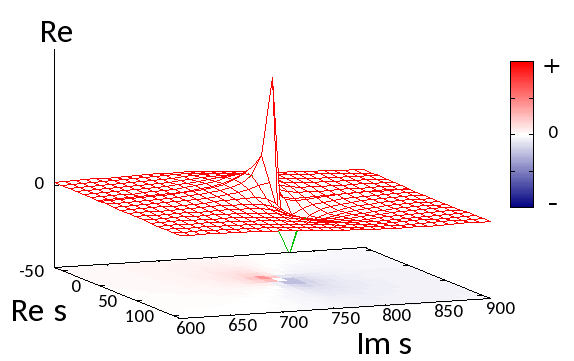}
   \end{center}
  \end{minipage} &
  \begin{minipage}{0.5\hsize}
   \begin{center}
    \includegraphics[bb = 0 0 574 361, width = 80mm]{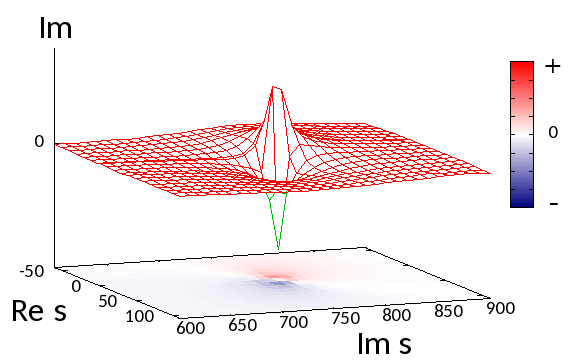}
   \end{center}
  \end{minipage}
 \end{tabular}
\caption{Close-up of sampled data points of a function $\delta r_\mathrm{sh}^*/r_\mathrm{sh}$ near a pole. Left and right panel show the real and imaginary part respectively.
The $z$-axes are in linear scale and a color map is projected on the bottom. }
\label{peakReIm}
\end{figure*}

The eigen modes, both stable or unstable, are found in the following way. 
We first assume the functional form of the perturbed shock radius as
\begin{equation}
\frac{\delta r_\mathrm{sh}}{r_\mathrm{sh}}  = \sum _{j} a_je^{\Omega_j t}\sin(\omega _jt +\phi _j),
\end{equation}
where $\Omega _j$ and $\omega _j$ are the growth/damping and oscillation frequency of the $j$-th mode ($j = 1, 2,3,\cdots $), respectively, and $a_j$ is the amplitude. To ensure the uniqueness of this expansion, we assume that $\omega _j \ge 0$ and $-\pi/2 \le \phi _j < \pi /2$ for any $j$.
Performing Laplace transform, we obtain
\begin{eqnarray}
\label{shock_evolution}
\frac{\delta r_\mathrm{sh}^*}{r_\mathrm{sh}}  = \sum _j a_j\frac{(s -\Omega _j)\sin\phi _j+\omega _j\cos \phi _j}{(s-\Omega _j)^2+\omega _j^2},
\end{eqnarray}
which has single poles at $s = \Omega _j \pm i\omega_j$. Here we used a relation:
\begin{eqnarray}
\label{Lsin}
\mathcal{L} [ e^{\Omega t}\sin(\omega t +\phi ) ] = \frac{(s -\Omega)\sin\phi +\omega \cos \phi}{(s-\Omega)^2+\omega^2}.
\end{eqnarray}
Hence we know the growth/damping rate and eigen frequency of a mode from the position of the corresponding pole.
The poles of $\delta r_\mathrm{sh}^*/r_\mathrm{sh}$ are found, on the other hand, by solving the initial-boundary value problem (\ref{eq.L-linearized})-(\ref{IB}) for a series of $s$. Since the functional value changes near a pole as show in Fig.~\ref{peakReIm}, for example,  we can easily find poles by observing the behavior of the function as long as the grid on the complex plane is sufficiently fine.\footnote{One of the strategies that work well is to find a local maximum of the absolute value of  $\delta r_\mathrm{sh}/r_\mathrm{sh}$.}

Incidentally, we obtain the time evolution of the shock radius by performing inverse Laplace transform for $\delta r_\mathrm{sh}^*/r_\mathrm{sh}$ numerically. Note that the inverse transform should be done by an integral along such a path that is parallel to the imaginary axis and all singularities lie to the left of the line \citep[][see also Appendix \ref{app.inverse}]{Text}. Furthermore, the obtained function is necessarily $0$ for $t<0$. If inappropriate path is chosen for the integral, i.e., if some poles are missed, the time evolution will not remain stationary at $t<0$. We would hence never miss the pole with the maximal growth rate if we look into the time evolution at $t<0$, which is one of the advantages of the Laplace-transform method.

\subsection{Models and parameters}\label{sec.model}
The initial-boundary value problem (\ref{eq.L-linearized})-(\ref{IB}) is solved for various background flows and perturbations that accrete onto the shock surface. A background state is characterized by the following parameters: Throughout the paper, a constant mass of the PNS, $M = 1.4$ M$_\odot$, is used. The mass accretion rate and neutrino luminosities are fixed as $\dot{M} = 0.6$ M$_\odot$  s$^{-1}$ and $T_{\nu_e} = T_{\bar{\nu}_e} = 4.5$ MeV. The accretion flow ahead of the shock is assumed to be a free fall of irons whose entropy and electron fraction are given as $S = 3k_B$ and $Y_e = 0.5$, where $k_B$ is the Boltzmann constant.  We employ Shen's EoS \citep{Shen}, which takes into account the contributions from nucleons, nuclei, photons, electron and positrons. For neutrino heating and cooling, the reaction rates of \citet{Bruenn} are adopted.
The inner boundary is assumed to coincide with the neutrino sphere, where the density is fixed to $10^{11}$ g cm$^{-3}$. Neutrino luminosities are also model parameters and some value is given to $L_\nu := L_{\nu _e}= L _{\bar{\nu}_e}$ for each.

\begin{table*}
\begin{center}
\caption{Background Flow Models\label{tbl-model}}
\begin{tabular}{ccccccccccc}
\tableline\tableline
Model & $L_\nu$ & $r_\mathrm{sh} $ & $r_\nu $ & $\omega _\mathrm{aac} $ & $\omega _\mathrm{pac}$ &  $r_\mathrm{gain}$& $\bar{\omega} _{BV} $&$\chi$ & \\
& [$10^{52}$ erg s$^{-1}$] & [$10^6$ cm] & [$10^6$ cm] & [ms$^{-1}$] & [ms$^{-1}$] & [$10^6$ cm] & [ms$^{-1}$] & & \\ 
\tableline
M06L20 &2.0 & 4.75 & 2.94 & 1.56 &5.90 &4.75 & - &0 \\
M06L25 &2.5 & 5.79 & 3.28 & 0.973 &3.99 & 5.79 & - &0 \\
M06L30 &3.0 & 6.93 & 3.60 & 0.644 &2.84  &6.45 & 0.270 &0.143 \\
M06L35 &3.5 & 8.22 & 3.89 &  0.444 &2.07  &6.80  & 0.320 &0.603 \\
M06L40 &4.0 &  9.78 & 4.15 &   0.313 & 1.52 & 7.18 & 0.340  &1.41 \\
M06L45 &4.5 & 11.8  & 4.41 & 0.221 & 1.10  & 7.73 & 0.332 &2.56 \\
M06L50 &5.0 & 14.6 & 4.64 &  0.152 & 0.754 & 8.44 & 0.302 &4.31 \\
M06L55 &5.5 & 19.3 & 4.87 & 0.0974 & 0.470 & 9.42 & 0.238 &6.77 \\
M06L60 &6.0 & 29.5 & 5.09 & 0.0510 & 0.232 & 10.8 & 0.153 &11.1 \\
\tableline
\end{tabular}
\end{center}
\end{table*}

We list up in Table \ref{tbl-model} the background-flow models that are representatively used in the study. 
In the table, $\omega_\mathrm{aac}$ and $\omega _\mathrm{pac}$ are characteristic frequencies of advective-acoustic cycle and purely acoustic cycle \citep{Blondin}, respectively, which are given as
\begin{eqnarray}
\omega _\mathrm{aac} &=& 2\pi \left[ \int _{r_{\nu }}^{r_\mathrm{sh}} \left( \frac{1}{|v_r|} + \frac{1}{c_s - |v_r|}\right)\diff r \right]^{-1} , \\
\omega _\mathrm{pac} &=& 2\pi \left[ \int _{r_{ \nu }}^{r_\mathrm{sh}} \left( \frac{1}{c_s + |v_r|} + \frac{1}{c_s - |v_r|}\right)\diff r \right]^{-1} ,
\end{eqnarray}
where $c_s$ is the sound speed and $r_\nu = r_{ \nu _e} = {r_{\bar{ \nu }_e}}$ is the neutrino sphere.
We also gave the $\chi$-parameter \citep{Foglizzo06} in the list, which is defined as
\begin{equation}
\chi = \int _{r_\mathrm{gain}}^{r_\mathrm{sh}} \left| \frac{\omega _{BW}}{v_r}\right| \diff r,
\end{equation}
where $r_\mathrm{gain}$ is the gain radius, i.e., the bottom boundary of the region with negative entropy gradients ($r_\mathrm{gain} \le r \le r_\mathrm{sh}$), and $\omega _{BW}$ is the Brunt-V\"{a}is\"{a}l\"{a} frequency:
\begin{eqnarray}
\omega _{BV} &=& \sqrt{\frac{GM}{r^2}\left|\frac{1}{\Gamma _1p}\frac{\diff p}{\diff r} -\frac{1}{\rho}\frac{\diff \rho}{\diff r}\right|},
\end{eqnarray}
with
\begin{eqnarray}
\Gamma _1 &:=& \left( \frac{\partial \ln  p}{\partial \ln \rho}\right) _{S,Y_e}.
\end{eqnarray}
According to \citet{Foglizzo06}, $\chi \gtsim 3$ is the condition for the flow being convectively unstable. We also list the mean Brunt-V\"{a}is\"{a}l\"{a} frequency for models with a non-vanishing gain region, which is defined by
\begin{eqnarray}
\bar{\omega} _{BV} = \frac{1}{r_\mathrm{sh} -r_\mathrm{gain}}\int _{r_\mathrm{gain}}^{r_\mathrm{sh}}\omega _{BV} \diff r .
\end{eqnarray}

The linearized equations are solved for an inner boundary condition: $\delta v_r = 0$. 
As for the outer boundary condition, based on the numerical result of \citet{KT}, we set the perturbed flow that accretes onto the shock surface approximately as
\begin{eqnarray}
\label{EXP1}
\frac{\delta \rho}{\rho _0} &= \sin (\omega _\mathrm{up} t +\varphi ), \\
\frac{\delta v_r}{v_{r0}} &= -0.5\sin (\omega _\mathrm{up} t +\varphi ), \\
\label{EXP3}
\frac{\delta \varepsilon}{\varepsilon _0} &= \sin(\omega _\mathrm{up} t +\varphi).
\end{eqnarray}
\textcolor{black}{According to their results, the time evolutions of perturbed quantities at a fixed radius is not a simple harmonic oscillation. If we focus on the timescale of a few hundred milliseconds, which is relevant for the development of SASI and shock revival itself, however, the approximation by the sinusoidal function is reasonable, since the temporal variation of the perturbation at the shock front is dictated mostly by the structure of the unperturbed accretion flow and is rather insensitive to the initial fluctuation in the envelope. See \citet{KT} for more details. It was found in the same paper that the typical frequency is in the range of 10-100 s$^{-1}$ and tends to become larger with $l$. In this study we regard the oscillation frequency as a free parameter and vary it in this range.}

\section{Numerical results}\label{NumResults}
In this paper, we mainly investigate unstable eigen modes and the effects of upstream fluctuations on them. We start with the intrinsic instabilities, by which we mean the unstable eigen modes in the absence of the upstream perturbations. We then show the latter effects.

\subsection{Intrinsic instabilities}
\begin{figure*}
 \begin{tabular}{cc}
  \begin{minipage}{0.45\hsize}
   \begin{center}
    \includegraphics[bb = 0 0 640 480, width = 80mm]{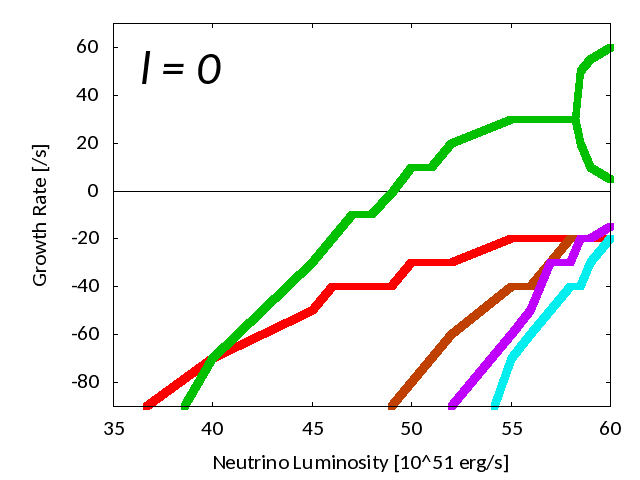}
   \end{center}
  \end{minipage} &
  \begin{minipage}{0.45\hsize}
   \begin{center}
    \includegraphics[bb = 0 0 640 480, width = 80mm]{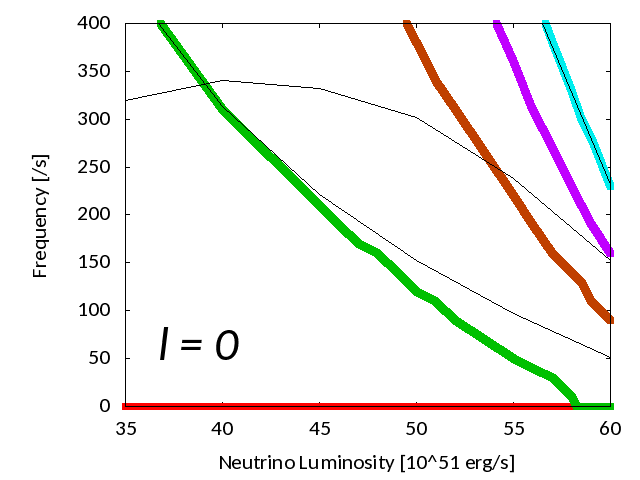}
   \end{center}
  \end{minipage}
 \end{tabular}
\caption{Plots of the eigenvalues of some spherically symmetric modes as functions of neutrino luminosity, $L_\nu$. Left: growth rates of the fundamental mode (red line) and higher overtones (green, brown, purple, light blue for first, second, third and fourth overtones, respectively). The horizontal black line shows $\Omega = 0$. Right: corresponding oscillation frequencies. Colors have the same meanings. Two thin black lines that are convex downward indicate $\omega _\mathrm{aac}$ (lower one) and $\omega _\mathrm{pac}$ (upper one) while the other black curve, which is convex upward, shows $\bar{\omega} _\mathrm{BV}$. The accretion rate is fixed to $\dot{M} = 0.6$ M$_\odot$ s$^{-1}$.} 
\label{Sphe}
\end{figure*}

\begin{figure*}
\begin{tabular}{cc}
\begin{minipage}{0.45\hsize}
\begin{center}
\includegraphics[bb = 0 0 640 480, width = 80mm]{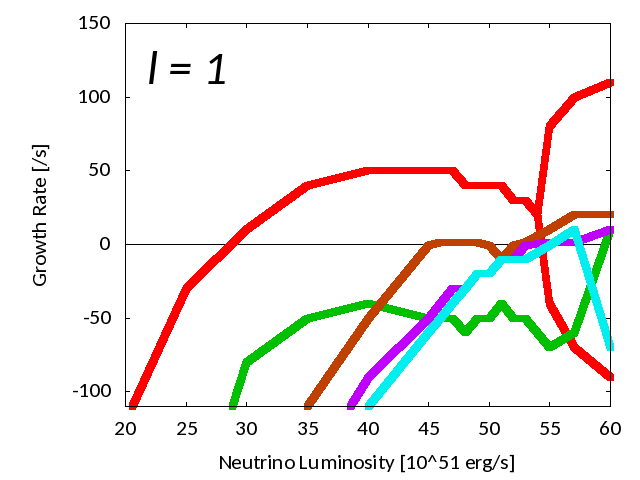}
\end{center}
\end{minipage} &
\begin{minipage}{0.45\hsize}
\begin{center}
\includegraphics[bb = 0 0 640 480, width = 80mm]{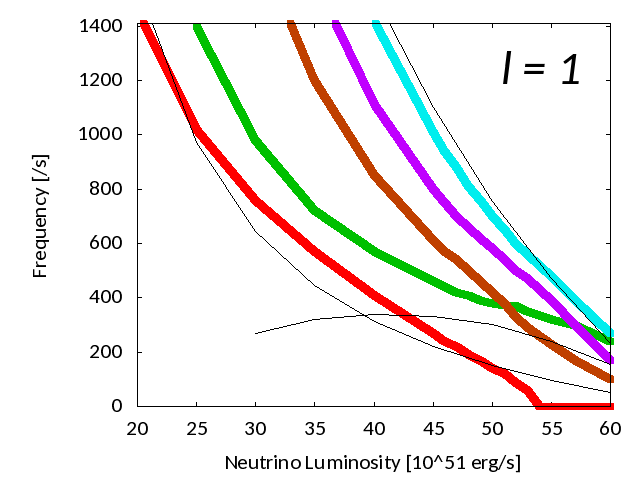}
\end{center}
\end{minipage} \\
\begin{minipage}{0.45\hsize}
\begin{center}
\includegraphics[bb = 0 0 640 480, width = 80mm]{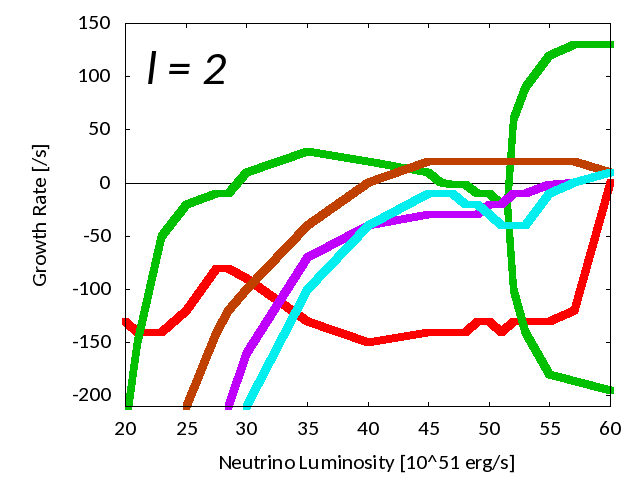}
\end{center}
\end{minipage} &
\begin{minipage}{0.45\hsize}
\begin{center}
\includegraphics[bb = 0 0 640 480, width = 80mm]{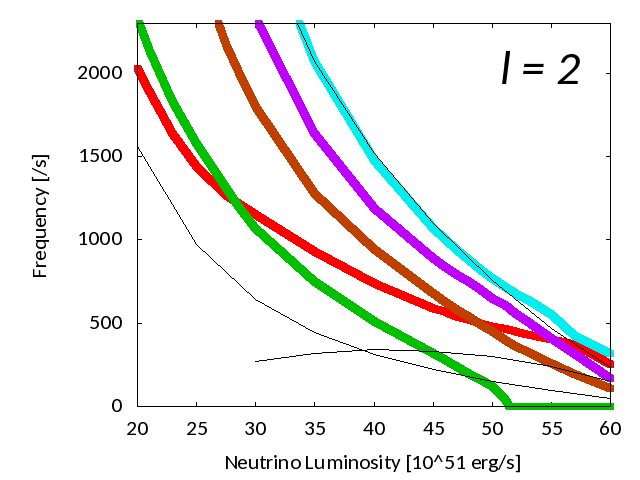}
\end{center}
\end{minipage} \\
\begin{minipage}{0.45\hsize}
\begin{center}
\includegraphics[bb = 0 0 640 480, width = 80mm]{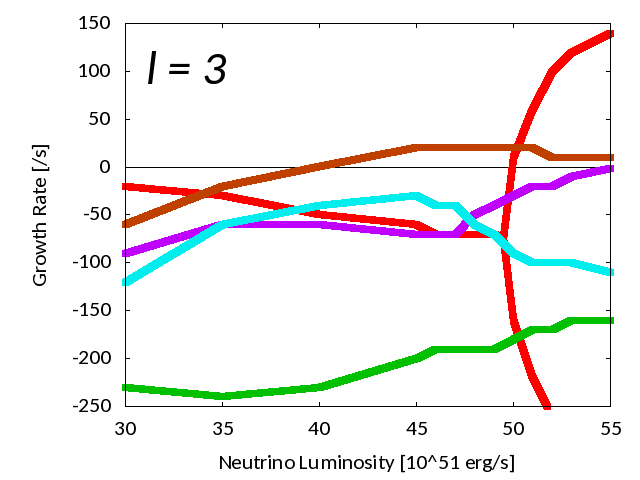}
\end{center}
\end{minipage} &
\begin{minipage}{0.45\hsize}
\begin{center}
\includegraphics[bb = 0 0 640 480, width = 80mm]{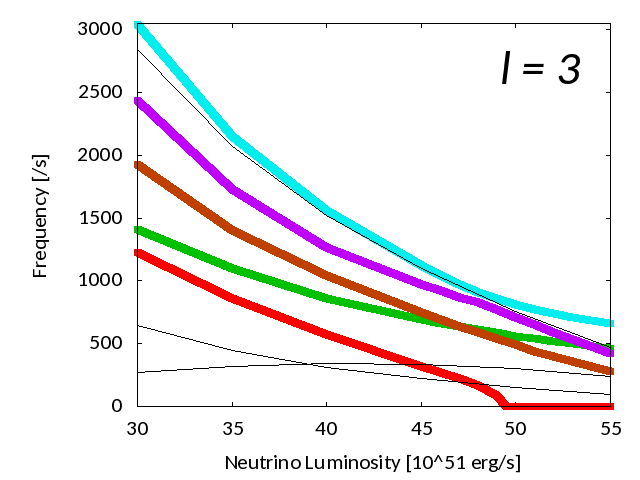}
\end{center}
\end{minipage} 
\end{tabular}
\caption{Same as Fig.~\ref{Sphe} but for some non-spherically symmetric modes with $l = 1,\ 2,\ 3$.}
\label{Higher}
\end{figure*}

\begin{figure*}
\begin{tabular}{cc}
\begin{minipage}{0.45\hsize}
\begin{center}
\includegraphics[bb = 0 0 640 480, width = 80mm]{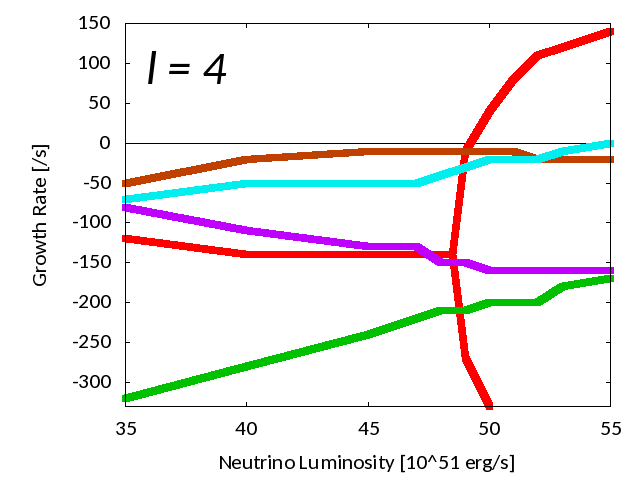}
\end{center}
\end{minipage} &
\begin{minipage}{0.45\hsize}
\begin{center}
\includegraphics[bb = 0 0 640 480, width = 80mm]{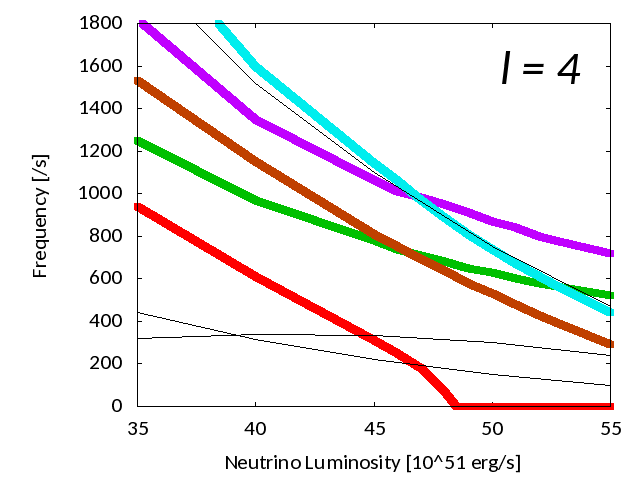}
\end{center}
\end{minipage} \\
\begin{minipage}{0.45\hsize}
\begin{center}
\includegraphics[bb = 0 0 640 480, width = 80mm]{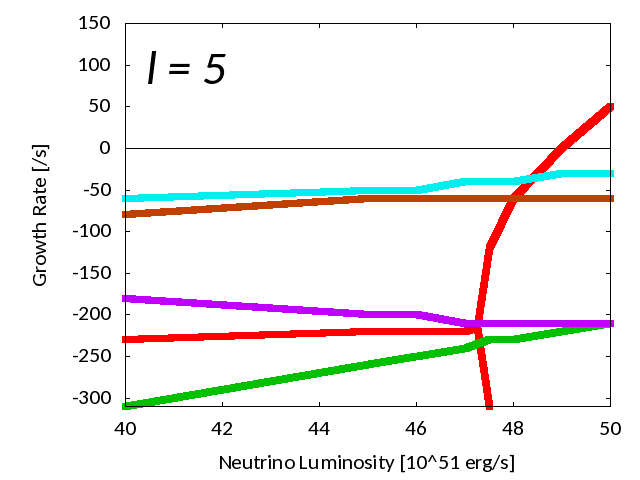}
\end{center}
\end{minipage} &
\begin{minipage}{0.45\hsize}
\begin{center}
\includegraphics[bb = 0 0 640 480, width = 80mm]{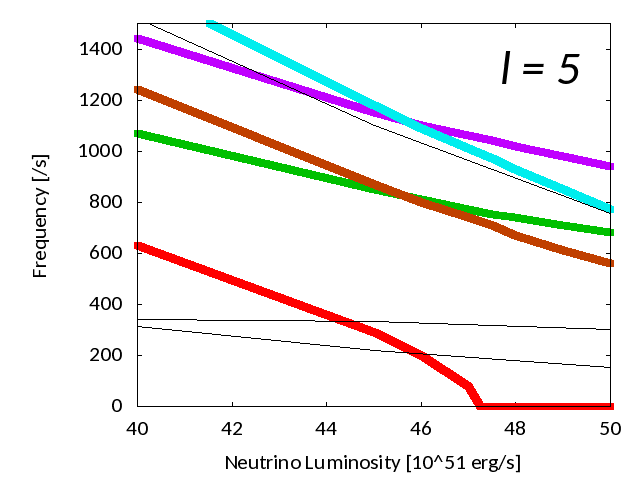}
\end{center}
\end{minipage} \\
\begin{minipage}{0.45\hsize}
\begin{center}
\includegraphics[bb = 0 0 640 480, width = 80mm]{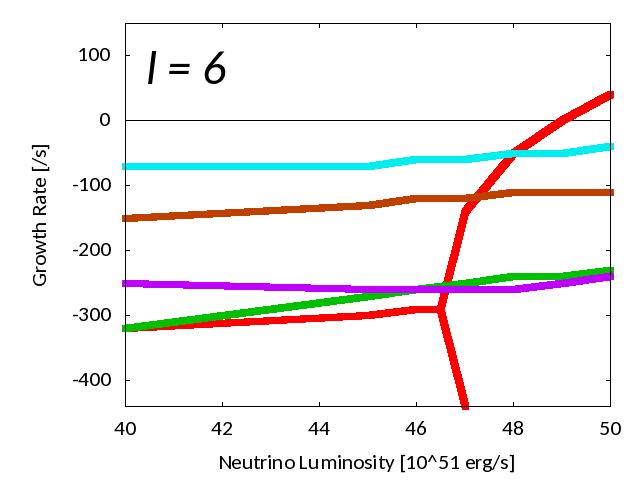}
\end{center}
\end{minipage} &
\begin{minipage}{0.45\hsize}
\begin{center}
\includegraphics[bb = 0 0 640 480, width = 80mm]{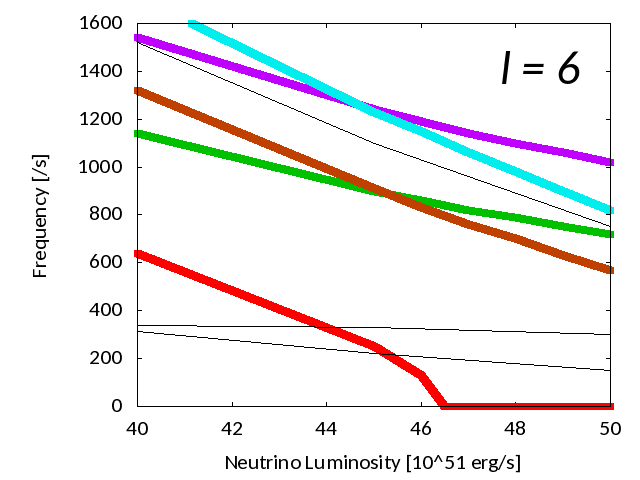}
\end{center}
\end{minipage} 
\end{tabular}
\caption{Same as Figs.~\ref{Sphe} and \ref{Higher} but for $l = 4,\ 5,\ 6$.}
\label{Higher2}
\end{figure*}

\begin{figure*}
 \begin{tabular}{cc}
  \begin{minipage}{0.45\hsize}
   \begin{center}
    \includegraphics[bb = 0 0 640 480, width = 80mm]{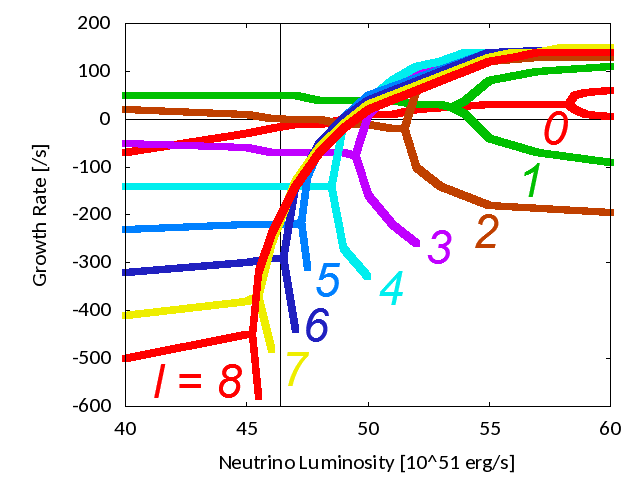}
   \end{center}
  \end{minipage} &
  \begin{minipage}{0.45\hsize}
   \begin{center}
    \includegraphics[bb = 0 0 640 480, width = 80mm]{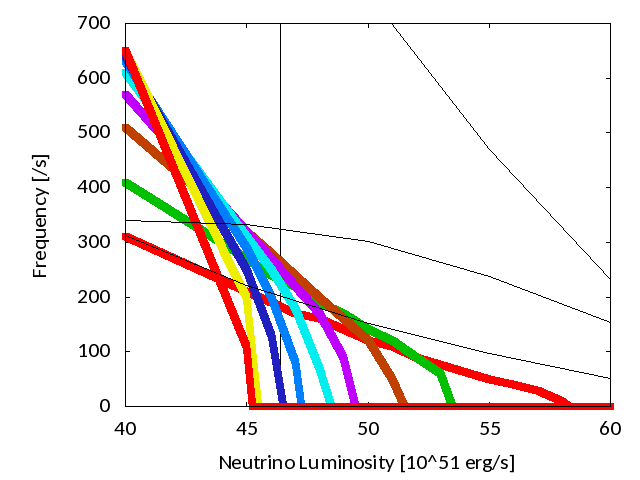}
   \end{center}
  \end{minipage}
 \end{tabular}
\caption{Growth rates (left) and oscillation frequencies (right) for bifurcated modes with $0 \le l \le 8$. In the left panel the horizontal black line shows $\Omega = 0$ and the vertical one indicates the luminosity where the $\chi$-parameter is equal to $3$. In the right panel the vertical black line is the same as in the left panel whereas the three black curves are the same as in the right panels in Figs.~\ref{Sphe}, \ref{Higher} and \ref{Higher2}.}
\label{Higher3}
\end{figure*}

As shown in the next section, the growth/damping rate and oscillation frequency of eigen modes are not changed by the upstream perturbations to the linear order. 
We can hence extract these information of the intrinsic modes by neglecting the upstream perturbation (the proof is given in the next section).
It is then equivalent to the ordinary normal mode analysis of SASI and/or convection with Fourier transform and it turns out that the results are qualitatively consistent with those of \citet{Yamasaki07} although they used different background flows with a higher mass accretion rate.

We first mention the spherically symmetric mode ($l = 0$), which is shown in Fig.~\ref{Sphe} for various neutrino luminosities. Different modes are distinguished with colors. They are normally classified according to the number of radial nodes in the corresponding eigen functions: the mode with the smallest number of nodes is called the fundamental mode while the others are referred to as the first-, second-, third-, $\cdots$, overtones as the number of nodes increases \citep{Yamasaki07}. Since we do not obtain eigen functions directly in our method, we refer to the mode with the lowest oscillation frequency as the fundamental mode and call other eigen modes as first-, second-, third overtones and so on in the ascending order of the oscillation frequency.

As is apparent in the left panel of Fig.~\ref{Sphe}, the background flow is stable for radial perturbations as long as the neutrino luminosity is low. It becomes unstable, however, once the luminosity reaches a threshold value, $L_\nu \sim 4.9 \times 10^{52}$ erg s$^{-1}$, where the first overtone becomes unstable. 
For much higher luminosities, the first overtone bifurcates into two branches and turns to be non-oscillatory. Bifurcations are also seen in higher-$l$ modes and we discuss the physical interpretation later. The fundamental mode of $l=0$ is always non-oscillatory irrespective of neutrino luminosities and was identified as a thermal mode by \citet{Yamasaki07}, which has no counterpart for higher-$l$ modes and will disappear if we turn off the perturbation of the heating rate.\footnote{This is what \citet{Yamasaki07} found in their paper. Although we have not confirmed it in this paper, we expect that it will be shared by the mode we found here.} 

Next we consider non-spherically symmetric modes in Figs.~\ref{Higher} and \ref{Higher2}, where $l = 1,\cdots,6$ modes are shown. 
We note that the magnitudes of oscillation frequencies for different modes can interchange one another as the neutrino luminosity increases. This was not observed in \citet{Yamasaki07}.\footnote{We apply the same naming rule here as for $l = 0$ at small $L_\nu$ although we do not know if it really corresponds to what it suggests at high luminosities. It is stressed, however, that the number of radial nodes is never essential for our analysis and the name of each mode is employed just for the sake of convenience.}
As demonstrated clearly, the modes with $l = 1,2$ become unstable first as the luminosity increases and the growth rate is highest for for the $l=1$ mode as long as the neutrino luminosity is less than $L_\nu \sim 5 \times 10^{52}$ erg s$^{-1}$.
They are SASI modes, which are likely to be driven by the advective-acoustic cycle, since the oscillation frequencies of these unstable modes follow more closely those of the advective-acoustic-cycle, $\omega _\mathrm{aac}$, rather than those of the purely-acoustic-cycle, $\omega _\mathrm{pac}$, or of the mean Brunt-V\"{a}is\"{a}l\"{a} frequencies, $\bar{\omega }_{BV}$.

As the neutrino luminosity becomes higher, the fundamental mode and the first overtone in some cases bifurcates into two branches and become non-oscillatory as does the first overtone with $l=0$. The growth rate in one of the two branches increases rapidly thereafter and higher-$l$ modes become dominant as a result. 
To compare these bifurcated modes, we compared them for $0 \le l \le 8$ in Fig.~\ref{Higher3}. These bifurcated non-spherically symmetric modes may be interpreted as convective modes \citep{Yamasaki07}. 
It is found that the value of $\chi$ parameter at the point when one of the bifurcated modes first becomes unstable is $\sim  4$ in our models, which seems to support the claim by \citet{Foglizzo06}.\footnote{There seems to have been some mistakes in \citet{Yamasaki07} in their evaluation of the $\chi$ parameter, since the rather minor difference between their models and ours does not appear to account for the discrepancy in the values of the $\chi$ parameter.}
It is also interesting to point out that the bifurcation seems to occur when the oscillation frequency falls below the frequency of the advective-acoustic cycle, $\omega _\mathrm{aac}$, as seen in Figs.~\ref{Sphe},  \ref{Higher}, \ref{Higher2} and \ref{Higher3}.\footnote{In fact, the bifurcation is observed not only in the fundamental mode but also in the first overtones with $l = 7,8$ at much higher luminosities, where their oscillation frequencies drop below $\omega _\mathrm{aac}$.}

Since the spherically symmetric mode becomes unstable at $L_\nu \sim 4.9 \times 10^{52}$ erg s$^{-1}$ and its growth rate is smaller than the convective modes, non-spherical modes are more important practically in the supernova explosion. SASI is dominant in the low luminosity regime while convection overwhelms SASI in the high $L_\nu$ regime. These results are in agreement qualitatively with what we have observed in many realistic simulations \citep[e.g.][]{Burrows12, Hanke13, Iwakami,IwakamiB} as well as with the previous linear analysis \citep{Yamasaki07}.

\subsection{Upstream perturbations}
\begin{figure*}
\begin{tabular}{cc}
\begin{minipage}{0.45\hsize}
\begin{center}
\includegraphics[bb = 0 0 640 380, width = 79mm]{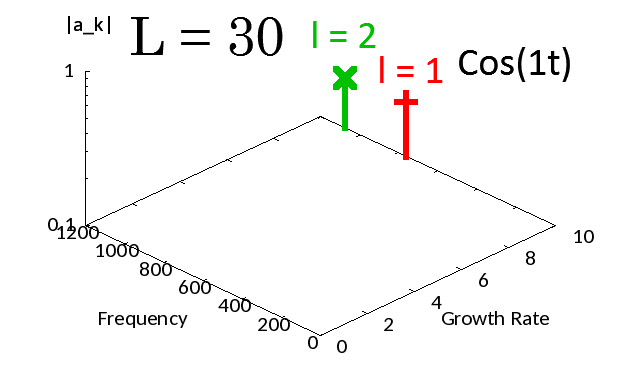}
\end{center}
\end{minipage} &
\begin{minipage}{0.45\hsize}
\begin{center}
\includegraphics[bb = 0 0 640 380, width = 79mm]{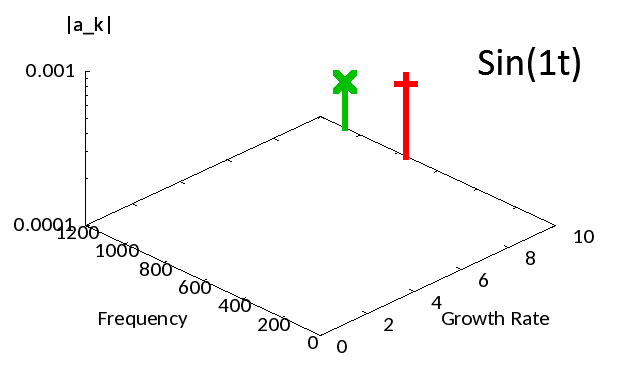}
\end{center}
\end{minipage} \\
\begin{minipage}{0.45\hsize}
\begin{center}
\includegraphics[bb = 0 0 640 380, width = 79mm]{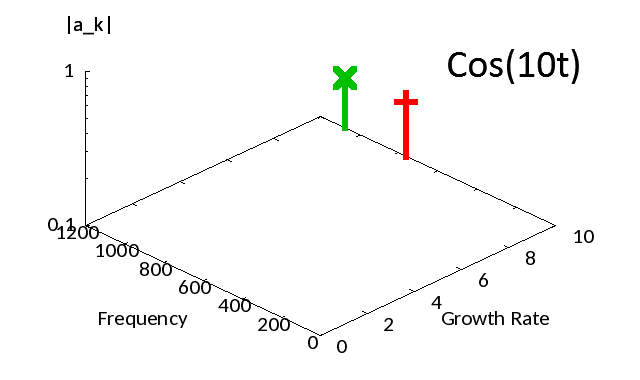}
\end{center}
\end{minipage} &
\begin{minipage}{0.45\hsize}
\begin{center}
\includegraphics[bb = 0 0 640 380, width = 79mm]{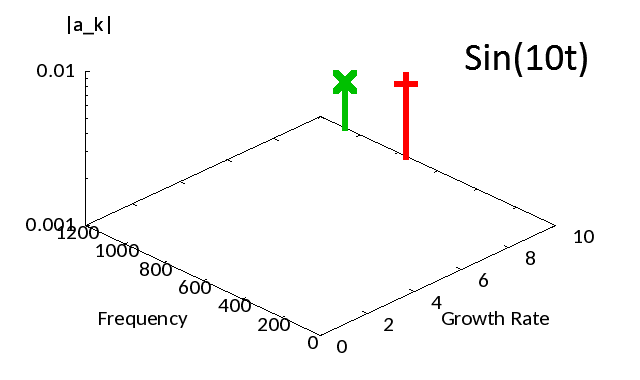}
\end{center}
\end{minipage} \\
\begin{minipage}{0.45\hsize}
\begin{center}
\includegraphics[bb = 0 0 640 380, width = 79mm]{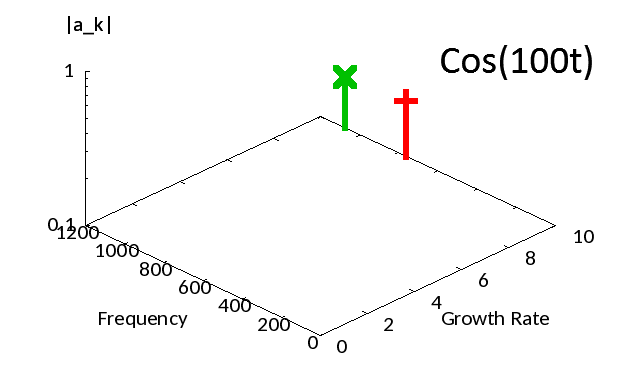}
\end{center}
\end{minipage} &
\begin{minipage}{0.45\hsize}
\begin{center}
\includegraphics[bb = 0 0 640 380, width = 79mm]{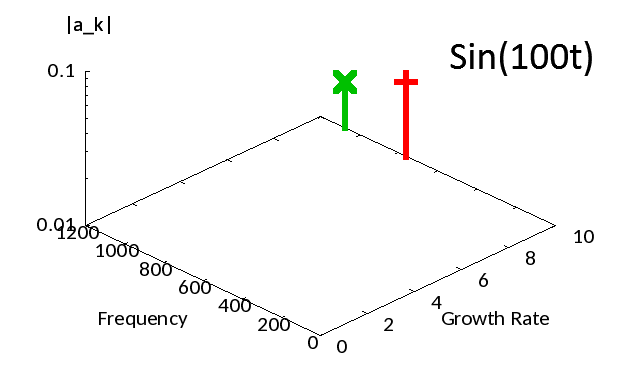}
\end{center}
\end{minipage} \\
\begin{minipage}{0.45\hsize}
\begin{center}
\includegraphics[bb = 0 0 640 380, width = 79mm]{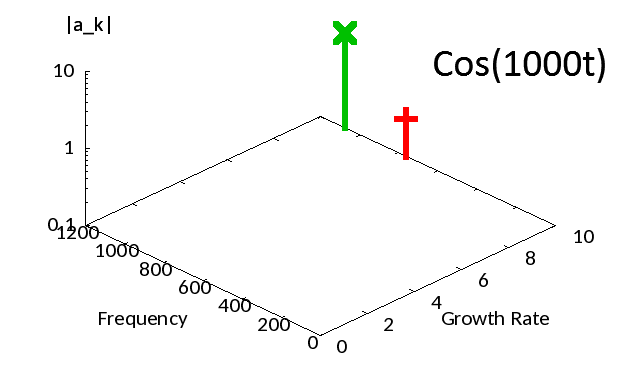}
\end{center}
\end{minipage} &
\begin{minipage}{0.45\hsize}
\begin{center}
\includegraphics[bb = 0 0 640 380, width = 79mm]{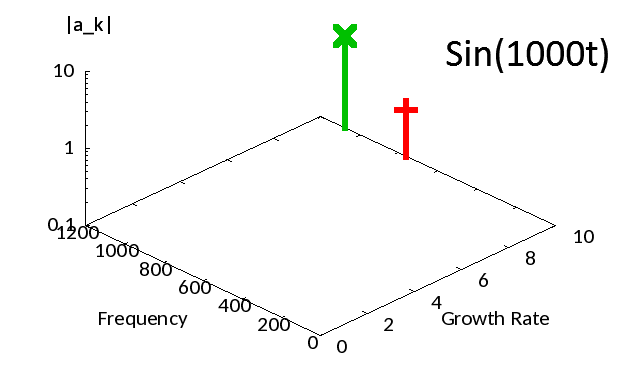}
\end{center}
\end{minipage} \\
\begin{minipage}{0.45\hsize}
\begin{center}
\includegraphics[bb = 0 0 640 380, width = 79mm]{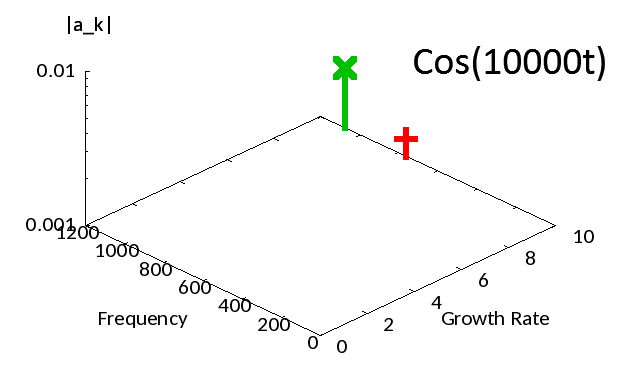}
\end{center}
\end{minipage} &
\begin{minipage}{0.45\hsize}
\begin{center}
\includegraphics[bb = 0 0 640 380, width = 79mm]{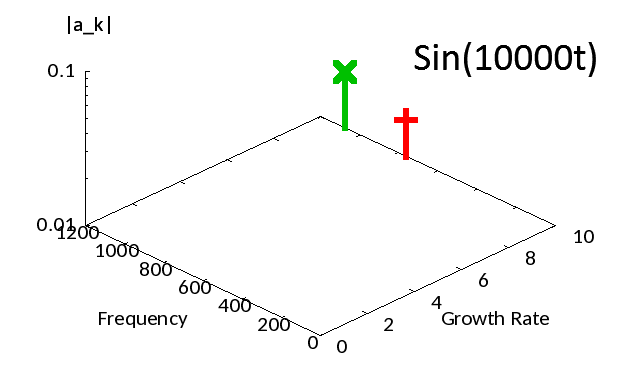}
\end{center}
\end{minipage} 
\end{tabular}
\caption{The absolute values of the excited amplitudes of the unstable SASI modes for $L_\nu = 3\times 10^{52}$ erg s$^{-1}$ that are excited by upstream fluctuations given in Eqs.~(\ref{EXP1})-(\ref{EXP3}). Left panels show the results for the cosine-type perturbations, i.e., $\varphi = \pi/2$, while the right panels are for the sine-type fluctuations, $\varphi = 0$. From top to bottom, the frequency of the upstream perturbation changes from $\omega _\mathrm{up}= 1$ s$^{-1}$ to $\omega _\mathrm{up}= 10^4$ s$^{-1}$ as indicated on the upper right corner of each panel. The vertical line below each plus/cross point indicates its position on the $(x,y)$ plane, which corresponds to the growth rate and oscillation frequency of the unstable mode. Different colors correspond to modes with different $l$.}
\label{SASIL30}
\end{figure*}

\begin{figure*}
\begin{tabular}{cc}
\begin{minipage}{0.45\hsize}
\begin{center}
\includegraphics[bb = 0 0 640 380, width = 79mm]{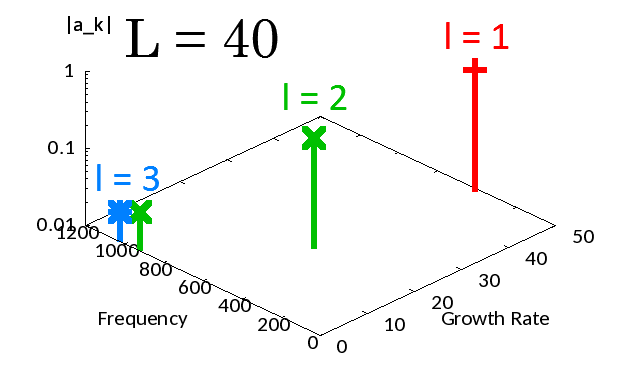}
\end{center}
\end{minipage} &
\begin{minipage}{0.45\hsize}
\begin{center}
\includegraphics[bb = 0 0 640 380, width = 79mm]{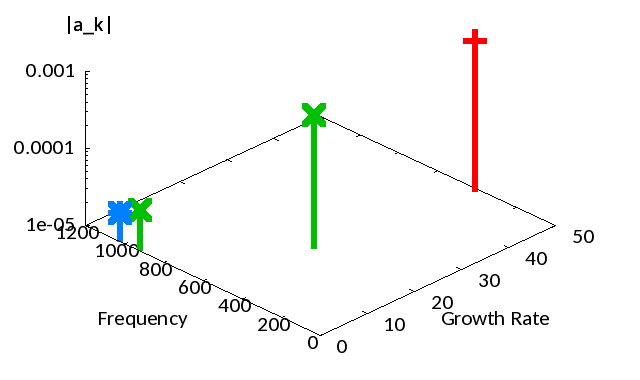}
\end{center}
\end{minipage} \\
\begin{minipage}{0.45\hsize}
\begin{center}
\includegraphics[bb = 0 0 640 380, width = 79mm]{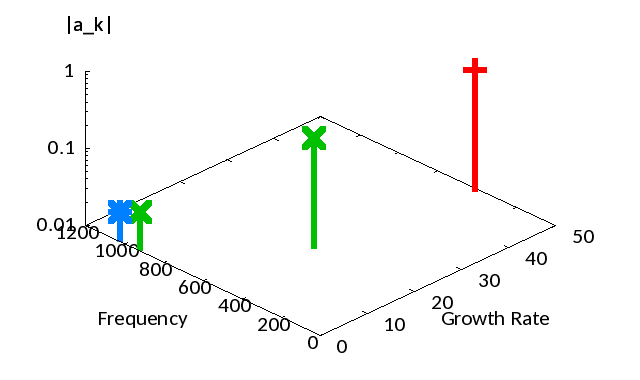}
\end{center}
\end{minipage} &
\begin{minipage}{0.45\hsize}
\begin{center}
\includegraphics[bb = 0 0 640 380, width = 79mm]{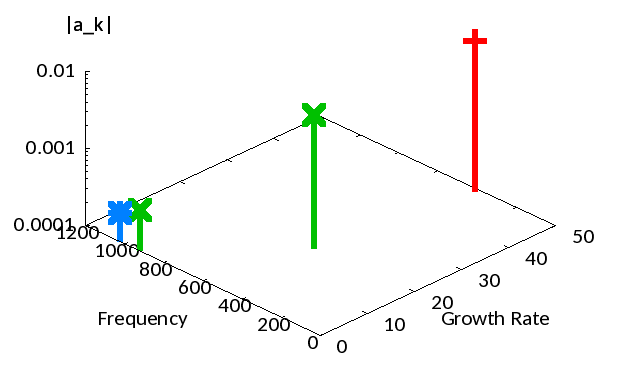}
\end{center}
\end{minipage} \\
\begin{minipage}{0.45\hsize}
\begin{center}
\includegraphics[bb = 0 0 640 380, width = 79mm]{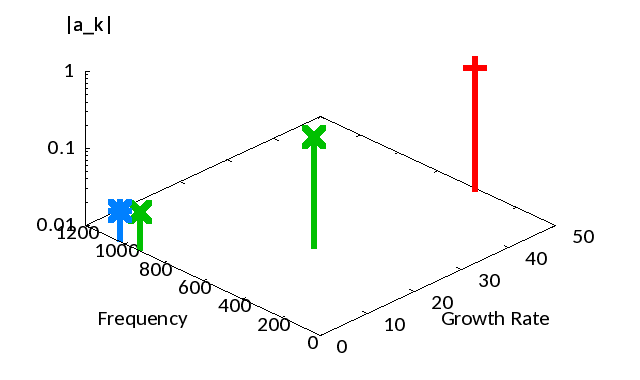}
\end{center}
\end{minipage} &
\begin{minipage}{0.45\hsize}
\begin{center}
\includegraphics[bb = 0 0 640 380, width = 79mm]{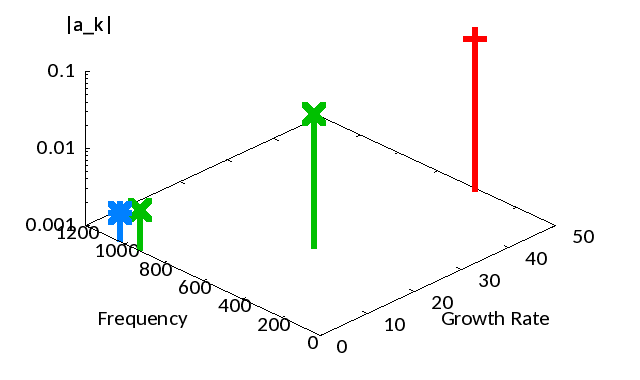}
\end{center}
\end{minipage} \\
\begin{minipage}{0.45\hsize}
\begin{center}
\includegraphics[bb = 0 0 640 380, width = 79mm]{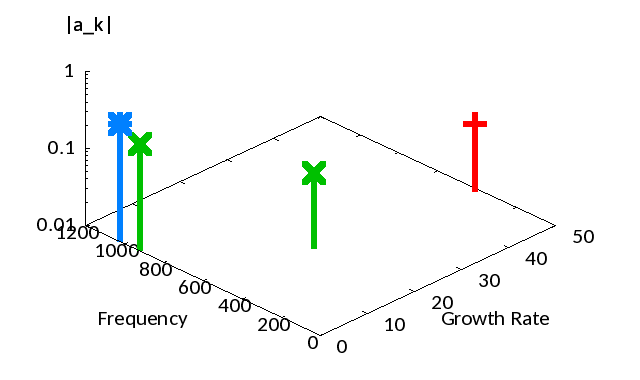}
\end{center}
\end{minipage} &
\begin{minipage}{0.45\hsize}
\begin{center}
\includegraphics[bb = 0 0 640 380, width = 79mm]{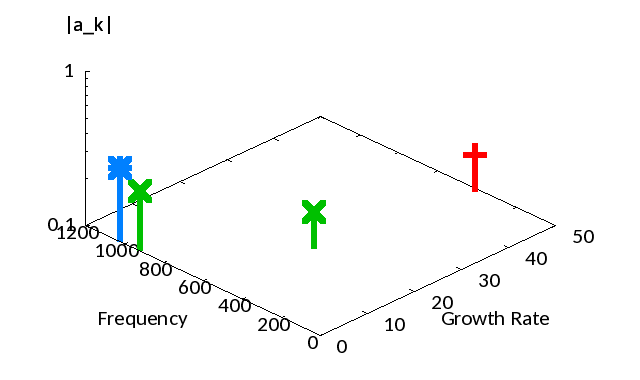}
\end{center}
\end{minipage} \\
\begin{minipage}{0.45\hsize}
\begin{center}
\includegraphics[bb = 0 0 640 380, width = 79mm]{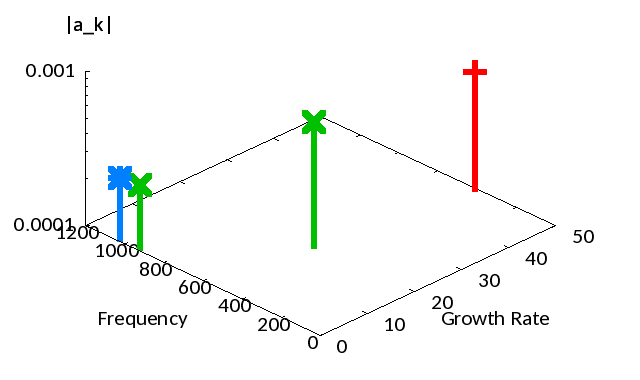}
\end{center}
\end{minipage} &
\begin{minipage}{0.45\hsize}
\begin{center}
\includegraphics[bb = 0 0 640 380, width = 79mm]{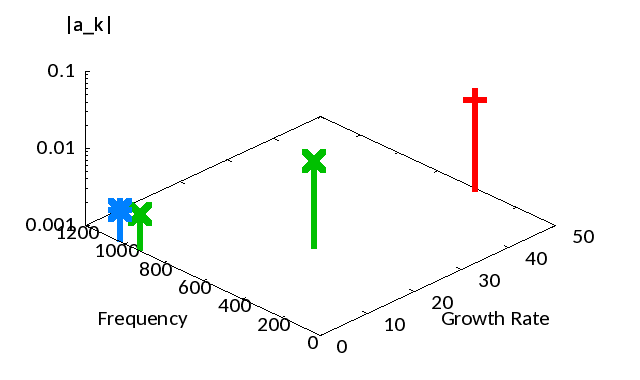}
\end{center}
\end{minipage} 
\end{tabular}
\caption{Same as Fig.~\ref{SASIL30} but for $L_\nu = 4\times 10^{52}$ erg s$^{-1}$. The two green plots are the different overtones with $l = 2$, respectively.}
\label{SASIL40}
\end{figure*}

\begin{figure*}
\begin{tabular}{cc}
\begin{minipage}{0.45\hsize}
\begin{center}
\includegraphics[bb = 0 0 640 380, width = 79mm]{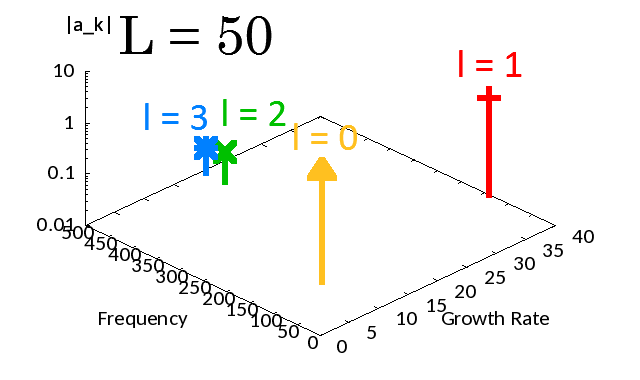}
\end{center}
\end{minipage} &
\begin{minipage}{0.45\hsize}
\begin{center}
\includegraphics[bb = 0 0 640 380, width = 79mm]{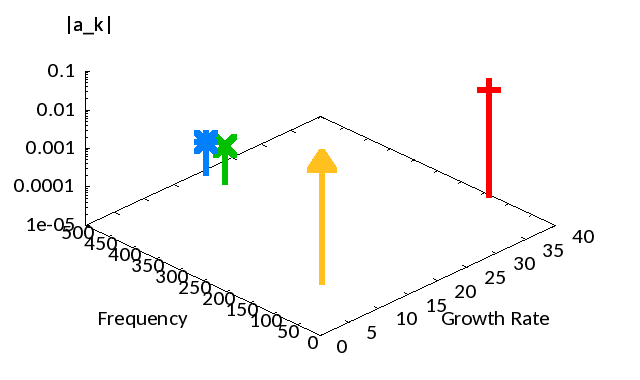}
\end{center}
\end{minipage} \\
\begin{minipage}{0.45\hsize}
\begin{center}
\includegraphics[bb = 0 0 640 380, width = 79mm]{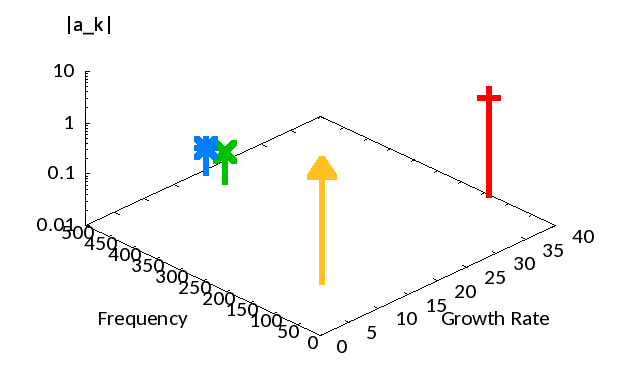}
\end{center}
\end{minipage} &
\begin{minipage}{0.45\hsize}
\begin{center}
\includegraphics[bb = 0 0 640 380, width = 79mm]{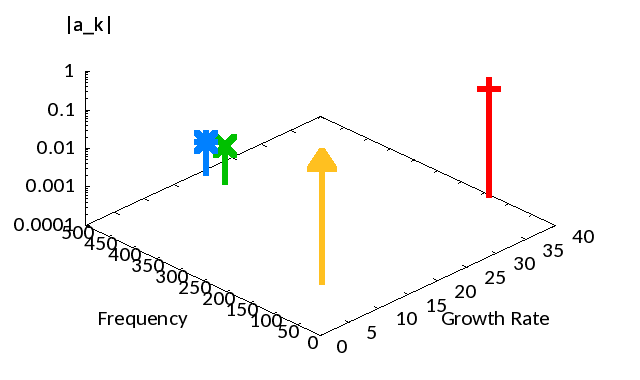}
\end{center}
\end{minipage} \\
\begin{minipage}{0.45\hsize}
\begin{center}
\includegraphics[bb = 0 0 640 380, width = 79mm]{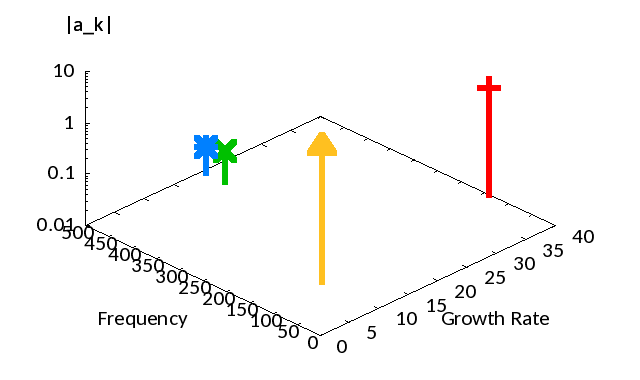}
\end{center}
\end{minipage} &
\begin{minipage}{0.45\hsize}
\begin{center}
\includegraphics[bb = 0 0 640 380, width = 79mm]{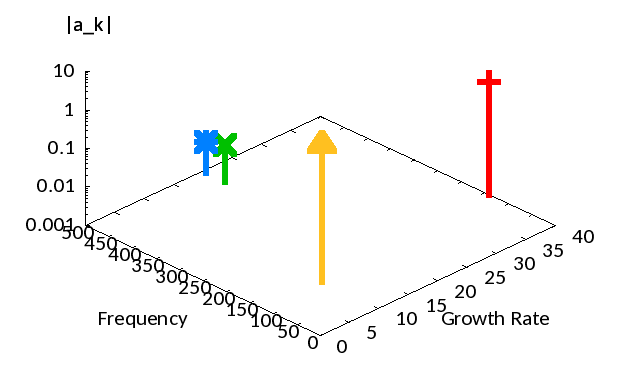}
\end{center}
\end{minipage} \\
\begin{minipage}{0.45\hsize}
\begin{center}
\includegraphics[bb = 0 0 640 380, width = 79mm]{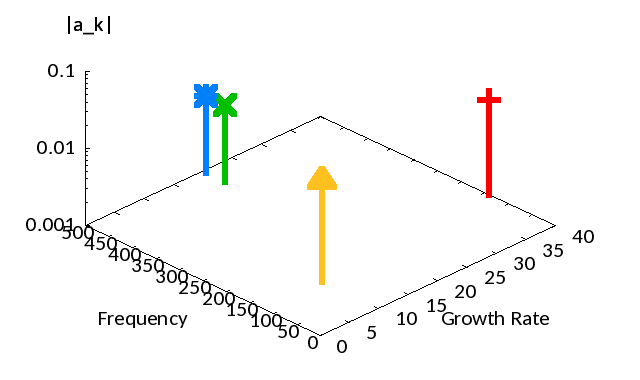}
\end{center}
\end{minipage} &
\begin{minipage}{0.45\hsize}
\begin{center}
\includegraphics[bb = 0 0 640 380, width = 79mm]{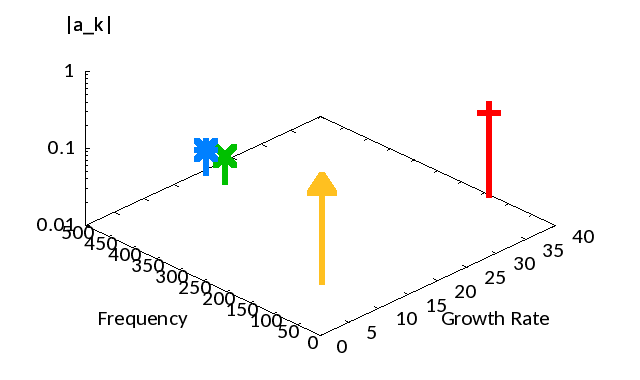}
\end{center}
\end{minipage} \\
\begin{minipage}{0.45\hsize}
\begin{center}
\includegraphics[bb = 0 0 640 380, width = 79mm]{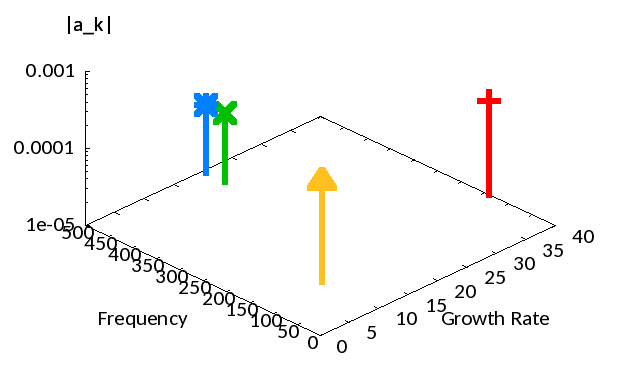}
\end{center}
\end{minipage} &
\begin{minipage}{0.45\hsize}
\begin{center}
\includegraphics[bb = 0 0 640 380, width = 79mm]{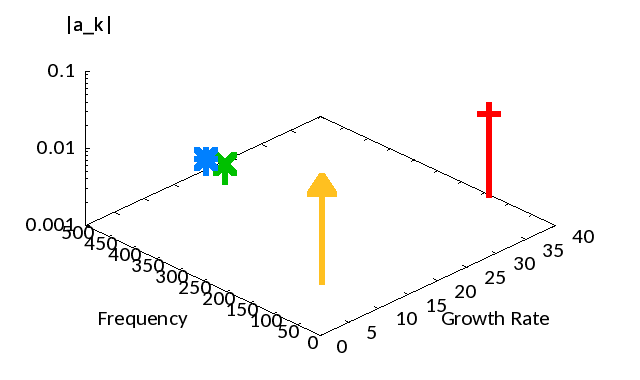}
\end{center}
\end{minipage} 
\end{tabular}
\caption{Same as Fig.~\ref{SASIL30} but for $L_\nu = 5\times 10^{52}$ erg s$^{-1}$.}
\label{SASIL50}
\end{figure*}

\begin{figure*}
\begin{tabular}{cc}
\begin{minipage}{0.45\hsize}
\begin{center}
\includegraphics[bb = 0 0 640 380, width = 79mm]{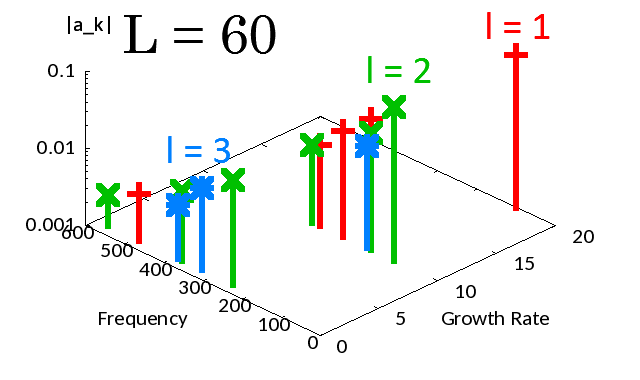}
\end{center}
\end{minipage} &
\begin{minipage}{0.45\hsize}
\begin{center}
\includegraphics[bb = 0 0 640 380, width = 79mm]{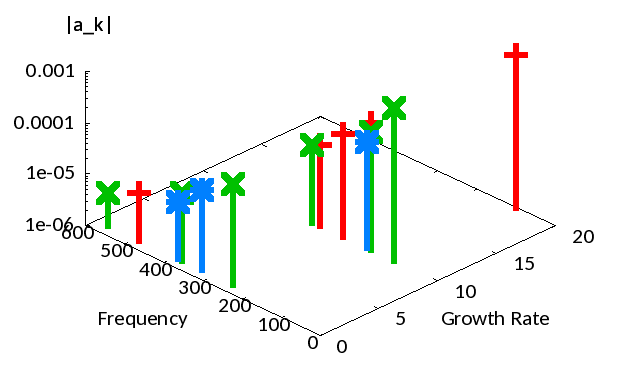}
\end{center}
\end{minipage} \\
\begin{minipage}{0.45\hsize}
\begin{center}
\includegraphics[bb = 0 0 640 380, width = 79mm]{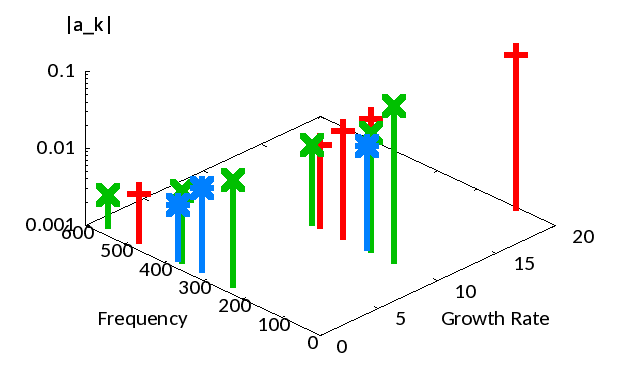}
\end{center}
\end{minipage} &
\begin{minipage}{0.45\hsize}
\begin{center}
\includegraphics[bb = 0 0 640 380, width = 79mm]{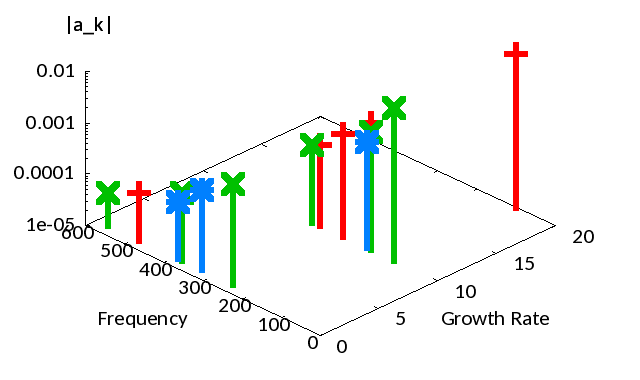}
\end{center}
\end{minipage} \\
\begin{minipage}{0.45\hsize}
\begin{center}
\includegraphics[bb = 0 0 640 380, width = 79mm]{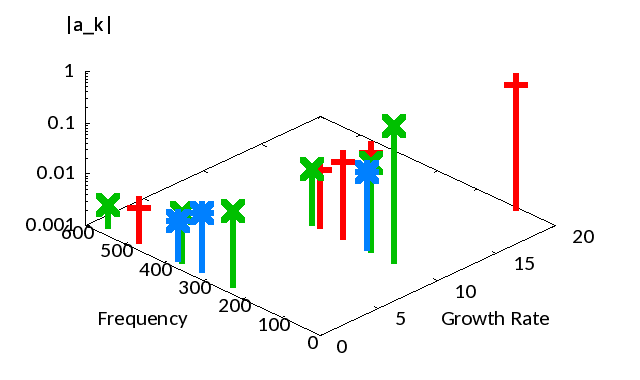}
\end{center}
\end{minipage} &
\begin{minipage}{0.45\hsize}
\begin{center}
\includegraphics[bb = 0 0 640 380, width = 79mm]{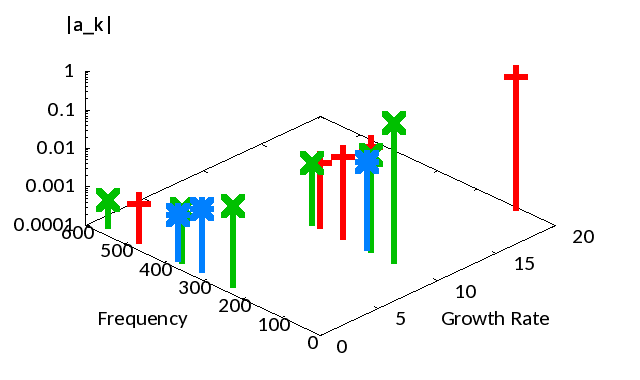}
\end{center}
\end{minipage} \\
\begin{minipage}{0.45\hsize}
\begin{center}
\includegraphics[bb = 0 0 640 380, width = 79mm]{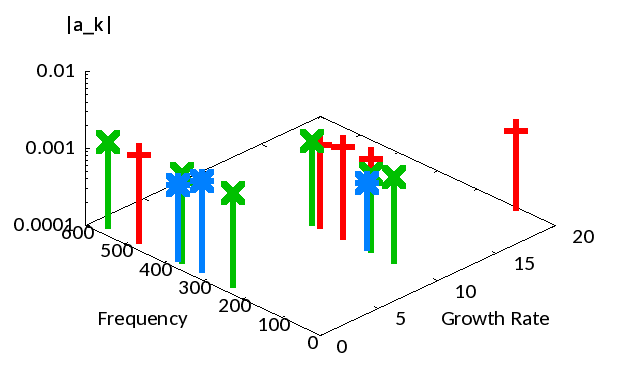}
\end{center}
\end{minipage} &
\begin{minipage}{0.45\hsize}
\begin{center}
\includegraphics[bb = 0 0 640 380, width = 79mm]{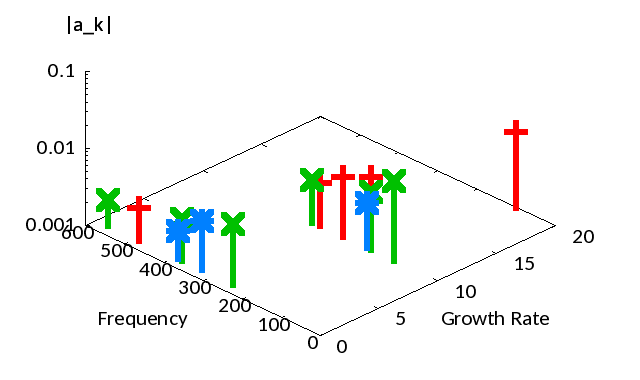}
\end{center}
\end{minipage} \\
\begin{minipage}{0.45\hsize}
\begin{center}
\includegraphics[bb = 0 0 640 380, width = 79mm]{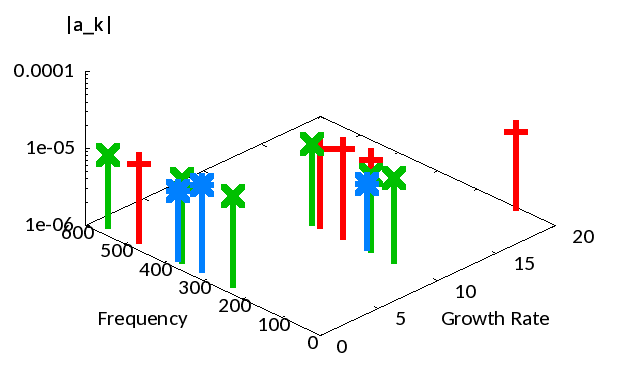}
\end{center}
\end{minipage} &
\begin{minipage}{0.45\hsize}
\begin{center}
\includegraphics[bb = 0 0 640 380, width = 79mm]{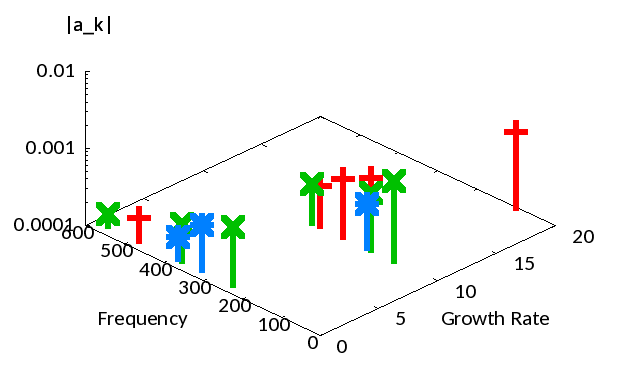}
\end{center}
\end{minipage} 
\end{tabular}
\caption{Same as Fig.~\ref{SASIL30} but for $L_\nu = 6\times 10^{52}$ erg s$^{-1}$. The plots with the same color are the different unstable modes with the same $l$, respectively.}
\label{SASIL60}
\end{figure*}

\begin{figure*}
\begin{tabular}{cc}
\begin{minipage}{0.45\hsize}
\begin{center}
\includegraphics[bb = 0 0 640 480, width = 63mm]{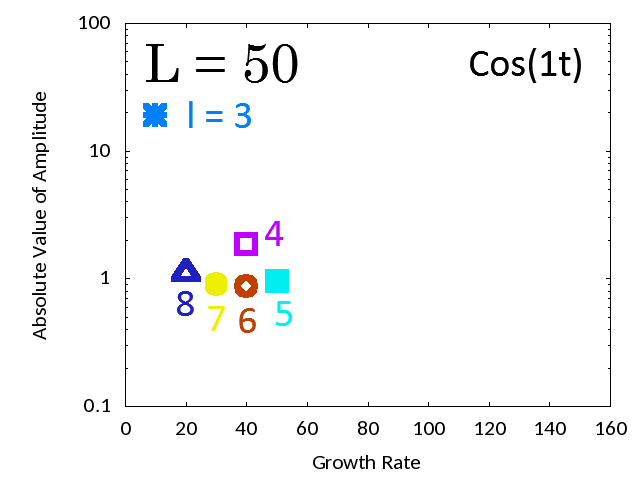}
\end{center}
\end{minipage} &
\begin{minipage}{0.45\hsize}
\begin{center}
\includegraphics[bb = 0 0 640 480, width = 63mm]{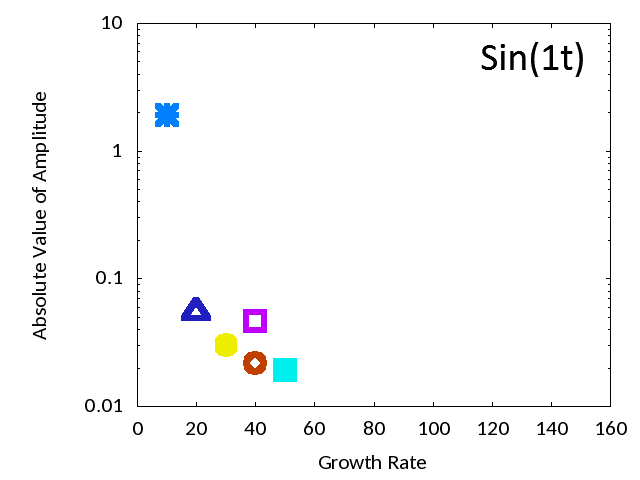}
\end{center}
\end{minipage} \\
\begin{minipage}{0.45\hsize}
\begin{center}
\includegraphics[bb = 0 0 640 480, width = 63mm]{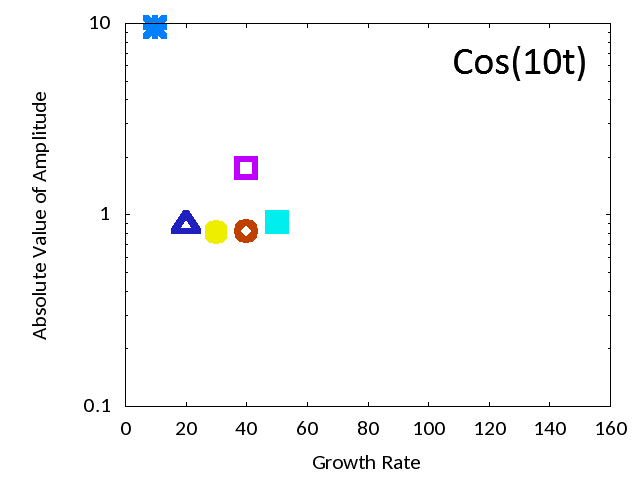}
\end{center}
\end{minipage} &
\begin{minipage}{0.45\hsize}
\begin{center}
\includegraphics[bb = 0 0 640 480, width = 63mm]{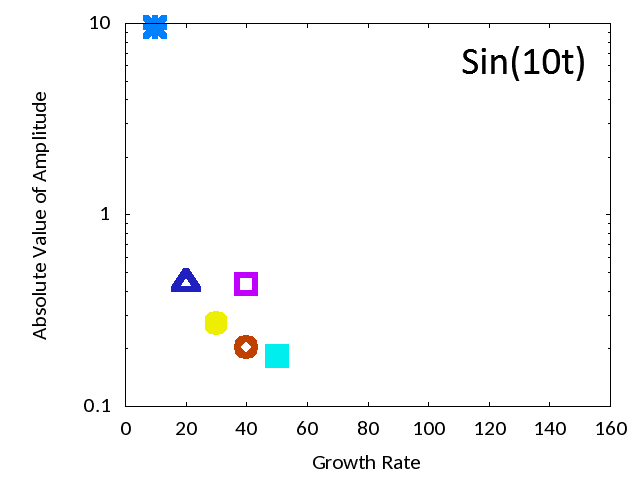}
\end{center}
\end{minipage} \\
\begin{minipage}{0.45\hsize}
\begin{center}
\includegraphics[bb = 0 0 640 480, width = 63mm]{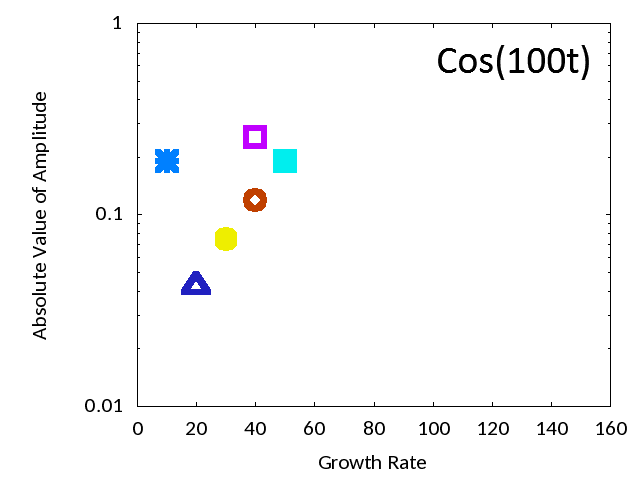}
\end{center}
\end{minipage} &
\begin{minipage}{0.45\hsize}
\begin{center}
\includegraphics[bb = 0 0 640 480, width = 63mm]{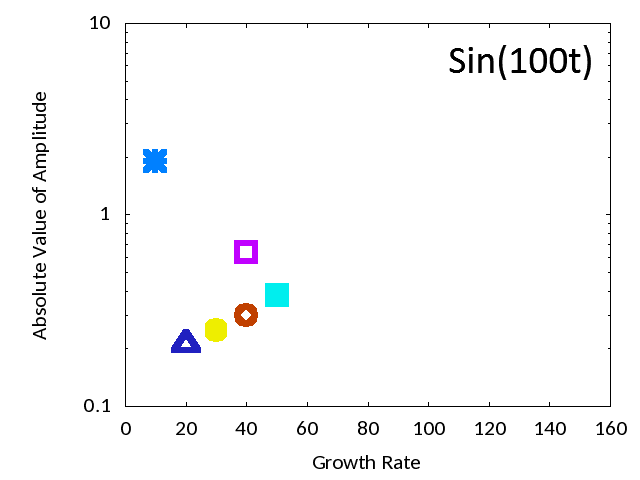}
\end{center}
\end{minipage} \\
\begin{minipage}{0.45\hsize}
\begin{center}
\includegraphics[bb = 0 0 640 480, width = 63mm]{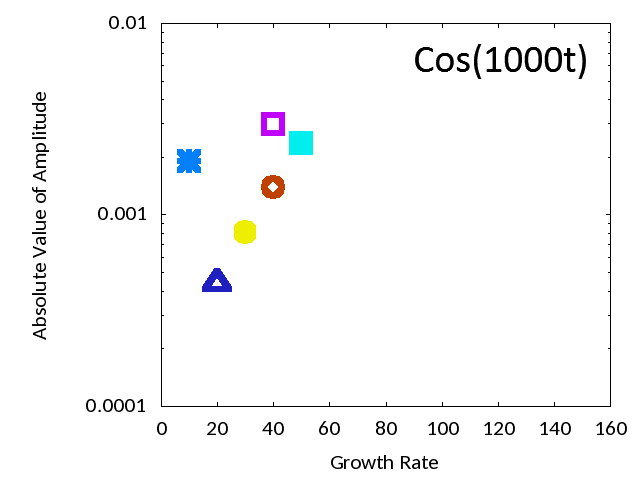}
\end{center}
\end{minipage} &
\begin{minipage}{0.45\hsize}
\begin{center}
\includegraphics[bb = 0 0 640 480, width = 63mm]{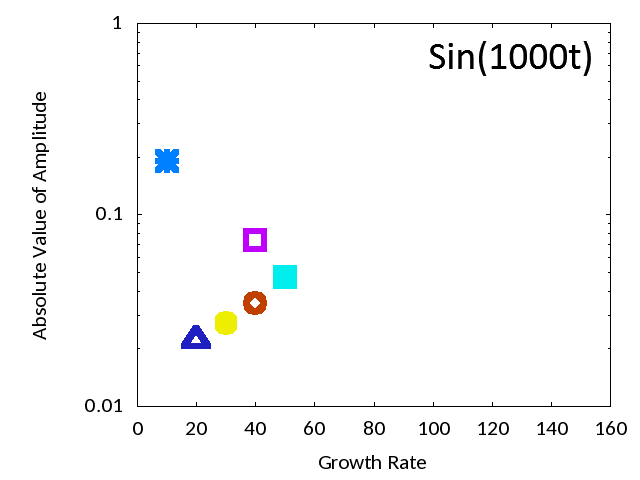}
\end{center}
\end{minipage} \\
\begin{minipage}{0.45\hsize}
\begin{center}
\includegraphics[bb = 0 0 640 480, width = 63mm]{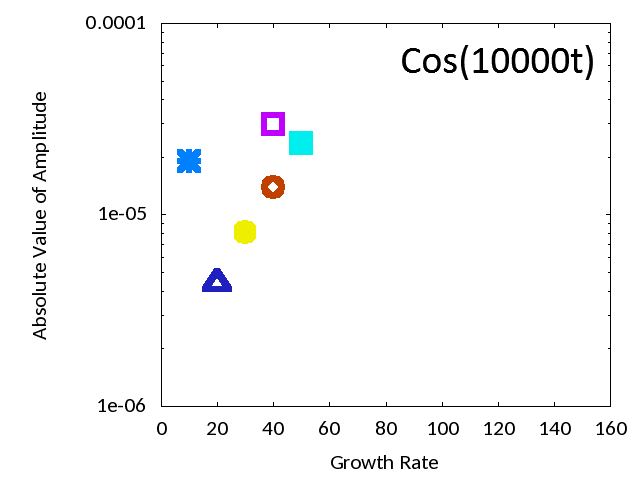}
\end{center}
\end{minipage} &
\begin{minipage}{0.45\hsize}
\begin{center}
\includegraphics[bb = 0 0 640 480, width = 63mm]{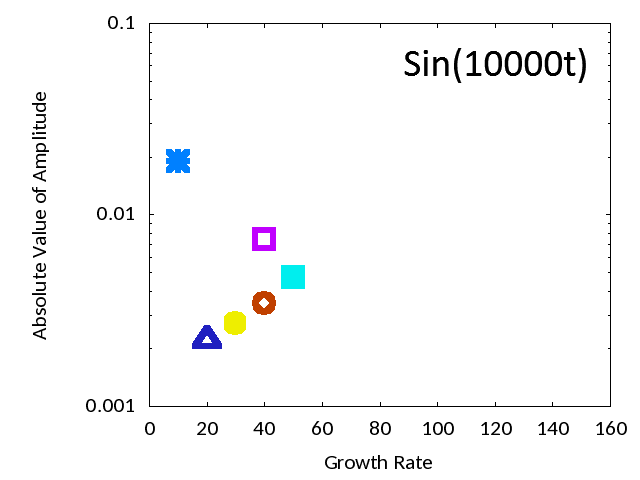}
\end{center}
\end{minipage} 
\end{tabular}
\caption{The absolute values of the excited amplitudes of the unstable convective modes for $L_\nu = 5\times 10^{52}$ erg s$^{-1}$ that are excited by the upstream fluctuations given in Eqs.~(\ref{EXP1})-(\ref{EXP3}). Left panels show the results for the cosine-type perturbations, i.e., $\varphi = \pi /2$, while right panels for the sine-type fluctuations, $\varphi = 0$. From top to bottom, the frequency of the upstream perturbation changes from $\omega _\mathrm{up}= 1$ s$^{-1}$ to $\omega _\mathrm{up}= 10^4$ s$^{-1}$ as indicated on the upper right corner of each panel. Different colors correspond to modes with different $l$.}
\label{ConvL50}
\end{figure*}

\begin{figure*}
\begin{tabular}{cc}
\begin{minipage}{0.45\hsize}
\begin{center}
\includegraphics[bb = 0 0 640 480, width = 63mm]{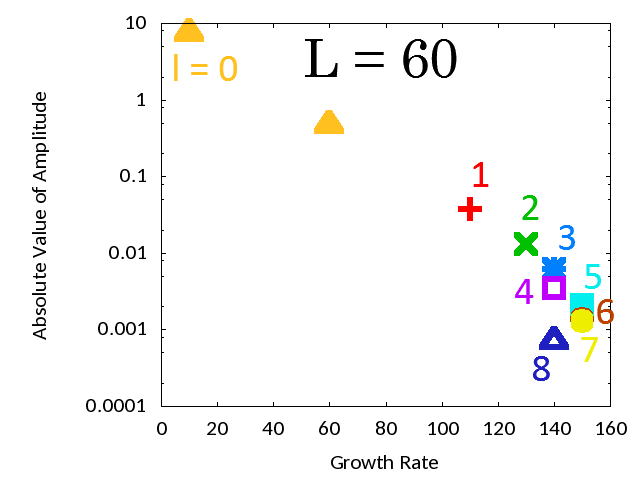}
\end{center}
\end{minipage} &
\begin{minipage}{0.45\hsize}
\begin{center}
\includegraphics[bb = 0 0 640 480, width = 63mm]{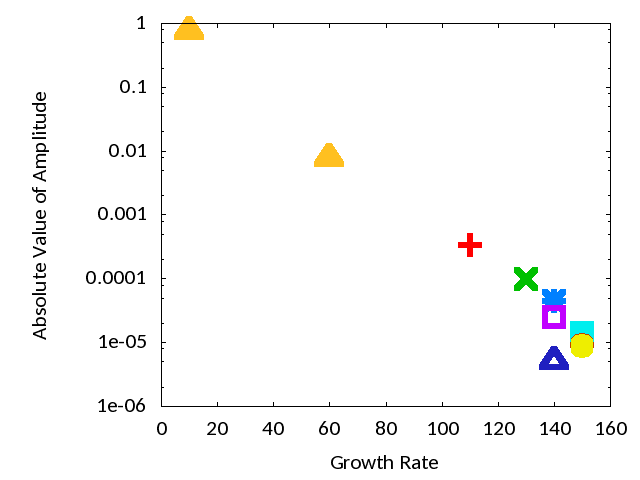}
\end{center}
\end{minipage} \\
\begin{minipage}{0.45\hsize}
\begin{center}
\includegraphics[bb = 0 0 640 480, width = 63mm]{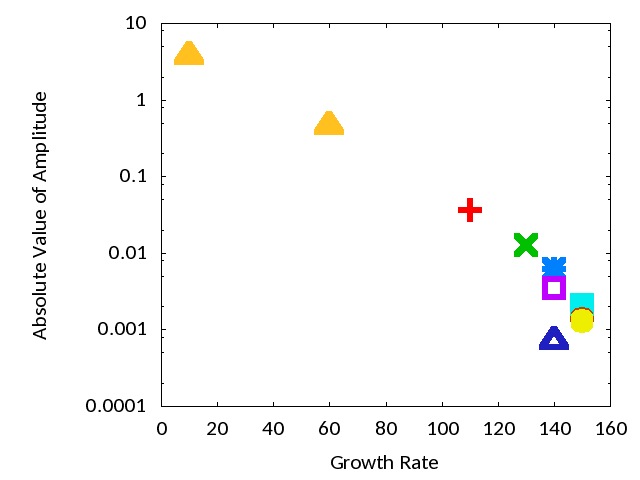}
\end{center}
\end{minipage} &
\begin{minipage}{0.45\hsize}
\begin{center}
\includegraphics[bb = 0 0 640 480, width = 63mm]{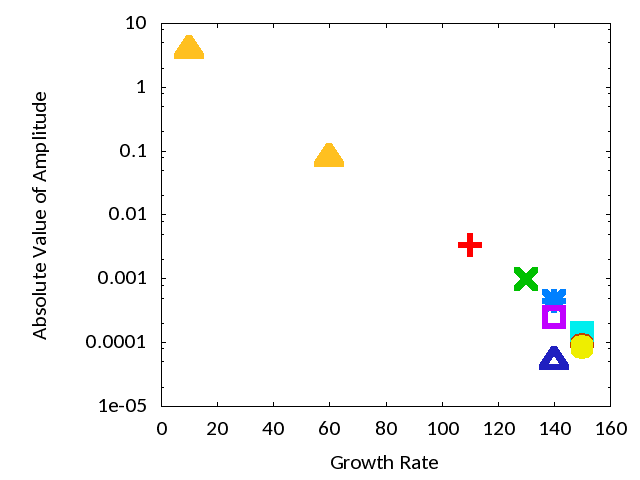}
\end{center}
\end{minipage} \\
\begin{minipage}{0.45\hsize}
\begin{center}
\includegraphics[bb = 0 0 640 480, width = 63mm]{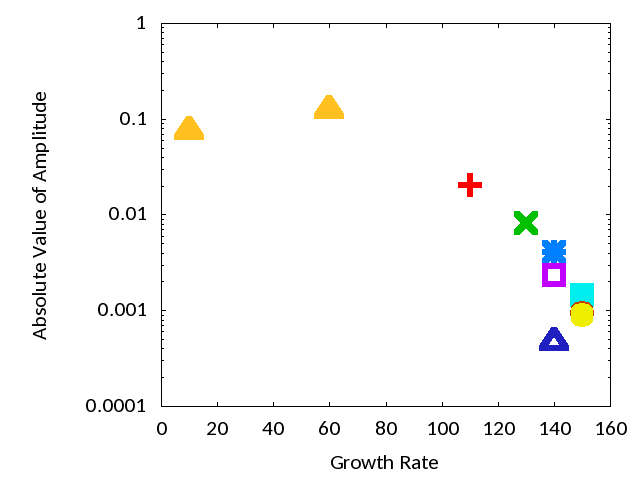}
\end{center}
\end{minipage} &
\begin{minipage}{0.45\hsize}
\begin{center}
\includegraphics[bb = 0 0 640 480, width = 63mm]{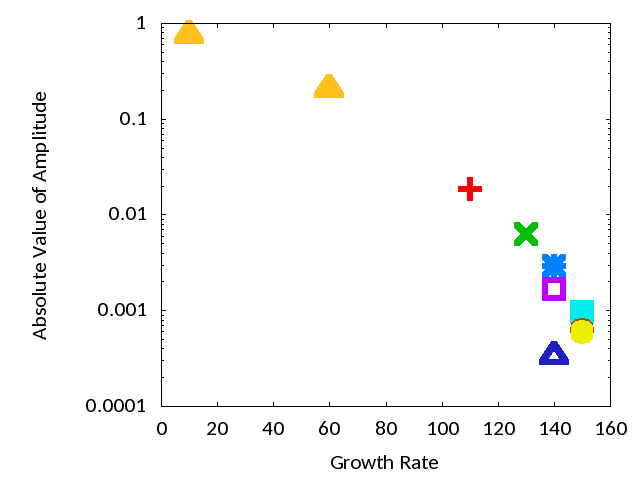}
\end{center}
\end{minipage} \\
\begin{minipage}{0.45\hsize}
\begin{center}
\includegraphics[bb = 0 0 640 480, width = 63mm]{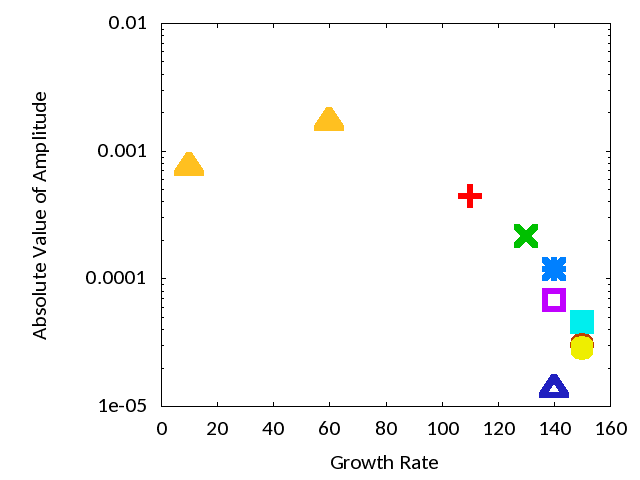}
\end{center}
\end{minipage} &
\begin{minipage}{0.45\hsize}
\begin{center}
\includegraphics[bb = 0 0 640 480, width = 63mm]{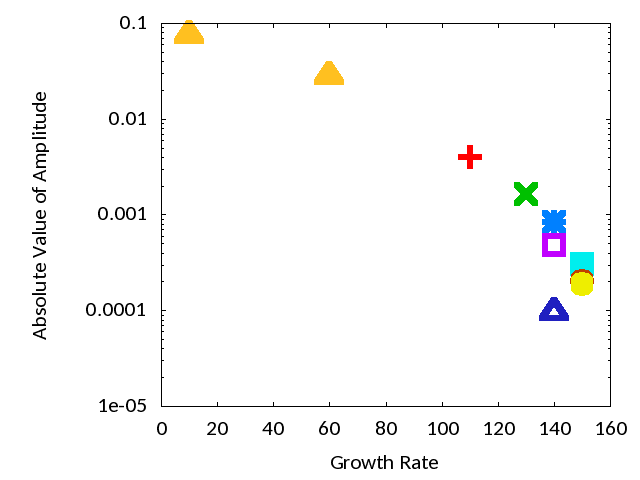}
\end{center}
\end{minipage} \\
\begin{minipage}{0.45\hsize}
\begin{center}
\includegraphics[bb = 0 0 640 480, width = 63mm]{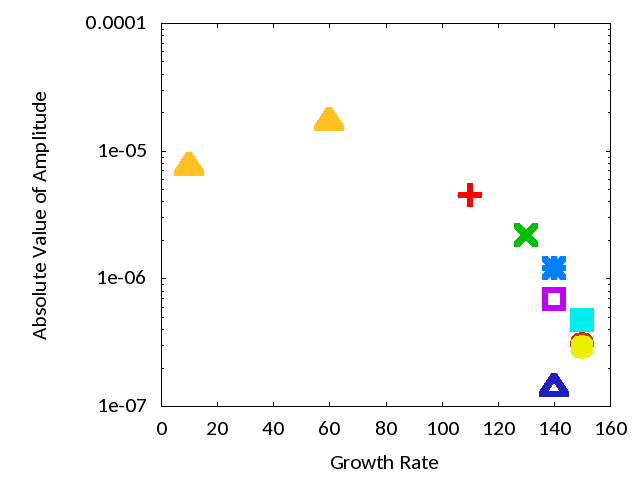}
\end{center}
\end{minipage} &
\begin{minipage}{0.45\hsize}
\begin{center}
\includegraphics[bb = 0 0 640 480, width = 63mm]{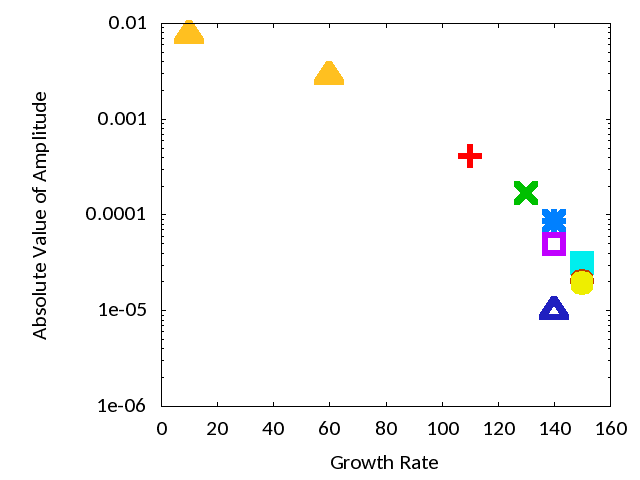}
\end{center}
\end{minipage} 
\end{tabular}
\caption{Same as Fig.~\ref{ConvL50} but for $L_\nu = 6\times 10^{52}$ erg s$^{-1}$.} 
\label{ConvL60}
\end{figure*}

\begin{figure*}
\begin{tabular}{cc}
\begin{minipage}{0.45\hsize}
\begin{center}
\includegraphics[bb = 0 0 640 480, width = 80mm]{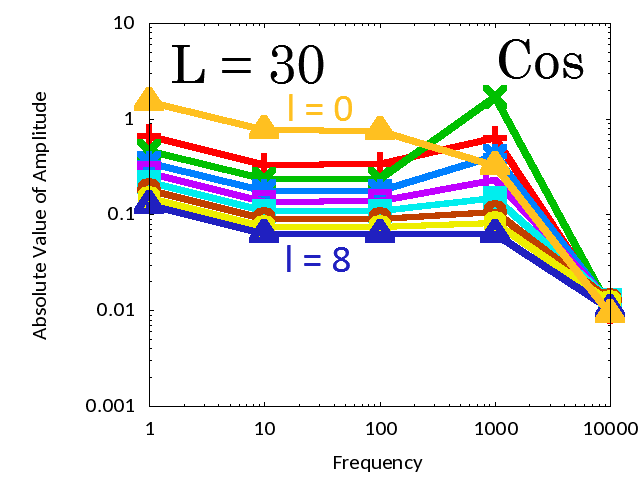}
\end{center}
\end{minipage} &
\begin{minipage}{0.45\hsize}
\begin{center}
\includegraphics[bb = 0 0 640 480, width = 80mm]{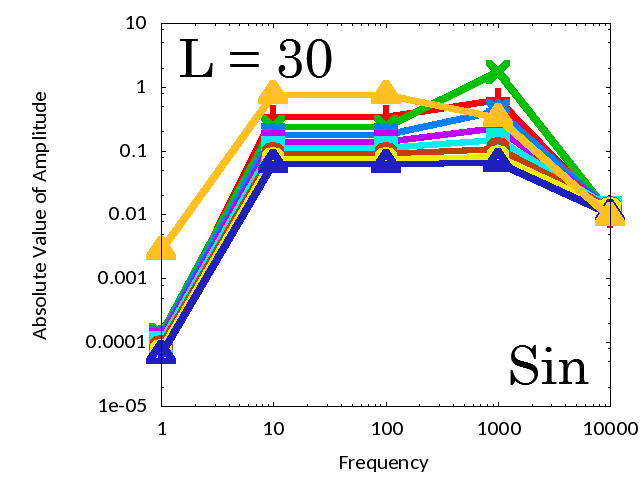}
\end{center}
\end{minipage} \\
\begin{minipage}{0.45\hsize}
\begin{center}
\includegraphics[bb = 0 0 640 480, width = 80mm]{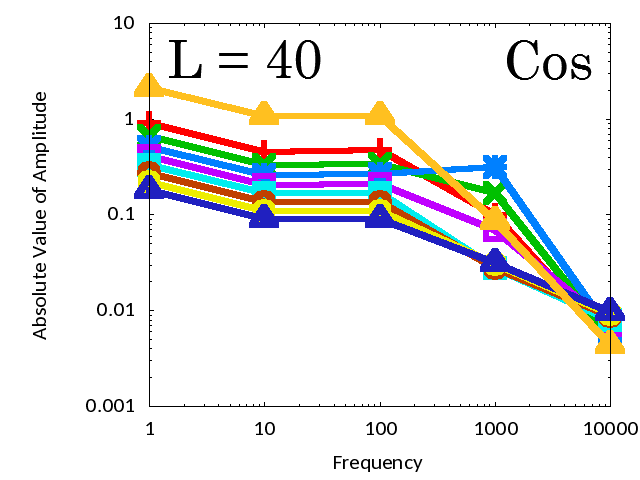}
\end{center}
\end{minipage} &
\begin{minipage}{0.45\hsize}
\begin{center}
\includegraphics[bb = 0 0 640 480, width = 80mm]{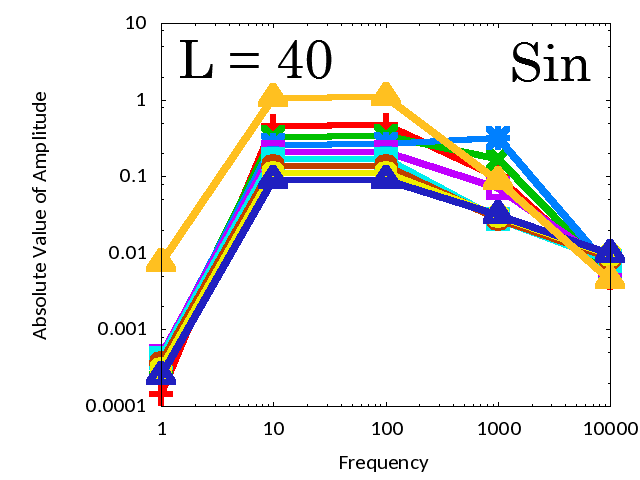}
\end{center}
\end{minipage} \\
\begin{minipage}{0.45\hsize}
\begin{center}
\includegraphics[bb = 0 0 640 480, width = 80mm]{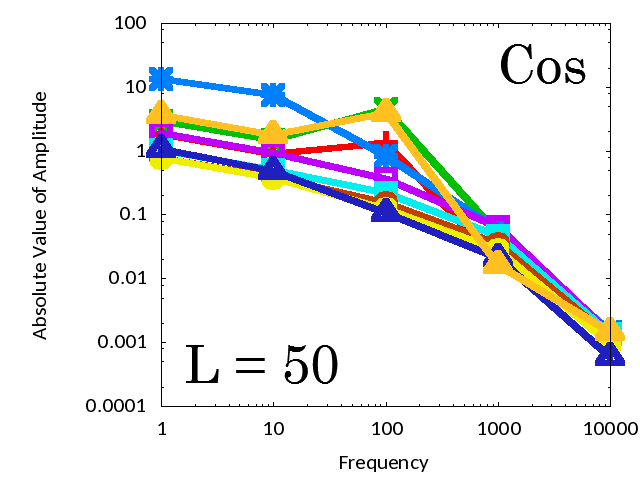}
\end{center}
\end{minipage} &
\begin{minipage}{0.45\hsize}
\begin{center}
\includegraphics[bb = 0 0 640 480, width = 80mm]{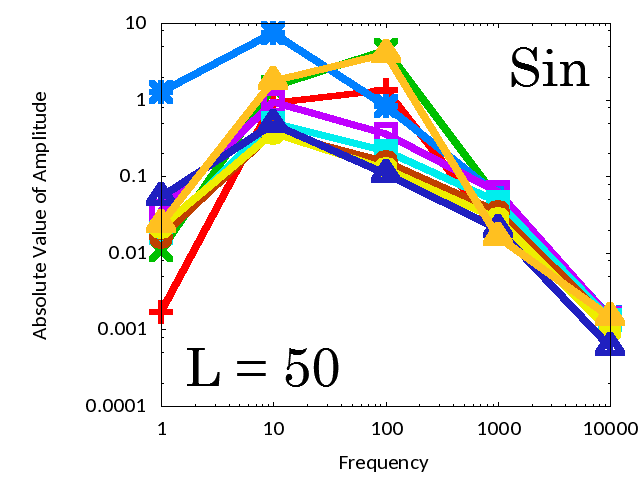}
\end{center}
\end{minipage}\\
\begin{minipage}{0.45\hsize}
\begin{center}
\includegraphics[bb = 0 0 640 480, width = 80mm]{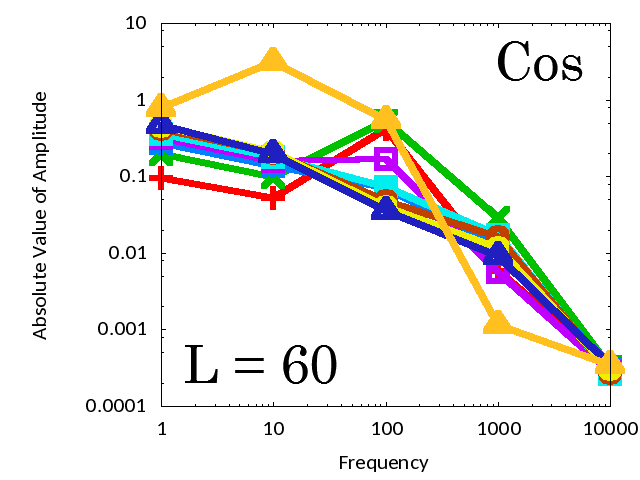}
\end{center}
\end{minipage} &
\begin{minipage}{0.45\hsize}
\begin{center}
\includegraphics[bb = 0 0 640 480, width = 80mm]{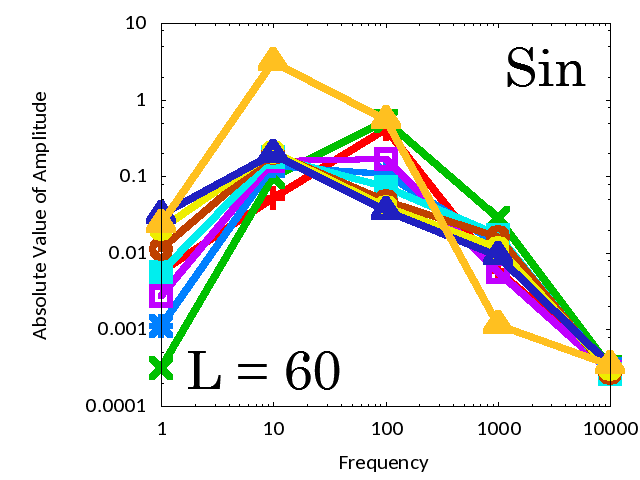}
\end{center}
\end{minipage}
\end{tabular}
\caption{The absolute values of the excited amplitudes of the forced-oscillation modes for various neutrino luminosities. Left panels show the results for the cosine-type perturbations, i.e., $\varphi = \pi/2$, while right panels for the sine-type fluctuations, $\varphi = 0$. From top to bottom, the neutrino luminosity changes from $L_\nu= 3\times 10^{52}$ erg s$^{-1}$ to $L_\nu= 6\times 10^{52}$ erg s$^{-1}$ as indicated in the figure. Different colors correspond to modes with different $l$, having the same meaning as in Figs.~\ref{SASIL30}-\ref{ConvL60}.}
\label{Ptrb}
\end{figure*}

Now we proceed to the case, in which we impose non-spherical perturbations also in the flow ahead of the shock surface. As proved in the next section, the upstream perturbations excite the intrinsic instabilities while they do not change the eigen frequencies. Put more specifically, the amplitude of an intrinsically unstable mode is turned out to be given as a linear combination of the contributions from the upstream perturbations, initial fluctuations in the shocked region, and fluctuations at the inner boundary (See Appendices \ref{sec.analyresults} and \ref{app.amplitude}).

In this paper, w pay attention to the first contribution and evaluate numerically the resultant amplitudes of the modes excited these upstream fluctuations.
They are given by Eq.~(\ref{EqAmpS}) or equivalently (\ref{EqAmpS2}) as product of the residue at the corresponding pole of the Laplace-transformed shock radius for the impulsive perturbation and the functional value of the Laplace-transformed upstream perturbation at the pole position.

We consider both SASI and convective instabilities with $0 \le l \le 8$, which have been obtained in the previous sub-section, and discuss them separately discussed below. We assume the upstream perturbation to have the functional forms given in Eqs.~(\ref{EXP1})-(\ref{EXP3}). We also change the frequency $\omega _\mathrm{up}$ and phase $\varphi$ included in them.

\subsubsection{effects on SASI}
We first study the excitation of intrinsically unstable SASI modes, which have $\Omega > 0$ and $\omega > 0$, by sinusoidal upstream perturbations with $\varphi = 0, \pi$/2 and $\omega _\mathrm{up} = 1-10^4$ s$^{-1}$. 

Fig.~\ref{SASIL30} shows the results for $L_\nu = 3\times 10^{52}$ erg s$^{-1}$, where the absolute values of the excited amplitudes are plotted. The amplitudes are normalized by that of the upstream perturbation and we note that the vertical scale differs from panel to panel. There are two unstable SASI modes with $l = 1,2 $ found for the neutrino luminosity. Similarly, Fig.~\ref{SASIL40} presents the results for $L_\nu = 4\times 10^{52}$ erg s$^{-1}$, where another $l=2$ mode becomes unstable as well as an $l=3$ mode. Figs.~\ref{SASIL50} and \ref{SASIL60} show the results for still higher luminosities, $5\times 10^{52}$ erg s$^{-1}$ and $6\times 10^{52}$ erg s$^{-1}$, respectively. We note that some convective modes also become unstable for these high luminosities, which are discussed separately later and not shown in the figures.

As a general trend of these results, we point out that the excited amplitude is larger for the cosine-type perturbation than for the sine-type if the frequency of the upstream perturbation is much smaller than the oscillation frequency of the eigen mode. As $\omega _\mathrm{up}$ gets larger, on the other hand, the amplitude of excitation for sine-type perturbations increases rapidly and, when $\omega _\mathrm{up}$ is comparable to the oscillation frequency of the eigen mode, the amplitudes for both types of perturbations are also similar to each other. For even larger $\omega_\mathrm{up}$ the order is reversed with the excitation by the sine-type perturbations being dominant.

The magnitude of the excited amplitudes is in the range from $10^{-6}$ to $10$. As shown in the figures, it peaks around $\omega _\mathrm{up} \sim \omega _j$ (the oscillation frequency of the $j$-th mode) and becomes smaller as $\omega _\mathrm{up}$ goes away from $\omega _j$.
At high luminosities, as the upstream frequency becomes higher, however, the amplitudes of all modes become comparable to one another, which is most clearly discernible in Fig.~\ref{SASIL60}.

\subsubsection{effects on convection}
The excitation of unstable convective modes by the upstream fluctuations are studied just in the same way as for the SASI modes. Since their oscillation frequency vanishes by definition, we show in Figs.~\ref{ConvL50} and \ref{ConvL60} the results in the two dimensional plane of the excited amplitude and growth rate. The former figure displays the results for $L_\nu = 5\times 10^{52}$ erg s$^{-1}$ while the latter for $L_\nu = 6\times 10^{52}$ erg s$^{-1}$.
We note that the convective modes with $1\le l \le 8$ are unstable for $L_\nu \gtsim 5\times 10^{52}$ erg s$^{-1}$ as indicated in Fig.~\ref{Higher3}.

As in the SASI modes, we find that the amplitude of excitation is larger for the cosine-type perturbations if $\omega _\mathrm{up}$ is much small while the excitation is dominated by the sine-type if $\omega _\mathrm{up}$ is large enough. The order change appears to occur at $\omega _\mathrm{up} \sim \Omega _j$ (the growth rate of the $j$-th mode).

The convective modes with smaller $\Omega$ are likely to be more strongly excited especially when $\omega _\mathrm{up}$ is very small. As the upstream frequency increases, the difference in the excited amplitudes among various modes tends to become smaller as seen in the figures with the modes with small $\Omega$ being suppressed more strongly.

\subsubsection{forced-oscillation modes}\label{forcedoscillation}
When the sinusoidal perturbation is added ahead of the shock, there naturally occurs a mode that oscillates stably with the same frequency as shown analytically in the next section. This is analogous to the forced oscillation of a spring.

The amplitude of this mode is also obtained from Eq.~(\ref{amplitude}). The residue at the corresponding pole is evaluated by integrating the Laplace-transformed function of the perturbed shock radius around the pole. We show the results in Fig.~\ref{Ptrb}.

For the cosine-type perturbations, the amplitude tends to increase as the upstream frequency decreases. In the case of the sine-type fluctuations, there appears a broad peak at $\omega_\mathrm{up}\sim$ 10-100 s$^{-1}$. 
As for the $l$-dependence, the amplitude becomes larger for smaller $l$ when the luminosity is small and hence SASI dominates convections ($L_\nu = 3, 4 \times 10^{52}$ erg s$^{-1}$) while it is reversed for larger luminosities where convections become dominant ($L_\nu = 5, 6 \times 10^{52}$ erg s$^{-1}$) although there are some outliers.

Since the growth rate of the mode vanishes, it will eventually be dominated by unstable SASI or convections. For upstream fluctuations of finite amplitudes, which may not be small indeed \citep{KT}, the forced-oscillation mode may not be negligible, either.

\section{Discussions}\label{sec.discussion}
To understand the numerical results presented above, we give some analytical considerations, which can be summarized as follows: (1) the growth/damping rates and oscillation frequencies of intrinsic modes are unaffected by the presence of the upstream perturbations, (2) these intrinsic modes are excited with different amplitudes by the upstream perturbations. In the following sub-sections, we discuss them in turn.

\subsection{Effects of upstream perturbations on the eigen frequencies of unstable modes} \label{HowChanged}
We here show that neither the growth rates nor the oscillation frequencies of intrinsic modes are altered by upstream fluctuations except for a special case.

The Laplace-transformed shock evolution is described by Eq.~(\ref{An3}), which is divided into three parts: the contribution from the upstream perturbations, Eq.~(\ref{eq.frac}), the one from the initial perturbations in the shocked region, Eq.~(\ref{ini.frac}), and the one from the inner boundary, Eq.~(\ref{ib.frac}).
Recalling that the growth rates and oscillation frequencies of eigen modes are obtained from the location of the corresponding poles, we conclude that zero points of the denominator in these equations give the eigen values unless cancellation occurs between the denominator and numerator, which is not expected in general. It is also possible that the numerators have their own poles, which is discussed later. Since the denominator is dependent only on the background quantities and the inner boundary condition, the poles originated from the denominator are unaffected by the presence of the upstream perturbation. This is of course true up to the linear order.

The poles of the numerator correspond to the forced-oscillation mode (see section \ref{forcedoscillation}). For example, if a sinusoidally oscillating density perturbation $\delta \rho (r_\mathrm{sh},t)\propto \sin (\omega _\mathrm{up}t)$  exists just ahead of the shock front, then the Laplace-transformed shock radius, Eq.~(\ref{eq.frac}), has a pole at $s = \pm \omega _\mathrm{up}$ (see Eq.~(\ref{Lsin})) This correspond to an 'eigen' mode that oscillates stably at the same frequency as the original upstream perturbation. It is stressed that this mode is stable and has nothing to do with other intrinsic modes in general.

An exceptional case is the resonance, however. It occurs only when the upstream perturbation has coincidentally the same growth rate and oscillation frequency as one of the eigen values, i.e., when the upstream perturbation is given by 
\begin{equation}
{\bf z}(t) \propto e^{\Omega _n t}\sin(\omega _nt + \varphi), \\
\end{equation}
where $\Omega _n$ and $\omega _n$ are the growth rate and oscillation frequency of the $n$-th intrinsic eigen mode, respectively.
Then the denominator of Eq.~(\ref{eq.frac}) is factorized by $(s - \Omega _n \mp \omega _n)^2$, that is, $s = \Omega _n\pm i \omega _n$ becomes a double pole. Note that the inverse Laplace transform of a multi-pole is
\begin{equation}
\mathcal{L}^{-1}\left[ \frac{1}{(s-s_0)^\nu}\right] = \frac{t^{\nu -1} e^{s_0 t}}{\Gamma (\nu)},
\end{equation}
where $\nu > 0$ is a real number, $s_0$ a complex constant, and $\Gamma (\nu)$ the gamma function. Applying the relation to the double pole, we find the evolution of the shock radius for the eigen mode gets proportional to $t\exp (\Omega _nt)\sin (\omega _nt)$, i.e., it  acquires an extra power of $t$. This is particularly important for $\Omega _n = 0$. We emphasize again, however, that the resonance occurs only when both the oscillation frequencies (and the growth rates if any) are identical for the upstream perturbation and one of the intrinsic modes, which may not be expected in general.

\subsection{Amplitudes of intrinsic modes excited by upstream perturbations}\label{AIEUP}
\begin{figure*}
\begin{tabular}{cc}
\begin{minipage}{0.45\hsize}
\begin{center}
\includegraphics[bb = 0 0 640 380, width = 80mm]{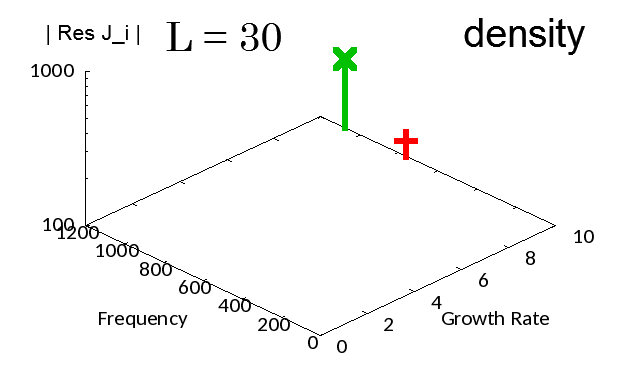}
\end{center}
\end{minipage} &
\begin{minipage}{0.45\hsize}
\begin{center}
\includegraphics[bb = 0 0 640 380, width = 80mm]{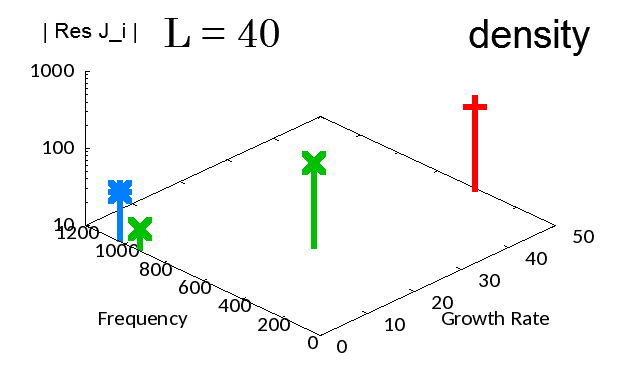}
\end{center}
\end{minipage} \\
\begin{minipage}{0.45\hsize}
\begin{center}
\includegraphics[bb = 0 0 640 380, width = 80mm]{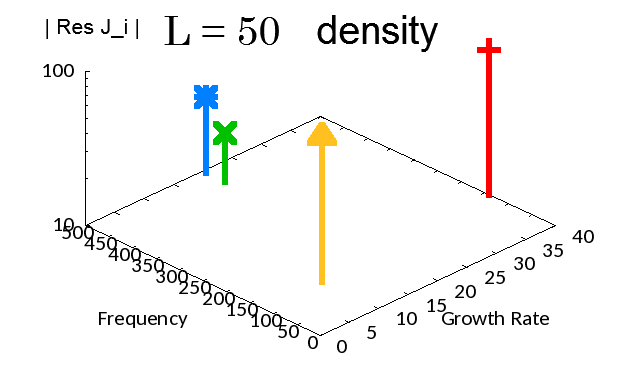}
\end{center}
\end{minipage} &
\begin{minipage}{0.45\hsize}
\begin{center}
\includegraphics[bb = 0 0 640 380, width = 80mm]{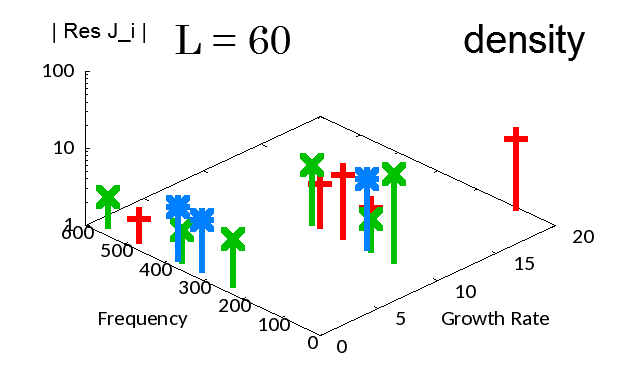}
\end{center}
\end{minipage} \\
\begin{minipage}{0.45\hsize}
\begin{center}
\includegraphics[bb = 0 0 640 480, width = 80mm]{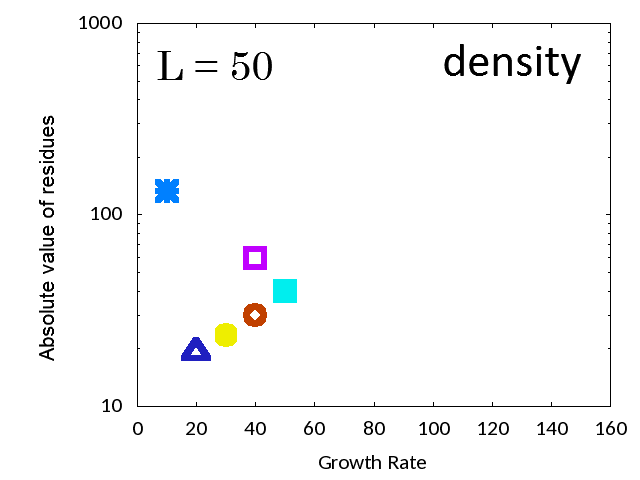}
\end{center}
\end{minipage} &
\begin{minipage}{0.45\hsize}
\begin{center}
\includegraphics[bb = 0 0 640 480, width = 80mm]{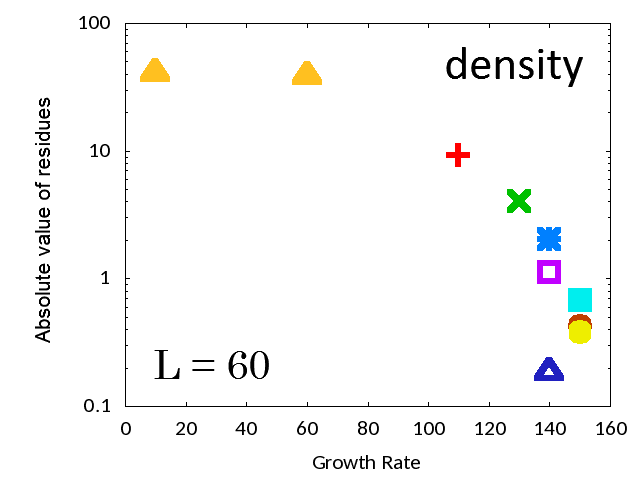}
\end{center}
\end{minipage} 
\end{tabular}
\caption{The absolute values of  $\mathop{\Res} \mathcal{J}_i^*$ for different modes. Only the density is perturbed impulsively here.  
The upper four panels show the results for SASI modes and the bottom two panels are counterparts for convective modes.}
\label{Delta1}
\end{figure*}

\begin{figure*}
\begin{tabular}{cc}
\begin{minipage}{0.45\hsize}
\begin{center}
\includegraphics[bb = 0 0 640 380, width = 80mm]{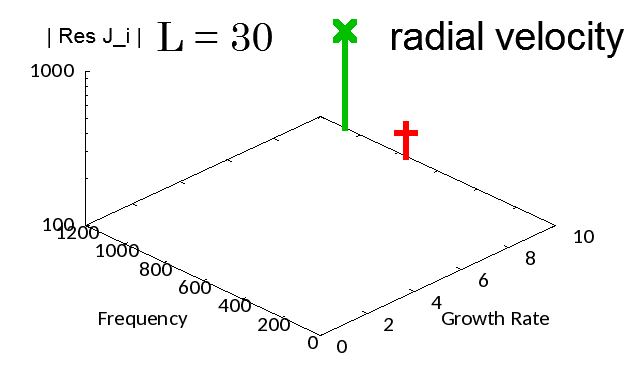}
\end{center}
\end{minipage} &
\begin{minipage}{0.45\hsize}
\begin{center}
\includegraphics[bb = 0 0 640 380, width = 80mm]{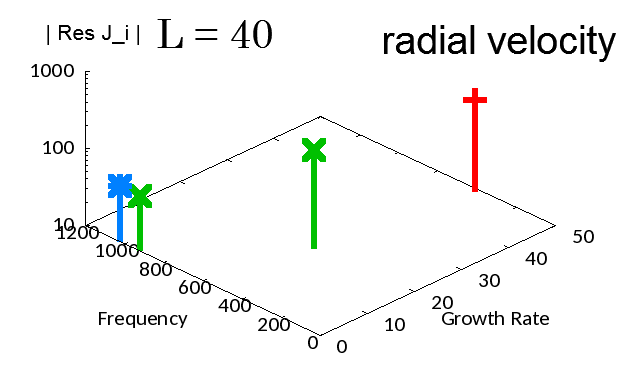}
\end{center}
\end{minipage} \\
\begin{minipage}{0.45\hsize}
\begin{center}
\includegraphics[bb = 0 0 640 380, width = 80mm]{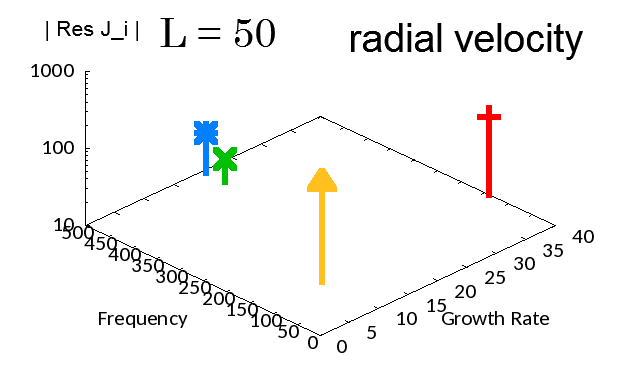}
\end{center}
\end{minipage} &
\begin{minipage}{0.45\hsize}
\begin{center}
\includegraphics[bb = 0 0 640 380, width = 80mm]{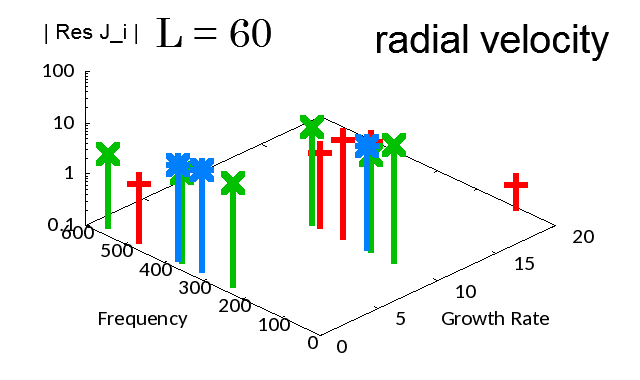}
\end{center}
\end{minipage} \\
\begin{minipage}{0.45\hsize}
\begin{center}
\includegraphics[bb = 0 0 640 480, width = 80mm]{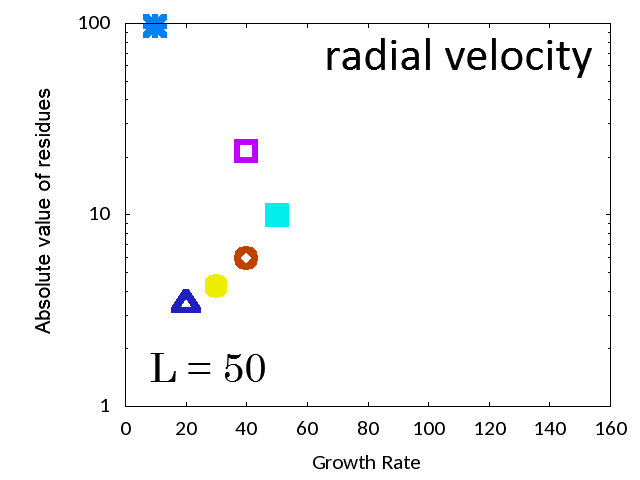}
\end{center}
\end{minipage} &
\begin{minipage}{0.45\hsize}
\begin{center}
\includegraphics[bb = 0 0 640 480, width = 80mm]{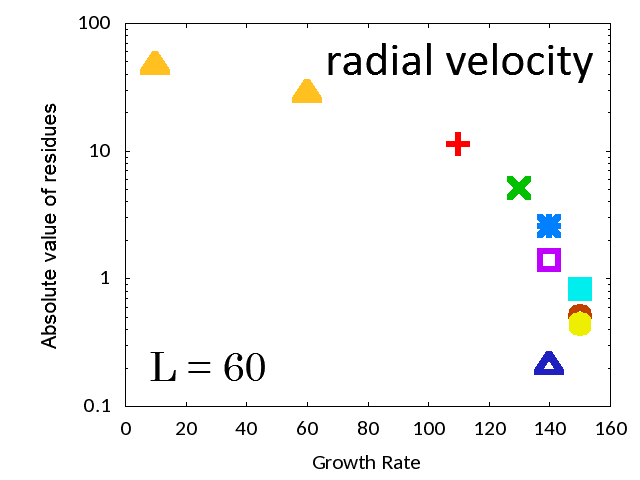}
\end{center}
\end{minipage} 
\end{tabular}
\caption{Same as Fig.~\ref{Delta1} but for the impulsive radial-velocity perturbation.}
\label{Delta2}
\end{figure*}

\begin{figure*}
\begin{tabular}{cc}
\begin{minipage}{0.45\hsize}
\begin{center}
\includegraphics[bb = 0 0 640 380, width = 80mm]{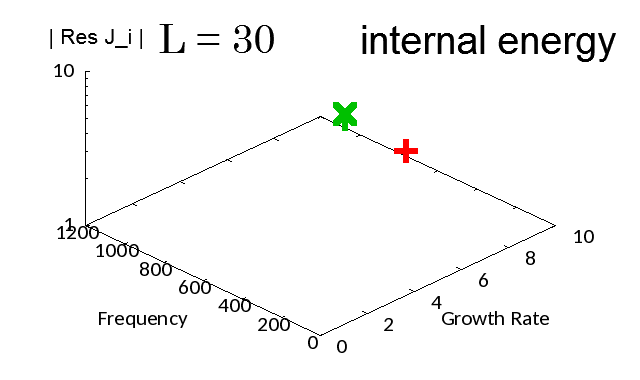}
\end{center}
\end{minipage} &
\begin{minipage}{0.45\hsize}
\begin{center}
\includegraphics[bb = 0 0 640 380, width = 80mm]{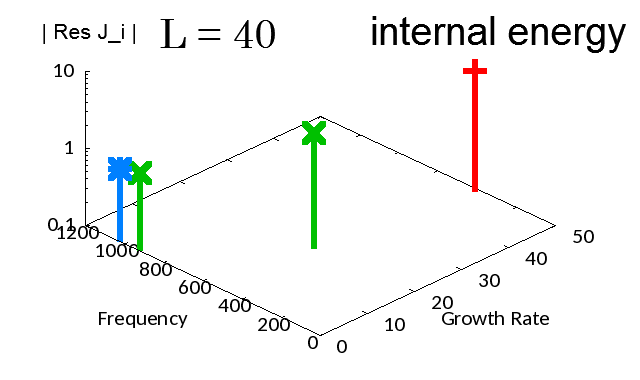}
\end{center}
\end{minipage} \\
\begin{minipage}{0.45\hsize}
\begin{center}
\includegraphics[bb = 0 0 640 380, width = 80mm]{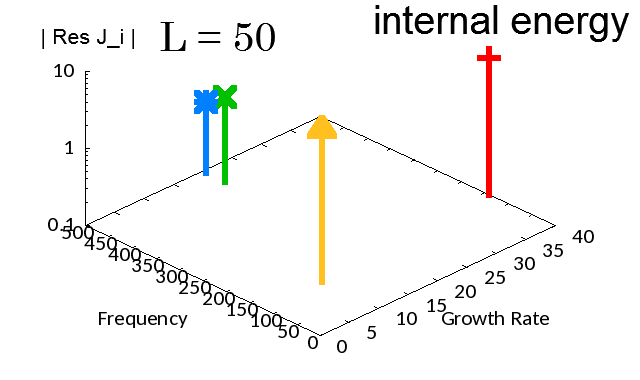}
\end{center}
\end{minipage} &
\begin{minipage}{0.45\hsize}
\begin{center}
\includegraphics[bb = 0 0 640 380, width = 80mm]{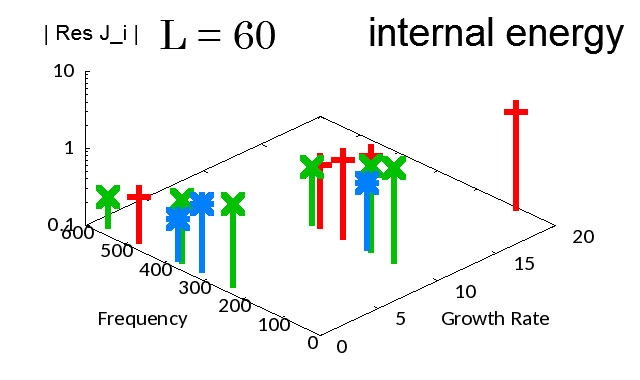}
\end{center}
\end{minipage} \\
\begin{minipage}{0.45\hsize}
\begin{center}
\includegraphics[bb = 0 0 640 480, width = 80mm]{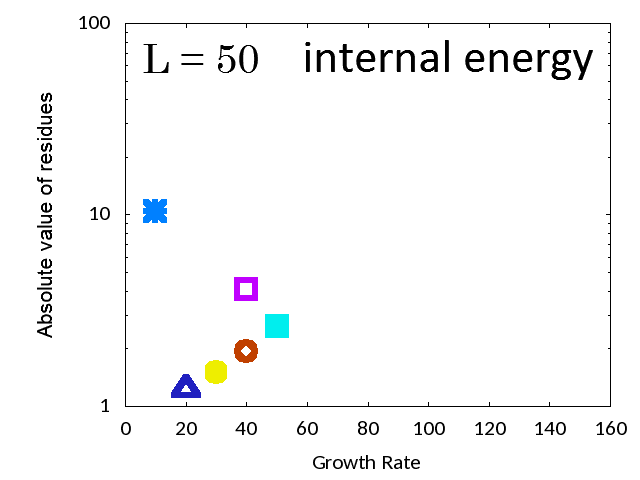}
\end{center}
\end{minipage} &
\begin{minipage}{0.45\hsize}
\begin{center}
\includegraphics[bb = 0 0 640 480, width = 80mm]{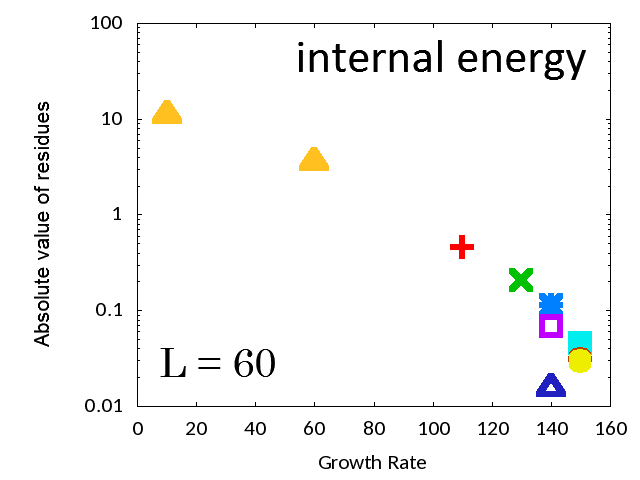}
\end{center}
\end{minipage} 
\end{tabular}
\caption{Same as Fig.~\ref{Delta1} but for the impulsive internal-energy perturbation.}
\label{Delta4}
\end{figure*}

\begin{figure*}
\begin{tabular}{cc}
\begin{minipage}{0.45\hsize}
\begin{center}
\includegraphics[bb = 0 0 650 510, width = 75mm]{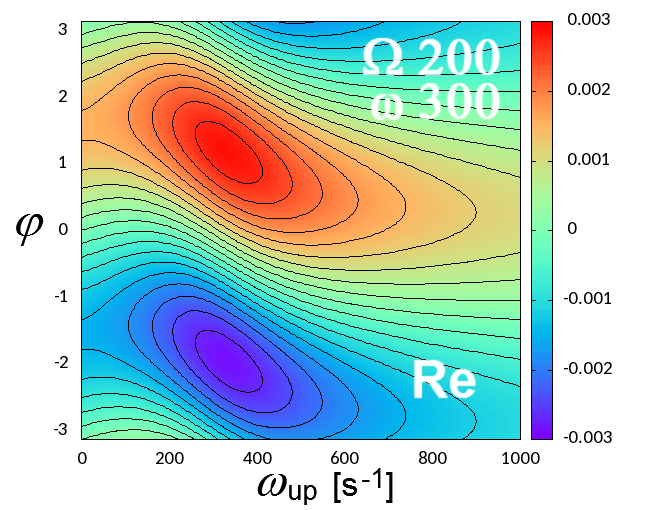}
\end{center}
\end{minipage} &
\begin{minipage}{0.45\hsize}
\begin{center}
\includegraphics[bb = 0 0 650 510, width = 75mm]{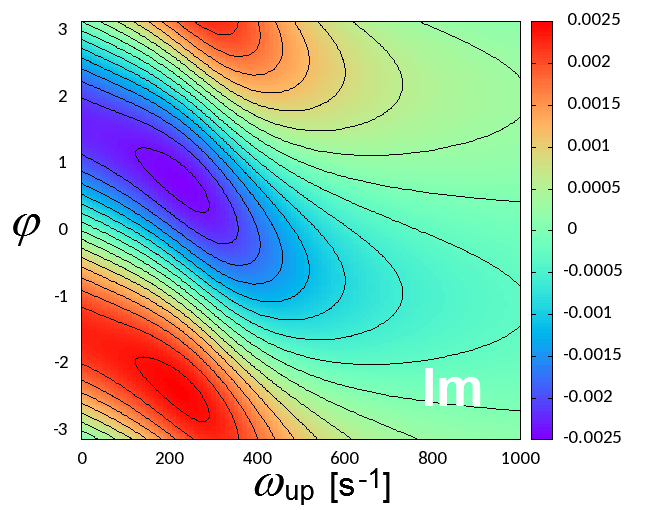}
\end{center}
\end{minipage} \\
\begin{minipage}{0.45\hsize}
\begin{center}
\includegraphics[bb = 0 0 650 510, width = 75mm]{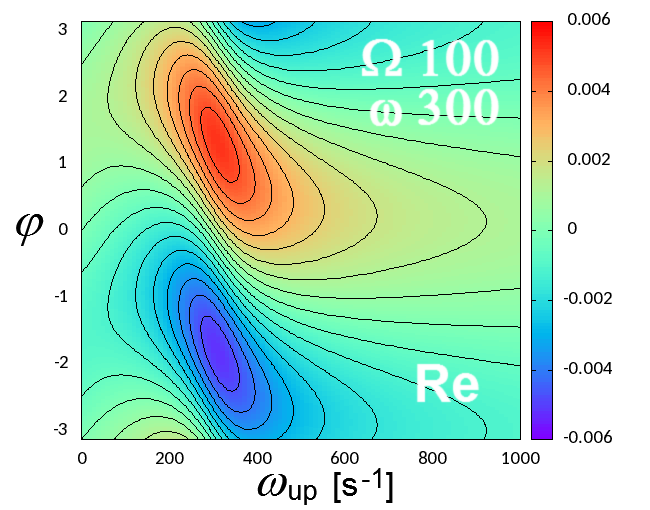}
\end{center}
\end{minipage} &
\begin{minipage}{0.45\hsize}
\begin{center}
\includegraphics[bb = 0 0 650 510, width = 75mm]{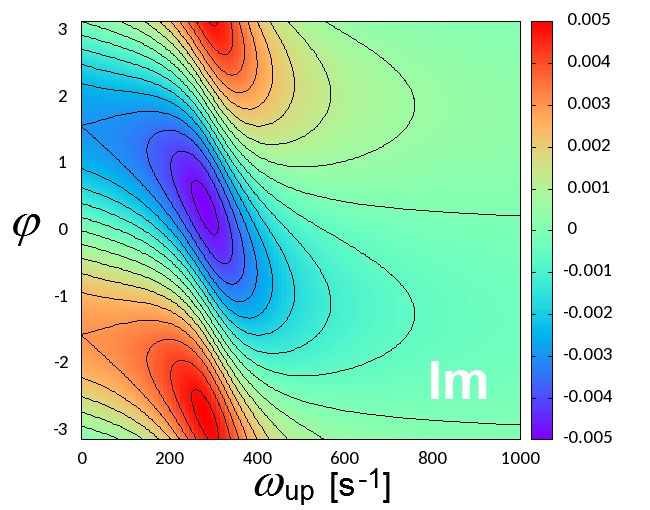}
\end{center}
\end{minipage} \\
\begin{minipage}{0.45\hsize}
\begin{center}
\includegraphics[bb = 0 0 650 510, width = 75mm]{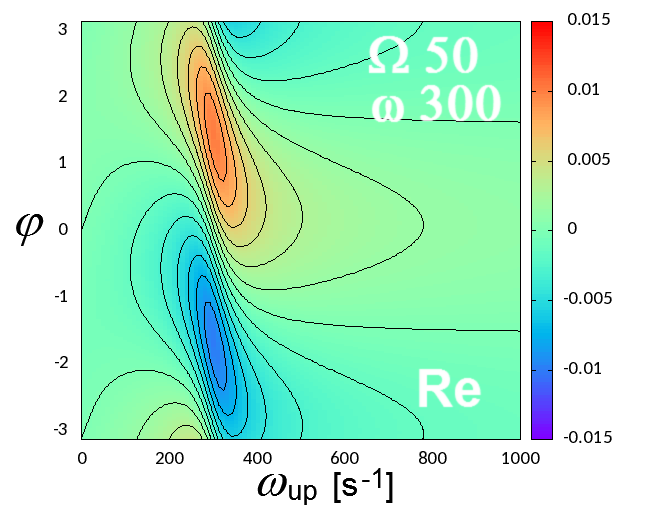}
\end{center}
\end{minipage} &
\begin{minipage}{0.45\hsize}
\begin{center}
\includegraphics[bb = 0 0 650 510, width = 75mm]{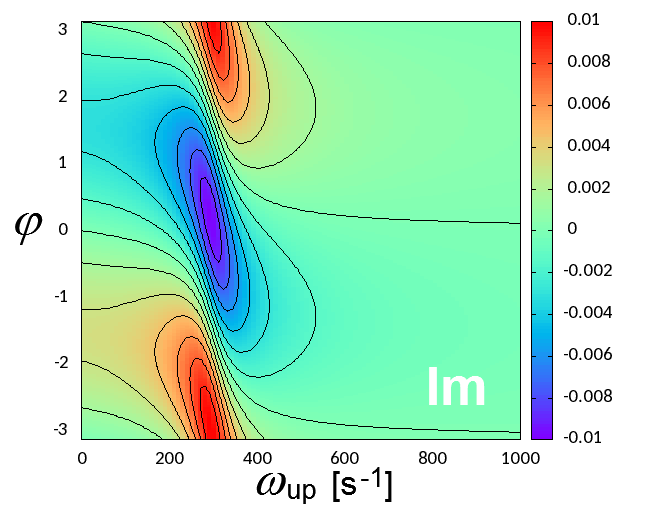}
\end{center}
\end{minipage} \\
\begin{minipage}{0.45\hsize}
\begin{center}
\includegraphics[bb = 0 0 650 510, width = 75mm]{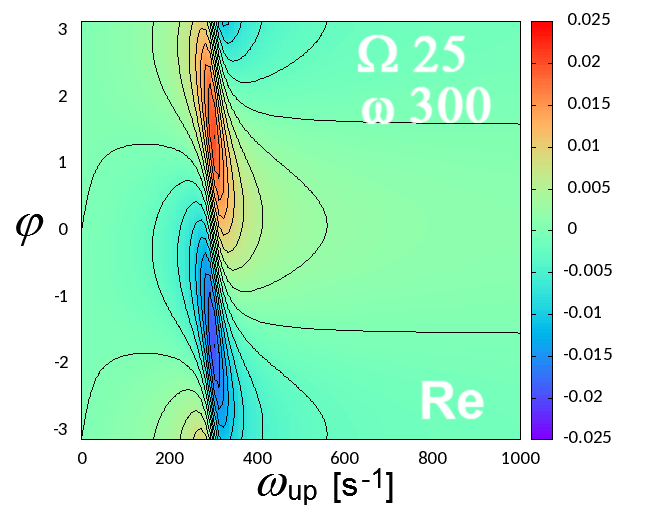}
\end{center}
\end{minipage} &
\begin{minipage}{0.45\hsize}
\begin{center}
\includegraphics[bb = 0 0 650 510, width = 75mm]{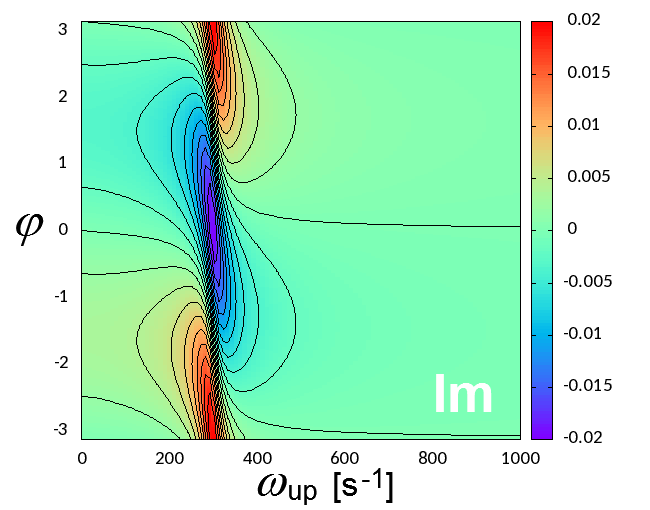}
\end{center}
\end{minipage}
\end{tabular}
\caption{Color maps with contour lines of the values of $z_i^*$ in Eq.~(\ref{AmpZ}) in the plane of the frequency and phase of the upstream perturbations, $\omega _\mathrm{up}$ and $\varphi$. The left and right panels show the real and imaginary parts, respectively. The oscillation frequency is fixed to $\omega = 300$ s$^{-1}$ and the growth rate is changed as $\Omega _j= 200, 100, 50$ and $25$ s$^{-1}$ from top to bottom.}
\label{Fig.zO}
\end{figure*}

\begin{figure*}
\begin{tabular}{cc}
\begin{minipage}{0.45\hsize}
\begin{center}
\includegraphics[bb = 0 0 650 510, width = 75mm]{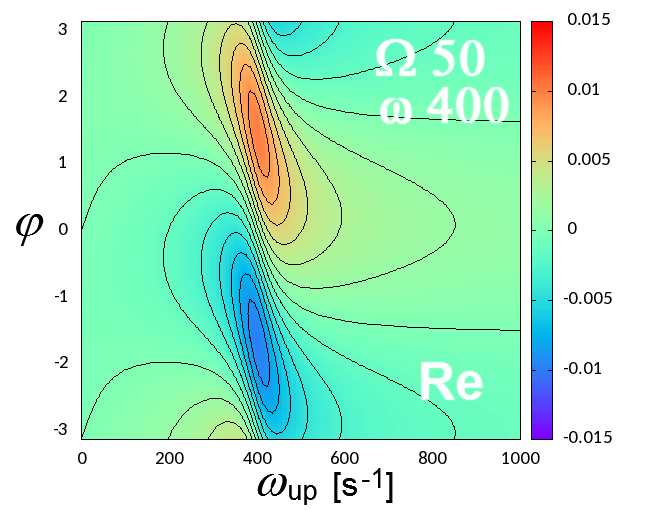}
\end{center}
\end{minipage} &
\begin{minipage}{0.45\hsize}
\begin{center}
\includegraphics[bb = 0 0 650 510, width = 75mm]{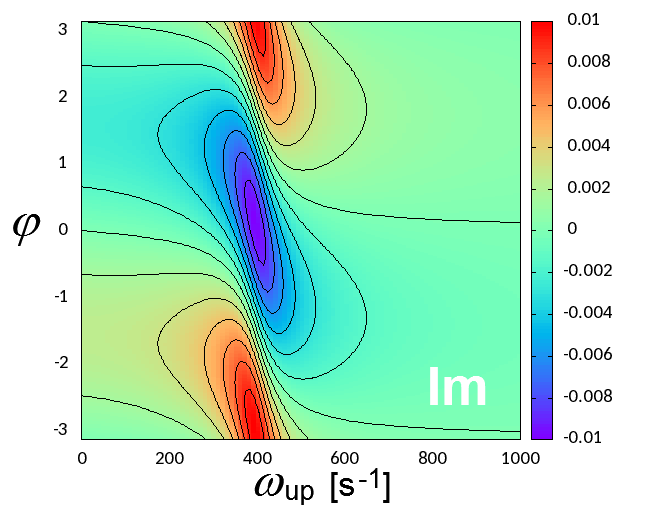}
\end{center}
\end{minipage} \\
\begin{minipage}{0.45\hsize}
\begin{center}
\includegraphics[bb = 0 0 650 510, width = 75mm]{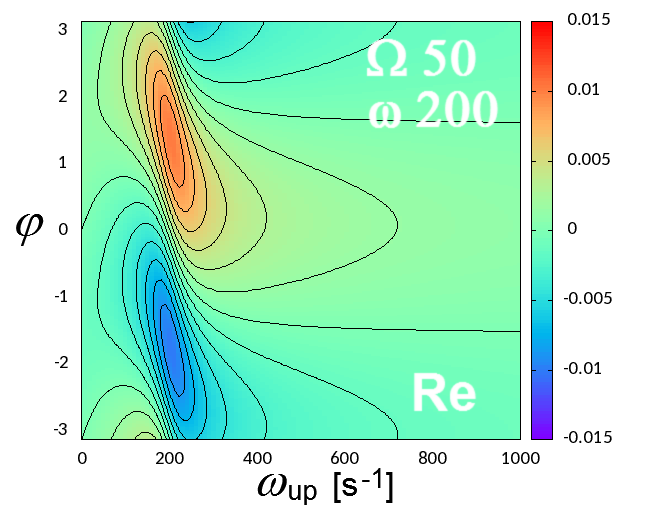}
\end{center}
\end{minipage} &
\begin{minipage}{0.45\hsize}
\begin{center}
\includegraphics[bb = 0 0 650 510, width = 75mm]{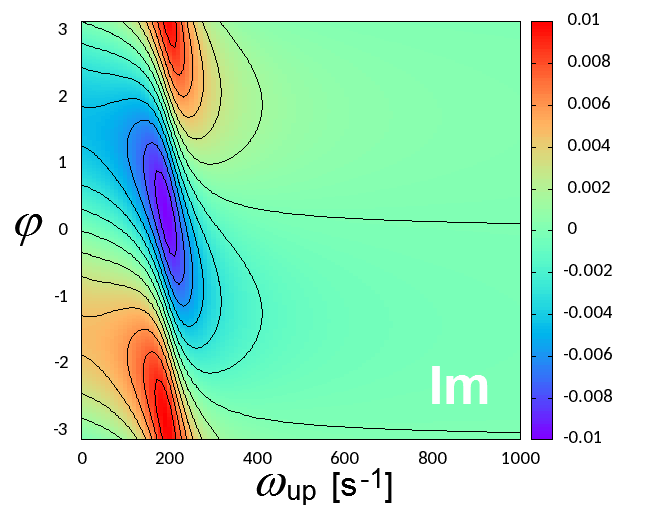}
\end{center}
\end{minipage} \\
\begin{minipage}{0.45\hsize}
\begin{center}
\includegraphics[bb = 0 0 650 510, width = 75mm]{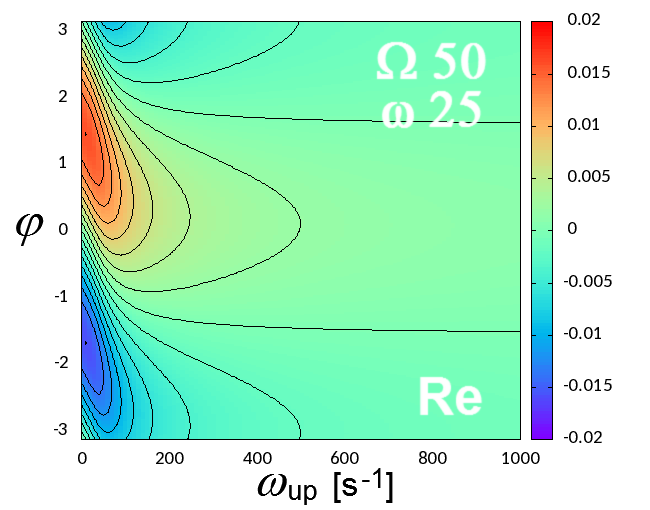}
\end{center}
\end{minipage} &
\begin{minipage}{0.45\hsize}
\begin{center}
\includegraphics[bb = 0 0 650 510, width = 75mm]{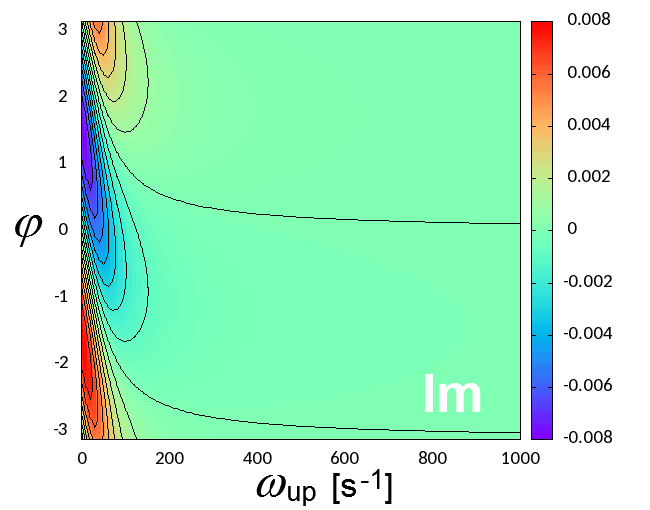}
\end{center}
\end{minipage} \\
\begin{minipage}{0.45\hsize}
\begin{center}
\includegraphics[bb = 0 0 650 510, width = 75mm]{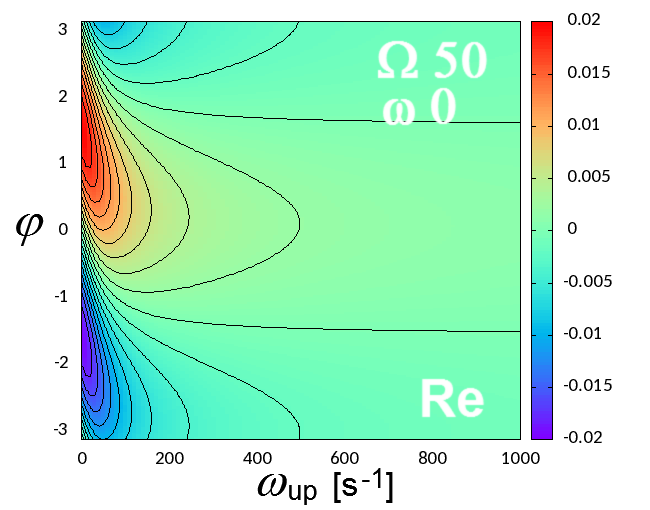}
\end{center}
\end{minipage} &
\begin{minipage}{0.45\hsize}
\begin{center}
\includegraphics[bb = 0 0 650 510, width = 75mm]{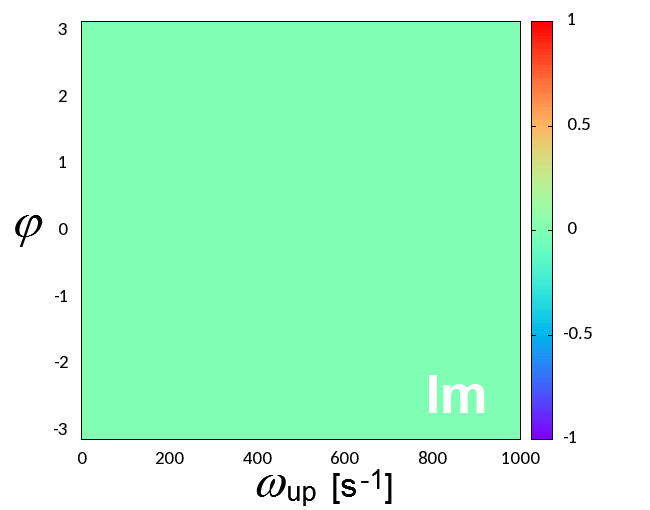}
\end{center}
\end{minipage}
\end{tabular}
\caption{Same as Fig.~\ref{Fig.zO} but the growth rate is fixed to $\Omega = 50$ s$^{-1}$ and the oscillation frequency is changed as $\omega _j = 400, 200, 25$ and $0$ s$^{-1}$ from the top to the bottom.}
\label{Fig.zw}
\end{figure*}

\begin{figure*}
\begin{tabular}{cc}
\begin{minipage}{0.45\hsize}
\begin{center}
\includegraphics[bb = 0 0 650 510, width = 85mm]{zO50w300Re.png}
\end{center}
\end{minipage} &
\begin{minipage}{0.45\hsize}
\begin{center}
\includegraphics[bb = 0 0 650 510, width = 85mm]{zO50w300Im.png}
\end{center}
\end{minipage} \\
\begin{minipage}{0.45\hsize}
\begin{center}
\includegraphics[bb = 0 0 650 510, width = 85mm]{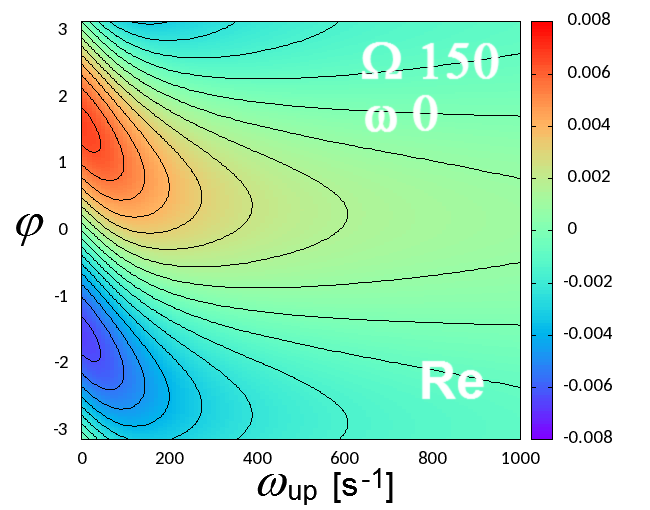}
\end{center}
\end{minipage} &
\begin{minipage}{0.45\hsize}
\begin{center}
\includegraphics[bb = 0 0 650 510, width = 85mm]{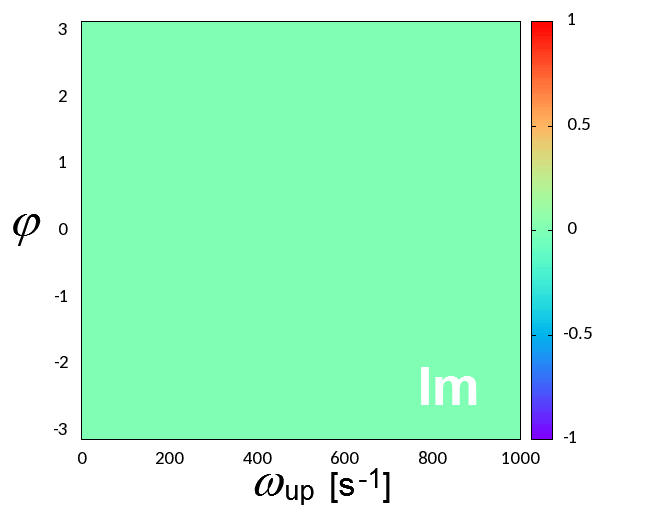}
\end{center}
\end{minipage}
\end{tabular}
\caption{Same as Figs.~(\ref{Fig.zO}) and (\ref{Fig.zw}) but here for the values of $\Omega $ and $\omega $ typical for SASI and convection. In the upper panels $\Omega _j = 50$ s$^{-1}$ and $\omega _j = 300$ s$^{-1}$ for SASI whereas in the lower panels $\Omega _j = 150$ s$^{-1}$ and $\omega _j = 0$ s$^{-1}$ for convection.}
\label{Fig.z}
\end{figure*}

More important is the fact that the upstream perturbation, upon hitting the shock, excites the intrinsic modes even without initial fluctuations in the down stream or the shocked region.

As discussed in Appendix \ref{app.amplitude}, the excited amplitude of the $j$-th mode is found to be roughly proportional to $z^*_i (s = s_j )\ (i=1,\cdots,n)$ as shown in (\ref{EqAmpP}), where $s_j=\Omega _j +i\omega _j$ denotes the position of the corresponding. 
More precisely speaking, the amplitude is given by the absolute value of a complex number, which is a sum of the products of $z_i^*(s = s_j)$ and the residue at $s=s_j$ of $\mathcal{J}_i^*$, where $\mathcal{J}_i$ is the response of the shock radius to the impulsive perturbation of the $i$-th component.

It will be informative to give $\mathop{\Res} \mathcal{J}_j^*$, since it is one of the elements to determine the excited amplitudes as mentioned above. 
We present their absolute values for the impulsive perturbations to density, radial velocity and internal energy separately in Figs.~\ref{Delta1}-\ref{Delta4}. As seen in the figures, the response to the perturbation in the internal energy is a few orders of magnitude smaller than those for the other cases, which are comparable to each other. Hence the density and/or velocity perturbations play more important roles to excite intrinsic modes than the internal energy perturbation.

Assuming that the upstream perturbation is a sinusoidal function in time, i.e., $z_i (t) \propto \sin (\omega _\mathrm{up} t+ \varphi)$ and hence $z_i^*(s) \propto (s\sin\varphi + \omega _\mathrm{up} \cos\varphi) /(s^2+\omega _\mathrm{up} ^2)$, we obtain
\begin{eqnarray}
\label{AmpZ}
z_i^*(s_j ) = \frac{(\Omega _j + i\omega _j)\sin \varphi + \omega _\mathrm{up}\cos \varphi}{(\Omega _j + i\omega _j)^2 +\omega _\mathrm{up}^2}.
\end{eqnarray}
We show the values of Eq.~(\ref{AmpZ}) in the plane of $\omega _\mathrm{up}$ and $\varphi$ as a color contour in Figs.~\ref{Fig.zO}, \ref{Fig.zw} and Fig.~\ref{Fig.z}. The former two figures illuminate the dependence on $\Omega _j$ and $\omega _j$. Fig.~\ref{Fig.z}, on the other hand, compares SASI and convection, choosing the typical values of $\Omega _j$ and $\omega _j$ for these cases: $\Omega _j = 50$ s$^{-1}$ and $\omega _j = 300$ s$^{-1}$ for SASI in the upper panel and $\Omega _j = 150$ s$^{-1}$ and $\omega _j = 0$ s$^{-1}$ for convection in the lower panel. 
As seen in the figure, $z_i^*$ takes maximum and minimum for the $\omega _\mathrm{up}$'s that are comparable to $\omega _j$ and is reduced to zero rather quickly as $\omega _\mathrm{up}$ departs from $\omega _j$, which is a common feature to SASI and convection.

\subsubsection{excitation of SASI}
In order to explain the amplitude distributions of excited SASI modes in the previous section, we focus on the case with $\varphi = 0$ or $\pi /2$. Paying attention to the fact that the growth rate $\Omega \sim 10$ s$^{-1}$ of SASI is in general much smaller than its oscillation frequency $\omega \sim 100$ s$^{-1}$, we consider some limiting cases below. 

We first take the high frequency limit of $\omega _\mathrm{up}$: $\Omega_j/\omega _\mathrm{up} \ll \omega _j/\omega _\mathrm{up} \ll 1$. Neglecting higher order terms, we find for high-frequency sine-type/cosine-type perturbations (HFSPs/HFCPs), respectively, the following results:
\begin{eqnarray}
\label{HFSASI}
z_i^*(s_j )&\sim & \left( i \frac{\omega _j}{\omega _\mathrm{up}}  \sin \varphi + \cos \varphi \right) \omega _\mathrm{up}^{-1},\\
&=& \left\{
\begin{array}{cc}
\omega_\mathrm{up} ^{-1} &\ (\varphi = 0: \mathrm{HFSPs} ) \\
\displaystyle i\frac{\omega _j}{\omega _\mathrm{up}}\omega _\mathrm{up}^{-1} &\ (\varphi = \pi /2 :\mathrm{HFCPs})
\end{array}\right. .
\end{eqnarray}
Note that HFCPs give much smaller amplitudes than HFSPs.

If we consider low-frequency upstream perturbations, on the other hand, $\omega _j/\omega _\mathrm{up} \gg \Omega_j/\omega _\mathrm{up}$ and $\omega _j/\omega _\mathrm{up} \gg 1$, we obtain the following:
\begin{eqnarray}
\label{LFSASI}
z_i^*(s_j ) &\sim & - \left( i \sin \varphi +\frac{\omega _\mathrm{up}}{\omega _j} \cos \varphi \right) \omega _j^{-1},\\
&=&\left\{
\begin{array}{cc}
- \displaystyle \frac{\omega _\mathrm{up}}{\omega _j} \omega_j ^{-1} &\ (\varphi = 0:\mathrm{LFSPs}) \\
-i\omega _j^{-1} &\ (\varphi = \pi /2:\mathrm{LFCPs})
\end{array}\right. ,
\end{eqnarray}
where LFSPs/LFCPs are the abbreviations of low-frequency sine-type/cosine-type perturbations. We note that the sine-type perturbation gives much smaller amplitudes than the cosine-type perturbation this time. 

The last case we consider is the one, in which the frequency of upstream perturbation is close to the oscillation frequency of SASI: $\omega _j/\omega _\mathrm{up} \sim 1 \gg \Omega _j/\omega _\mathrm{up}$.
We introduce here an index, $\nu > 0$, to measure how close these two frequencies are to each other:
\begin{equation}
\frac{\omega _j}{\omega _\mathrm{up}} = 1 + O\left( \left(\frac{\Omega_j}{\omega _\mathrm{up}} \right)^\nu \right).
\end{equation}
For $\nu \gg 1$, $\omega_\mathrm{up}$ is very close to $\omega_j$ and we obtain 
\begin{eqnarray}
\label{ExSFSASI}
z_i^*(s_j ) &\sim & \frac{1}{2} \left( \sin \varphi -i \cos \varphi \right) \Omega _j^{-1},\\
&=&\left\{
\begin{array}{cc}
- \displaystyle \frac{i}{2} \Omega_j ^{-1} &\ (\varphi = 0:\mathrm{ExSFSPs}) \\
\displaystyle \frac{1}{2} \Omega_j ^{-1} &\ (\varphi = \pi /2:\mathrm{ExSFCPs})
\end{array}\right. ,
\end{eqnarray}
where the prefix ExSF means `extremely similar frequency'.
For $\nu   \ll 1$, on the other hand, $\omega _\mathrm{up}$ is a bit more different from $\omega _j$ and we obtain
\begin{eqnarray}
\label{SFSASI}
z_i^*(s_j ) &\sim & \left( \frac{\omega _j}{\Omega_j} \right)^{\nu } \left( i \sin \varphi +\cos \varphi \right) \omega _j^{-1},\\
&=&\left\{
\begin{array}{cc}
\displaystyle \left( \frac{\omega _j}{\Omega _j} \right)^{\nu } \omega _j^{-1} &\ (\varphi = 0:\mathrm{SFSPs}) \\
\displaystyle i\left( \frac{\omega _j}{\Omega _j} \right)^{\nu} \omega _j^{-1} &\ (\varphi = \pi /2:\mathrm{SFCPs})
\end{array}\right. ,
\end{eqnarray}
where the prefix SF means 'similar frequency'.

From the comparison of the results given above, we can deduce which type of perturbations can drive what modes efficiently. In fact, we can easily get the following inequalities: ExSFCPs $\sim$ ExSFSPs $\propto (\omega _j/\Omega_j) \omega _j^{-1}\gg$ SFCPs $\sim$ SFSPs $\propto (\omega _j/\Omega _j)^{\nu} \omega_j^{-1} \gtsim$ LFCPs $\propto \omega _j^{-1} \gg $ LFSPs $\propto (\omega_\mathrm{up}^L / \omega  _j) \omega _j^{-1} \sim$ HFSPs $\propto (\omega _j/\omega _\mathrm{up}^H)\omega _j^{-1} \gg$ HFCPs $\propto (\omega _j/\omega _\mathrm{up}^H)^2\omega _j^{-1}$. We added here the superscripts L and H to emphasize that $\omega _\mathrm{up}^L \ll \omega _j \ll \omega _\mathrm{up}^H$. These inequalities indicate that the upstream perturbations with extremely similar frequencies are the most efficient to drive SASI. Even if the frequency is not extremely similar, $\nu \ll 1$, it is still effective. If $\omega _\mathrm{up}$ is substantially different from $\omega _j$, on the other hand, LFCPs are the most efficient to excite SASI and HFCPs are the most inefficient. The sine-type perturbations are always in the middle, being independent of $\omega _\mathrm{up}$. 

It makes sense that the upstream perturbations with the frequencies that are similar to that of an intrinsic mode are good at exciting the mode. As mentioned earlier, the resonance, which has an extra power dependence of time as $t\exp(\Omega_j t)\sin(\omega _jt)$, occur only if both the growth rate and oscillation frequency of upstream perturbations coincide those of an intrinsic mode. This is practically impossible and what occurs in reality is the large excitation amplitude of the intrinsic mode that has the oscillation frequency very close to that of the upstream perturbation.

If such a {\it quasi-resonance} condition is not satisfied, LFCPs are more effective because they are almost step functions perturbations, which have broad spectra in frequency. In fact, the amplitude excited by LFCPs is independent of $\omega _\mathrm{up}$, since there is no characteristic timescale in the step function. On the other hand, LFSPs go to zero in the limit of $\omega _\mathrm{up}\rightarrow 0$, which is obviously ineffective to drive intrinsic modes. As a matter of fact, $z_i^*$ is proportional to $\omega _\mathrm{up}$, which, too, goes to zero in the same limit.
It is also understandable that intrinsic modes with low oscillation frequencies are more strongly amplified by LFCPs and LFSPs.

High-frequency perturbations are inefficient in exciting intrinsic modes as is evident from the fact that the excited amplitudes go to zero in the limit of $\omega _\mathrm{up} \rightarrow \infty$. This may be interpreted as a consequence of the phase cancellation in rapid oscillation of the upstream perturbation.

\subsubsection{excitation of convections}
Next we consider the excited amplitudes of convective modes. It is noted here again that they are characterized by the vanishing oscillation frequency, $\omega _j= 0$. 
Neglecting higher order terms in Eq.~(\ref{AmpZ}) again, we obtain for the high frequency limit, i.e., $\Omega _j/\omega_\mathrm{up} \ll 1$, 
\begin{eqnarray}
\label{HFConv}
z_i^*(s_j ) &\sim & \left(\frac{\Omega _j}{\omega _\mathrm{up}}\sin \varphi + \cos \varphi \right)  \omega _\mathrm{up} ^{-1} , \\
&=&\left\{
\begin{array}{cc}
\omega_\mathrm{up}^{-1} &\ (\varphi = 0:\mathrm{HFSPs}) \\
\displaystyle \frac{\Omega _j}{\omega _\mathrm{up}} \omega _\mathrm{up}^{-1} &\ (\varphi = \pi /2:\mathrm{HFCPs})
\end{array}\right. .
\end{eqnarray}
For the low-frequency limit ($\Omega _j /\omega _\mathrm{up} \gg 1 $), on the other hand, we find
\begin{eqnarray}
\label{LFConv}
z_i^*(s_j ) &\sim & \left( \sin \varphi + \frac{\omega _\mathrm{up}}{\Omega _j}\cos \varphi\right) \Omega _j^{-1}, \\
&= & \left\{
\begin{array}{cc}
\displaystyle \frac{\omega _\mathrm{up}}{\Omega _j} \Omega _j^{-1}  &\ (\varphi = 0:\mathrm{LFSPs}) \\
\Omega_j^{-1} &\ (\varphi = \pi /2:\mathrm{LFCPs})
\end{array}\right. .
\end{eqnarray}
For the last case of $\Omega _j/\omega _\mathrm{up} \sim 1$, we obtain
\begin{eqnarray}
z_i^*(s_j ) &\sim & \frac{1}{2}(\sin \varphi + \cos \varphi) \Omega _j^{-1}, \\
&= & \left\{
\begin{array}{cc}
\displaystyle \frac{1}{2} \Omega _j^{-1}  &\ (\varphi = 0:\mathrm{SFSPs}) \\
\displaystyle \frac{1}{2} \Omega _j^{-1} &\ (\varphi = \pi /2:\mathrm{SFCPs})
\end{array}\right. .
\end{eqnarray}

Taking the derivative of $z_i^*$ with respect to $\omega _\mathrm{up}$, we see that the maximum value of $z_i^*$ is obtained at $\omega _\mathrm{up}/\Omega _j = \cos\varphi /(1+\sin \varphi)$ as 
\begin{eqnarray}
z_i^*(s_j ) &=& \frac{(1+\sin \varphi)}{2\Omega_j}, \\
&= & \left\{
\begin{array}{cc}
\displaystyle \frac{1}{2} \Omega _j^{-1}  &\ (\varphi = 0 \Longleftrightarrow \omega_\mathrm{up} = \Omega _j:\mathrm{SFSPs}) \\
\Omega_j^{-1} &\ (\varphi = \pi/2 \Longleftrightarrow \omega_\mathrm{up} = 0:\mathrm{LFCPs})
\end{array}\right. .
\end{eqnarray}
It is interesting that $\omega _\mathrm{up} = \Omega _j$ for the sine-type perturbations, since the former is an oscillation frequency whereas the latter is a growth rate.
For the cosine-type perturbations, on the other hand, the maximum is obtained in the low-frequency limit: $\omega _\mathrm{up}$ = 0. 

From the comparison of the results given above, we obtain the following inequalities:  SFCP $\sim$ SFSPs $\sim$ LFCPs $\propto \Omega_j ^{-1} \gg$ LFSPs $\propto (\omega _\mathrm{up}^L/\Omega _j)\Omega _j^{-1} \sim$ HFSPs $\propto (\Omega _j/\omega _\mathrm{up}^H)\Omega _j^{-1} \gg$ HFCPs $\propto (\Omega _j/\omega _\mathrm{up}^H)^2\Omega _j^{-1}$.
It is hence concluded that LFCPs, SFCPs and SFSPs are more efficient in the excitation of convections as for SASI. The excited amplitudes are inversely proportional to $\Omega _j$ and independent of $\omega _\mathrm{up}$, which is similar to the SASI case again. 

\subsection{\textcolor{black}{General perturbations}}\textcolor{black}{
So far we focused on the sinusoidal perturbations given in Eqs.~(\ref{EXP1})-(\ref{EXP3}) alone. Note that any perturbation can be expressed in principle as a superposition of sinusoidal ones according to the Fourier expansion. In the linear analysis, the result for non-sinusoidal perturbations is also a sum of the individual results for sinusoidal one with different frequencies. This is demonstrated explicitly in Appendix \ref{G}, where we present the excited amplitudes for an arbitrary superposition of sinusoidal perturbations. As shown there, the linear combination may enhance or suppress the excited amplitudes of intrinsic modes depending on their phases.}

\subsection{Implications for the shock dynamics}
Recent multi-dimensional numerical simulations of O- and Si-burnings in massive stars \citep[e.g.][]{Meakin06, Chat,Couch3,Muller16} demonstrated that the fluctuations in these layers may reach as large as a few to ten percent. On the other hand, \citet{KT} showed by linear analysis that the fluctuations can be amplified during the super-sonic accretion by a factor of  several to ten, being proportional to $l$. In this paper we have demonstrated that these upstream perturbations can excite intrinsic SASI or convective modes when they hit the shock surface. We have observed that the excited amplitudes can be as high as $\sim 1$-$10$ times those of the original upstream perturbations if the frequency of the upstream perturbation is appropriate.

Fortunately, this seems to be the case indeed. \citet{KT} reported that Si/O layers acquire temporal variations of about $10^2$ s$^{-1}$ during the accretion, which is not very different from the oscillation frequencies of SASI or the growth rates of convections. It is hence likely that these intrinsic modes are excited with large amplitudes by the fluctuations in the accreting Si/O layers, which also seems consistent with recent numerical studies \citep{Couch, Couch2,MJ14,Couch3}.

\textcolor{black}{Recently \citet{Abd2} investigated the amplification of the turbulent kinetic energy by the passage through a planar shock in the linear interaction approximation. Although their formulation cannot treat the shock instabilities such as SASI and convection, they found that the turbulent energy can be amplified indeed, provided a vorticity wave enters the shock at some appropriate angles and the phase lag of an entropy wave, if exists, is not so large. As a result, they also concluded that the fluctuations in the envelopes can help shock revival.}

\section{Summary \& Conclusion}\label{sec.summary2}
We investigated the links between the fluid instabilities in the iron core of CCSNe and the non-spherical fluctuations in the accreting envelopes, which are expected to be generated by violent convections in the Si/O layers. We formulated the problem as an initial-boundary-value problem and employed Laplace transform to find the intrinsic modes such as SASI and convection and calculate their excitations by the upstream perturbations.

We first sought the intrinsic modes, especially unstable ones, which are eigen modes in the absence of the upstream perturbations, for background flows with different model parameters. The results are consistent with those of the previous normal mode analyses \citep{Foglizzo06,Yamasaki07,Guilet12}: SASI, which seems to be driven by the advective-acoustic cycle, prevails in low neutrino luminosities whereas convection becomes dominant as the luminosity increases.
Then we investigated the amplitudes of the modes that are excited by the perturbations in the matter accreting on shock. Based on our previous work on the amplification of fluctuations during the accretion \citep{KT}, we approximated the perturbations just ahead of the shock by the functional form given in Eqs.~(\ref{EXP1})-(\ref{EXP3}). Changing the parameters, i.e., phase and frequencies of the perturbations, rather arbitrarily, we systematically studied the couplings between the upstream perturbations with the intrinsic modes for different background flows. We also gave analytic expressions to the excited amplitudes in some limiting cases.

We showed analytically that the resonance, in which the growth rate of an unstable mode acquire an extra power of $t$, occurs only when both of their oscillation frequencies and growth rates coincide exactly between the intrinsic mode and the upstream perturbation. Hence it does not happen practically.
What occurs actually is, as we demonstrated both analytically and numerically, that the excited amplitude becomes larger when the upstream frequency is close either to one of the oscillation frequencies for SASI or to one of the growth rates for convection. If these frequencies are not very close to each other, the excitation efficiency declines rather quickly. It is also demonstrated that the discontinuous perturbation given as a step function of time can excite various modes thanks to its broad spectrum in frequency. Fortunately, the upstream perturbations acquire temporal variations during accretion, whose timescales are not much different from those of SASI or convection as reported in \citet{KT}.

The magnitude of the excitations may not be small. In fact, the violent convections in O- and Si-burnings are supposed to generate fluctuations at several percent level \citep{Bazan, Asida, Meakin06, Meakin07,Arnett, Chat,Muller16}, which will then be amplified during accretion by a factor of a few to ten \citep{KT}. They in turn may excite some intrinsic modes with amplitudes that are larger by another factor of $1$-$10$.
We may hence expect that the shock radius can be perturbed at several tens percent initially when the upstream fluctuations hit the shock. Of course, our linear analysis will not be applicable to such large-amplitude fluctuations.  It should be also noted that even if the excited amplitude is not so large, the upstream perturbation shorten the time it takes the fluid instabilities to grow to non-linear phases. All these findings seem to be consistent what was observed in the recent numerical simulations \citep{Couch,Couch2,MJ14,Fernandez14,Couch3}. 
\textcolor{black}{The latest study by \citet{Abd2} also found that the turbulent kinetic energy can be amplified by the passage of through the shock wave although their formulation cannot treat the shock instabilities such as SASI and convection.}

Although the turbulent stellar structures are booming now, other multi-dimensional effects such as magnetic fields and stellar rotation may also play an important role in the links between the perturbations in the envelope and the development of instabilities in the core and should be studied next.

\acknowledgements
K.T.~thanks H. Nagakura for discussions on the dynamics of CCSNe. \textcolor{black}{We acknowledge the center for the Computational Astrophysics, CfCA, the National Astronomical Observatory of Japan for the use of the XC30 and the general common use computer system.} This work is partially supported by a Research Fellowship for Young Scientists from the Japan Society for the Promotion of Science as well as by the Grants-in-Aid for Scientiﬁc Research (A) (Nos. 24244036, 24740165), the Grants-in-Aid for Scientiﬁc Research on Innovative Areas, New Development in Astrophysics through Multi-messenger Observations of Gravitational Wave Sources (No. 24103006),

\appendix
\section{A. Matrices and vectors}\label{app.matrix}
In this section, we give the matrices and vectors that appeared in Section \ref{sec.2.method}.

The vector that describes perturbed states is defined by Eq.~(\ref{eq.perturbedstate}).
In this order of the components, the matrices in the basic equations in the form of 
\begin{eqnarray}
\label{eq.AppBasic}
M\frac{\partial \bf{y}}{\partial t} +A'\frac{\partial \bf{y}}{\partial r} +B'\bf{y} = {\bf 0}
\end{eqnarray}
are given as follows:
\begin{eqnarray}
M(r) &=&\left(
\begin{array}{cccccc}
1\ & 0\ & 0\ & 0\ & 0\ & 0 \ \\
0\ & 1\ & 0\ & 0\ & 0\ & 0 \ \\
0\ & 0\ & 1\ & 0\ & 0\ & 0 \ \\
\displaystyle -\frac{p}{\rho v_r^2}\ & 0\ & 0\ &\displaystyle \frac{\varepsilon }{v_r^2 } \ & 0 \ & 0 \ \\
0\ & 0\ & 0\ & 0\ & 1\ & 0 \ \\
0\ & 0\ & 0\ & 0\ & 0\ & 1 \
\end{array}\right), \\
A' (r) &=& \left(
\begin{array}{cccccc}
v_r\ & v_r\ & 0\ & 0\ & 0\ & 0 \ \\
\displaystyle \frac{1}{v_r}\frac{\partial p}{\partial \rho}\ & v_r\ & 0\ & \displaystyle \frac{\varepsilon }{\rho v_r}\frac{\partial p}{\partial \varepsilon} \ & \displaystyle \frac{Y_e}{\rho v_r}\frac{\partial p}{\partial Y_e}\ & 0 \ \\
0\ & 0\ & v_r\ & 0\ & 0\ & 0 \ \\
\displaystyle -\frac{p}{\rho v_r} \ & 0 \ & 0\ &\displaystyle \frac{\varepsilon }{v_r}\ & 0 \ & 0 \ \\
0\ & 0\ & 0\ & 0\ & v_r\ & 0 \ \\
0\ & 0\ & 0\ & 0\ & 0\ & v_r \
\end{array}\right), \\
B'(r) &=& \left(
\begin{array}{cccccc}
0\ & 0\ & \displaystyle -v_r\frac{l(l+1)}{r} \ & 0\ & 0\ & 0 \ \\
B'_{11} &\displaystyle 2\frac{\diff v_r}{\diff r}\ & 0\ & \displaystyle \frac{1}{\rho v_r} \frac{\diff }{\diff r}\left(\varepsilon \frac{\partial p}{\partial \varepsilon} \right) \ & \displaystyle \frac{1}{\rho v_r}\frac{\diff}{\diff r} \left( Y_e\frac{\partial p}{\partial Y_e} \right) \ & 0 \ \\
\displaystyle \frac{1}{r v_r}\frac{\partial p}{\partial \rho} \ & 0\ & \displaystyle \frac{v_r}{r} +\frac{\diff v_r}{\diff r} \ & \displaystyle \frac{\varepsilon}{r \rho v_r}\frac{\partial p}{\partial \varepsilon} \ & \displaystyle \frac{Y_e}{r\rho v_r}\frac{\partial p}{\partial Y_e}\ & 0 \ \\
B'_{41}\ & \displaystyle \frac{1}{v_r}\left(\frac{\diff \varepsilon}{\diff r} - \frac{p}{\rho ^2}\frac{\diff \rho}{\diff r}\right) \ & 0\ & B'_{44}\ & B'_{45}\ & 0 \ \\
\displaystyle \frac{v_r}{Y_e}\frac{\diff Y_e}{\diff r} -\frac{m_b}{Y_e}\frac{\partial \lambda}{\partial \rho}\ & \displaystyle \frac{v_r}{Y_e}\frac{\diff Y_e}{\diff r} \ & 0\ & \displaystyle -\frac{m_b \varepsilon}{\rho Y_e}\frac{\partial \lambda}{\partial \varepsilon}\ & \displaystyle \frac{v_r}{Y_e}\frac{\diff Y_e}{\diff r}-\frac{m_b}{\rho}\frac{\partial \lambda}{\partial Y_e}\ & 0 \ \\
0\ & 0\ & 0\ & 0\ & 0\ & \displaystyle \frac{v_r}{r} +\frac{\diff v_r}{\diff r} \
\end{array}\right), \nonumber \\
\left. \right.
\end{eqnarray}
with
\begin{eqnarray}
B'_{11} &=& \frac{1}{\rho v_r}\frac{\diff}{\diff r}\left( \rho\frac{\partial p}{\partial \rho} \right) -\frac{1}{\rho v_r}\frac{\diff p}{\diff r} ,\\
B'_{41} &=& \frac{1}{v_r}\left( \frac{p}{\rho^2}\frac{\diff \rho}{\diff r}   -\frac{1}{\rho}\frac{\diff \rho}{\diff r}\frac{\partial p}{\partial \rho} -\frac{\rho}{v_r}\frac{\partial q}{\partial \rho} \right), \\
B'_{44} &=& \frac{1}{v_r}\left( \frac{\diff \varepsilon}{\diff r} -\frac{\varepsilon }{\rho ^2}\frac{\diff \rho}{\diff r}\frac{\partial p}{\partial \varepsilon } -\frac{\varepsilon }{v_r}\frac{\partial q}{\partial \varepsilon} \right), \\
B'_{45} &=& -\frac{1}{v_r}\left( \frac{Y_e }{\rho ^2}\frac{\diff \rho}{\diff r}\frac{\partial p}{\partial Y_e} +\frac{Y_e }{v_r}\frac{\partial q}{\partial Y_e} \right) .
\end{eqnarray}
Note that we took $\rho, \varepsilon$ and $Y_e$ as independent thermodynamic quantities in the above expression and we did not express the fixed variables for notational simplicity when we take their partial derivatives. We obtain the basic equations in the form of  Eq.~(\ref{eq.linearized}) by defining $A := -A'^{-1}M$ and $B := -A'^{-1}B$ and multiplying Eq.~(\ref{eq.AppBasic}) with $A^{-1}$ from the left.

The linearized Rankine-Hugoniot relations are schematically given by 
\begin{equation}
\label{eq.AppRH}
P^{(d)}{\bf y}(r_\mathrm{sh},t) = P^{(u)}{\bf z}(t)  + \frac{\partial }{\partial t} \frac{\delta r_\mathrm{sh}}{r_\mathrm{sh}} {\bf c}' + \frac{\delta r_\mathrm{sh}}{r_\mathrm{sh}} {\bf d}',
\end{equation}
where ${\bf z}(t)$ is the vector that describes the perturbed quantities just in front of the shock surface.
The matrices $P$ and vectors ${\bf c}'$ and ${\bf d}'$ in the equation have the following forms:
\begin{eqnarray}
P &=& \left(
\begin{array}{cccccc}
1\ & 1\ & 0\ & 0\ & 0 \ & 0\ \\
\displaystyle v_{r} +\frac{1}{v_{r}} \frac{\partial p}{\partial \rho} \ & 2v_{r} \ & 0\ & \displaystyle \frac{\varepsilon }{j} \frac{\partial p}{\partial \varepsilon } \ & \displaystyle \frac{Y_{e}}{j} \frac{\partial p}{\partial Y_e} \ & 0\  \\
0\ & 0\ & v_r \ & 0\ & 0\ & 0\ \\
\displaystyle \frac{E}{\rho} +\frac{\partial p}{\partial \rho} \ & \displaystyle \frac{E +p + \rho v_r^2}{\rho} \ & 0 \ & \displaystyle \varepsilon +\frac{\varepsilon }{\rho}\frac{\partial p}{\partial \varepsilon} \ & \displaystyle \frac{Y_e}{\rho}\frac{\partial p}{\partial Y_e} \ & 0\ \\
1\ & 1\ & 0\ & 0\ & 1\ & 0\ \\
0\ & 0\ & 0\ & 0\ & 0\ & v_r\ 
\end{array}\right), \\
{\bf c}' &=& \left( \frac{r_\mathrm{sh} \jump{\rho }}{j}, \ 0, \ 0 ,\ \frac{r_\mathrm{sh}\jump{E}}{j},\ \frac{r_\mathrm{sh}\jump{\rho }}{j},\  0 \right)^T, \\
{\bf d}' &=& \left( 0,\ 2\jump{v_r}+\frac{GM\jump{\rho }}{r_\mathrm{sh}j} ,\ \frac{\jump{p}}{j },\  -\frac{r_\mathrm{sh}\jump{\rho q}}{j},\ -\frac{r_\mathrm{sh}m_b\jump{\lambda}}{j Y_e},\ 0 \right)^T, 
\end{eqnarray}
and the superscripts $\mathrm{(u)}$ and $\mathrm{(d)}$ of $P$ mean that it is evaluated with the background quantities above and below the shock wave, respectively. 
The mass flux is denoted by $j:= \rho _0 \up v_{r0}\up = \rho _0\down v_{r0}\down$ and the bracketed symbol, $\jump{X}:= X \down -X\up$, is a jump of a quantity $X$ across the shock.

Defining $R := P^{-1\mathrm{(d)}}P^\mathrm{(u)}$, ${\bf c} := P^{-1\mathrm{(d)}}{\bf c}'$ and ${\bf d} := P^{-1\mathrm{(d)}}{\bf d}'$ and multiplying Eq.~(\ref{eq.AppRH}) with $P^{-1\mathrm{(d)}}$ from the left, we obtain the linearized Rankine-Hugoniot relations in the form of Eq.~(\ref{eq.outerboundary}). Then, Laplace-transforming this equation with respect to $t$, we obtain Eq.~(\ref{OB}).

\section{B. Solution of the boundary-value problem}\label{app.solvetheeigenvalueproblem}
We show here that the boundary-value problem given in Eqs.~(\ref{eq.L-linearized})-(\ref{IB}) is easily solved, provided the function $f^*$ that gives the inner boundary condition in Eq.~(\ref{IB}) is linearly dependent on ${\bf{y}}^*$. In fact, we can then obtain the solution, $\delta r_\mathrm{sh}^*/r_\mathrm{sh}$, for a given $s$ by integrating the ordinary differential equation (\ref{eq.L-linearized}) twice as discussed below. 

The idea is quite simple: this is a special case of the Newton-Raphson method, in which an approximate solution $x^\star$ to $F(x^\star) = 0$ is improved by solving the linearized equation
\begin{eqnarray}
F(x^\star +\delta x) \sim F(x^\star) +J(x^\star)\delta x = 0,
\end{eqnarray}
where $J$ is the Jacobian, $\diff F/\diff x$. In fact, $F$ and $x$ can be replaced with $f^*$ and $\delta r_\mathrm{sh}^*/r_\mathrm{sh}$, respectively, in our problem. 
If $F \propto x$, the linearized equation is identical to the original equation and hence the improved approximation given by $x_\mathrm{sol} = x^\star +\delta x = x^\star -J^{-1}F$ is actually the exact solution. In the following, we show that our problem is indeed linear and obtain the Jacobian matrix.

The linearity is almost obvious since we are considering a linearized system: the inner boundary condition takes in general the form given in Eq.~(\ref{An2}) and ${\bf y}^*(r_{\nu _e},s)$ is proportional to $\delta r_\mathrm{sh}^*/r_\mathrm{sh}$, from which the linear dependence of the function $f^*$ on $\delta r_\mathrm{sh}^*/r_\mathrm{sh}$ follows:
\begin{eqnarray}
f^* = {\bf a}^*(s)\cdot \tilde{\Lambda}^*(s) \left[ (s{\bf c} +{\bf d})\frac{\delta r_\mathrm{sh}^*(s)}{r_\mathrm{sh}} +R{\bf z}^*(s)   \right] - {\bf a}^*(s)\cdot{\bf \tilde{h}}^*[{\bf y}_0](s) +b^*(s) = 0.
\end{eqnarray}

The Jacobian matrix is then given as
\begin{eqnarray}
J = \frac{\diff f^*}{\diff (\delta r_\mathrm{sh}^*/r_\mathrm{sh})} 
= {\bf a}^*(s)\cdot \tilde{\Lambda}^*(s)(s {\bf c} +{\bf d}),
\end{eqnarray}
where ${\bf a}^*(s)$ is known once the function $f^*$ is given; the other factor, $s\tilde{\Lambda}^*(s)(s {\bf c} +{\bf d})$, can be calculated by integrating the ordinary differential equation for the modified outer boundary condition:
\begin{eqnarray}
\label{MOB}
{\bf y}^*(r_\mathrm{sh},s) &=& s{\bf c} +{\bf d}.
\end{eqnarray}
We thus come to the conclusion we can obtain the solution of the boundary-value problem for a given $s$ by integrating the ordinary differential equation (\ref{eq.L-linearized}) twice: once for the outer boundary condition Eq.~(\ref{OB}) with an arbitrary guess for $\delta r_\mathrm{sh}/r_\mathrm{sh}$, and the other time for the modified boundary condition, Eq.~(\ref{MOB}).

\section{C. Inversion formula}\label{app.inverse}
\begin{figure}
\begin{tabular}{cc}
\begin{minipage}{0.45\hsize}
\begin{center}
\includegraphics[bb = 0 0 992 638, width = 80mm]{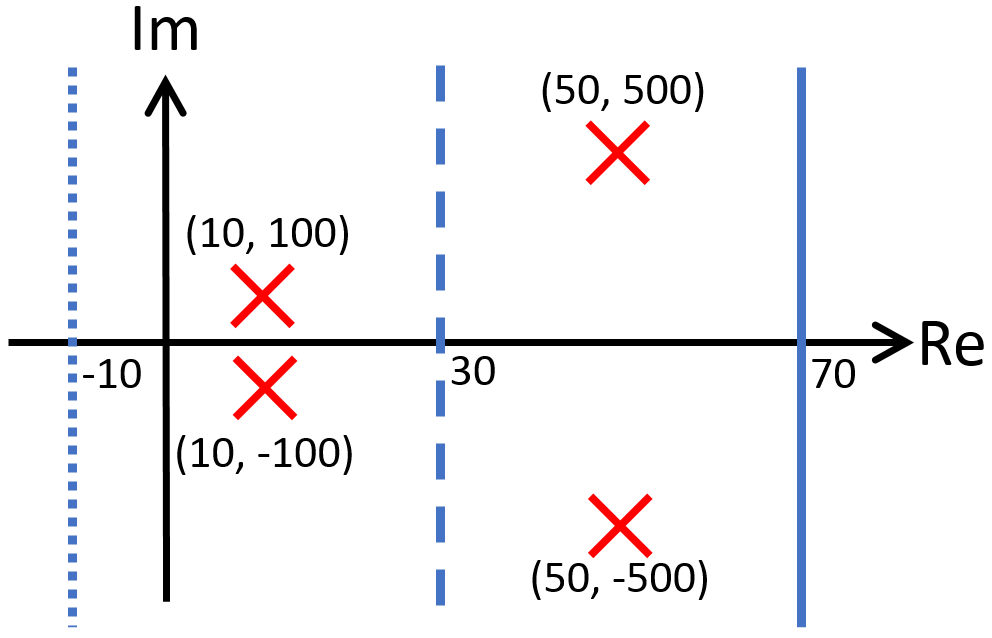}
\end{center}
\end{minipage} &
\begin{minipage}{0.45\hsize}
\begin{center}
\includegraphics[bb = 0 0 640 480, width = 80mm]{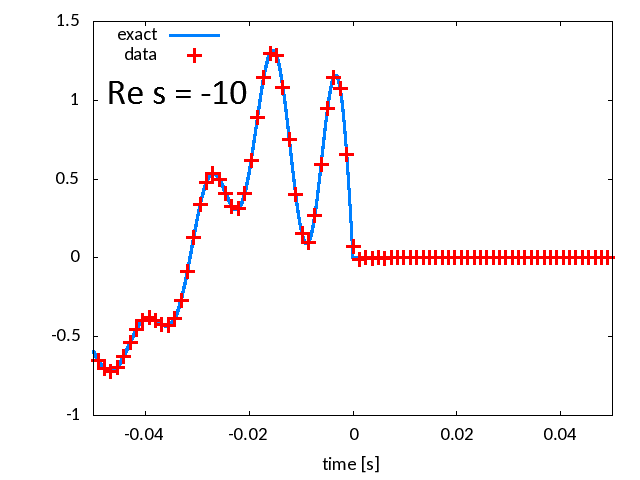}
\end{center}
\end{minipage} \\
\begin{minipage}{0.45\hsize}
\begin{center}
\includegraphics[bb = 0 0 640 480, width = 80mm]{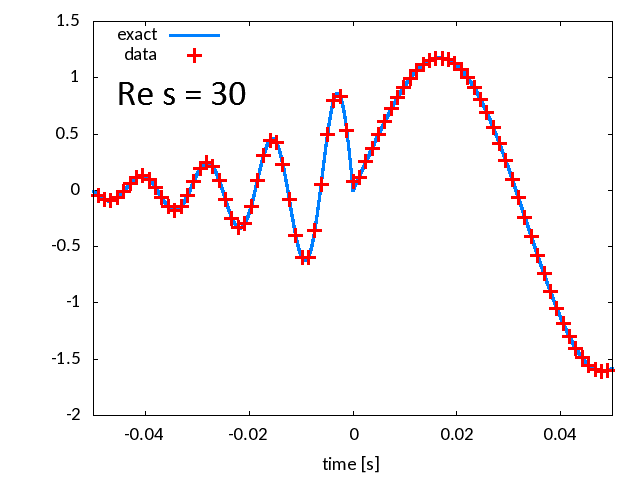}
\end{center}
\end{minipage} &
\begin{minipage}{0.45\hsize}
\begin{center}
\includegraphics[bb = 0 0 640 480, width = 80mm]{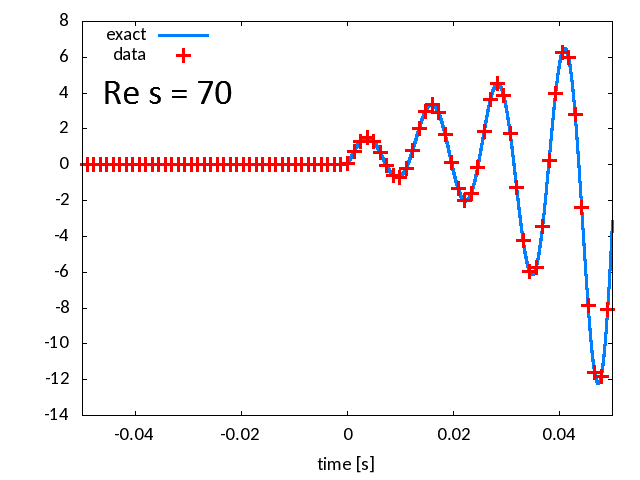}
\end{center}
\end{minipage}
\end{tabular}
\caption{The integral given in Eq.~(\ref{InverseInt}) for $f^*$ given in Eq.~(\ref{fstarex}) along three different paths with $\mathrm{Re}\ s = -10, 30, 70$. The upper left panel depicts the locations of the poles of $f^*$ (red crosses) and the three integral paths (three different types of blue lines) on the complex plane. 
The upper right panel shows the exact solution (blue line) and the results for numerical integration (red crosses) along the path with $\mathrm{Re}\ s = -10$. The lower panels are the same but for the integral path with $\mathrm{Re}\ s = 30$ (left) and $\mathrm{Re}\ s = 70$ (right), respectively.}
\label{Fig.Inv}
\end{figure}

Inverse Laplace transform is given by the Fourier-Mellin formula \citep{Text}:
\begin{equation}
\label{InverseInt}
f(t) = \frac{1}{2\pi i} \int _L e^{ts}\mathcal{L}[f](s) \mathrm{d}s,
\end{equation} 
where the integral path, $L$, is a straight line that is parallel to the imaginary axis and lies to the right of all the poles of $f^*(s) = \mathcal{L}[f](s)$.
We do not know {\it a priori} where those poles are, however, and hence it may happen that the path sits to the left of some poles.
We show below what happens if we employ such a wrong path.

Let us take a function $f(t) = \exp(\Omega t)\sin(\omega t +\phi)$ as an example. Its Laplace transform is
\begin{eqnarray}
f^*(s)  = \frac{1}{2i}\left( \frac{e^{i\phi }}{s-\Omega -i\omega } - \frac{e^{-i\phi }}{s-\Omega +i\omega }\right),
\end{eqnarray}
for which the poles are located at $s = \Omega +\pm i \omega$.
For appropriate paths with $\mathrm{Re}\ s > \Omega $, the integral, Eq.~(\ref{InverseInt}), correctly reproduces the original function as
\begin{eqnarray}
\left\{
\begin{array}{cc}
e^{\Omega t}\sin (\omega t+\phi) & (t > 0), \\
\displaystyle \frac{1}{2} \sin \phi& (t = 0), \\
0 & ( t < 0).
\end{array}\right.
\end{eqnarray}
Here the point is that the inversed function vanishes at $t < 0$, which is guaranteed by the fact that the integral path is running to the right of all poles.

If, on the other hand, the path is inappropriate with either $\mathrm{Re}\ s < \Omega$ or $\mathrm{Re}\ s = \Omega$, the integral gives incorrect results. In the former case, for example, the resultant function is
\begin{eqnarray}
\left\{
\begin{array}{cc}
0 & (t > 0), \\
\displaystyle -\frac{1}{2} \sin \phi& (t = 0), \\
-e^{\Omega t}\sin(\omega t) & ( t < 0).
\end{array}\right.
\end{eqnarray}
The latter case, on the other hand, gives
\begin{eqnarray}
\left\{
\begin{array}{cc}
\displaystyle \frac{1}{2} e^{\Omega t}\sin (\omega t)  & (t > 0), \\
\displaystyle 0 & (t = 0), \\
-\displaystyle \frac{1}{2} e^{\Omega t}\sin (\omega t) & ( t < 0).
\end{array}\right.
\end{eqnarray}
The important thing is that the integral gives non-zero values for $t<0$ in both cases. 

In Fig.~\ref{Fig.Inv}, we illustrate the situation for the following function, linear combination of two growing sinusoidal functions, which is somewhat closer to the actual problem we considered in this paper:
\begin{eqnarray}
f(t) = e^{10t}\sin(100t) +e^{50t}\sin(500t).
\end{eqnarray}
Its Laplace transform is given as
\begin{eqnarray}
\label{fstarex}
f^*(s) = \frac{100}{(s -10)^2 + 100^2} + \frac{500}{(s -50)^2 + 500^2},
\end{eqnarray}
which has four single poles as shown in the figure (marked as red-crossed points). We performed the integral, Eq.~(\ref{InverseInt}), for $f^*(s)$ along three different paths with $\mathrm{Re}\ s  = -10$, $\mathrm{Re}\ s  = 30$ and $\mathrm{Re}\ s  = 70$, respectively. The results are presented in the same figure. Note that only the last one is appropriate for the inverse Laplace transform, which is indeed vindicated in the figure with the functional value being vanishing at $t<0$. The integrals along the other two paths return the following results:
\begin{eqnarray}
-[1-\theta(t) ][e^{10t}\sin(100t) +e^{50t}\sin(500t)] &(\mathrm{Re}\ s = -10), \\
e^{10t}\sin(100t)\theta(t) -[1-\theta (t)]e^{50t}\sin(500t)&(\mathrm{Re}\ s = 30),
\end{eqnarray}
where $\theta (t)$ is the Heaviside function. They are evidently wrong, having non-vanishing values at $t<0$ in particular.

The fact that the inversed function should vanish at $t<0$ can be employed to judge if we have missed the pole with the greatest real part, i.e. the one corresponding to the most unstable mode.

\section{D. Formal solutions}\label{sec.analyresults}
In this section, we assume that the perturbation in the shocked region is initially non-vanishing in general. Then an additional term shows up in the Laplace-transformed linearized equations as follows:
\begin{eqnarray}
\label{An0}
\frac{\diff {\bf y}^*}{\diff r} &=& (sA +B){\bf y}^* -A{\bf y}_0(r),
\end{eqnarray}
where ${\bf y}_0(r) := {\bf y}(r,t=0)$ denotes the initial perturbation.
The outer and inner boundary conditions are unchanged and given as Eqs.~(\ref{OB}) and (\ref{IB}), respectively.

Using the path-ordering operator, $\mathcal{P}$, \citep{Peskin}, the solution of Eqs.~(\ref{An0}) with the outer boundary condition, Eq.~(\ref{OB}), is symbolically given as
\begin{eqnarray}
{\bf y}^*(r,s) &=& \mathcal{P} \left[ \exp \left( \int _{r_\mathrm{sh}} ^r \diff r' (sA +B) \right) \right] \left\{ {\bf y}^* (r_\mathrm{sh}, s) -\int _{r_\mathrm{sh}} ^r \diff r' \mathcal{P}\left[\exp \left( -\int _{r_\mathrm{sh}}^{r'} \diff r'' (sA+B)\right) \right] A(r'){\bf y}_0(r') \right\},
\\
&=& \Lambda^*(r,s) {\bf y}^* (r_\mathrm{sh}, s) -\Lambda^*(r,s)\int _{r_\mathrm{sh}}^r \diff r' \Lambda^{*-1}(r',s)A(r'){\bf y}_0(r'),
\\
\label{EqLast}
&=& \Lambda^*(r,s) {\bf y}^* (r_\mathrm{sh}, s) - {\bf h}^*[{\bf y}_0](r,s),
\end{eqnarray}
where we defined an $n \times n$ matrix, $\Lambda ^*$, with $n$ being the number of components in ${\bf y}$, and a vectorial functional, ${\bf h ^*}$, of ${\bf y}_0$ as
\begin{eqnarray}
\Lambda^*(r,s) &:=&\mathcal{P} \left[ \exp \left( \int _{r_\mathrm{sh}} ^r \diff r' (sA+B) \right) \right], 
\\
{\bf h}^*[{\bf y}_0](r,s) &:=& \Lambda^*(r,s)\int _{r_\mathrm{sh}}^r \diff r' \Lambda^{*-1}(r',s)A(r'){\bf y}_0(r').
\end{eqnarray}
Evaluating Eq.~(\ref{EqLast}) at the neutrino sphere and using the outer boundary condition, Eq.~(\ref{OB}), we obtain the following formal solution:
\begin{eqnarray}
{\bf y}^*(r_{\nu _e},s) &=& \Lambda^*(r_{\nu _e},s) \left[ (s{\bf c}+{\bf d}) \frac{\delta r_\mathrm{sh}^*(s)}{r_\mathrm{sh}}  +R{\bf z}^*(s)   \right] - {\bf h}^*[{\bf y}_0](r_{\nu _e},s), \\
\label{An1}
&=: &\tilde{\Lambda}^*(s) \left[(s{\bf c}+{\bf d}) \frac{\delta r_\mathrm{sh}^*(s)}{r_\mathrm{sh}} +R{\bf z}^*(s)   \right] - {\bf  \tilde {h}}^*[{\bf y}_0](s).
\end{eqnarray}
Here and hereafter we denote the function $\Lambda^*$ and ${\bf h}^*$ that are evaluated at the inner boundary as $\tilde{\Lambda}^*(s)$ and ${\bf \tilde{h}}^*[{\bf y}_0](s)$ for notational simplicity.

The inner boundary condition, Eq.~(\ref{IB}), should be linear with respect to $y_i^*$, since we are considering linear perturbations. It hence takes the following form in general:
\begin{eqnarray}
\label{An2}
{\bf a }^*(s)\cdot {\bf y}^* (r_{\nu _e},s) + b^*(s)= 0,
\end{eqnarray}
where ${\bf a}^*$ and $b^*$ are some functions of $s$. 
Then the Laplace-transformed shock radius can be formally solved from Eqs.~(\ref{An1}) and (\ref{An2}) as follows:
\begin{eqnarray}
\label{An3}
\frac{\delta r_\mathrm{sh}^*}{r_\mathrm{sh}} &=& -\frac{{\bf a}^*(s)\cdot  \tilde{\Lambda}^*(s)R{\bf z}^*(s) +{\bf a}^*(s)\cdot {\bf \tilde{h }}^*[{\bf y}_0](s) -b^*(s) }{{\bf a}^*(s)\cdot \tilde{\Lambda}^*(s)(s{\bf c}+{\bf d})} , \\
\label{An3-2}
&=:& \mathcal{I}^*[{\bf y}_0]({\bf z}^*,s).
\end{eqnarray}

Let us consider the case with ${\bf z}^* = {\bf 0}$, i.e., there is no upstream perturbation:
\begin{equation}
\label{EqD12}
\mathcal{I}^*[{\bf y}_0]({\bf 0},s) = -\frac{{\bf a}^*(s)\cdot {\bf \tilde{h}}^*[{\bf y}_0](s) -b^*(s) }{{\bf a}^*(s)\cdot \tilde{\Lambda}^*(s)(s{\bf c}+{\bf d})}.
\end{equation}
The inverse-transform of this function describes the time evolution of the perturbed shock radius that is induced by the initial perturbation given in the shocked region, ${\bf y}_0(r)$, as well as by the fluctuation at the inner boundary represented with $b^*(s)$.
Subtracting Eq.~(\ref{EqD12}) from Eq.~(\ref{An3}), we obtain the shock motion induced by the upstream perturbation alone, which is denoted by $(\delta r_\mathrm{sh}/r_\mathrm{sh})_\mathrm{ex}(t) = \mathcal{L}^{-1}[(\delta r_\mathrm{sh}/r_\mathrm{sh})_\mathrm{ex}^* ]$:
\begin{eqnarray}
\left(\frac{\delta r_\mathrm{sh}}{r_\mathrm{sh}} \right)_\mathrm{ex}^* &=& \frac{\delta r_\mathrm{sh}^*}{r_\mathrm{sh}} -\mathcal{I}^*[{\bf y}_0]({\bf 0},s), \\
\label{eq.frac}
&=& -\frac{{\bf a}^*(s)\cdot  \tilde{\Lambda}^*(s)R{\bf z}^*(s)}{{\bf a}^*(s)\cdot \tilde{\Lambda}^*(s)(s{\bf c} +{\bf d})}, \\
\label{eq.again}
&=:& \mathcal{J}^*({\bf z}^*,s), \\
&=& \mathcal{J}^*(z_1^*, \cdots, z_n^*,s),\\
\label{residueex}
&=&\sum _{k=1}^n \mathcal{J}_k^* z_k^*.
\end{eqnarray}
In the last equation we defined a set of functions,
\begin{eqnarray}
\mathcal{J}_k^*(s) &:=& \mathcal{J}^*(0, \cdots, 0, 1, 0, \cdots, 0,s) \ (k=1,\cdots,n),
\end{eqnarray}
where the arguments of $\mathcal{J}^*(\cdots)$ on the right hand side are set to be zero except for the $k$-th one, which is put to unity. Since $\mathcal{L}[\delta (t)] = 1$, $\mathcal{J}_k(t) = \mathcal{L}^{-1}[\mathcal{J}_k^*]$ describes $\delta r_\mathrm{sh}/r_\mathrm{sh}$ for the impulsive perturbation of the unit strength, i.e., $\delta (t)$, added only to the $k$-th component of ${\bf z}(t)$. This implies that $\mathcal{J}_k(t)$ can be regarded as a Green's function. In fact, recalling that the Laplace transform of the convolution $(f*g)(t)$ is the product of the Laplace transforms of the individual functions, $\Laplace [f]\Laplace[g]=f^*(s)g^*(s)$, \citep[e.g.][]{Text}, we obtain
\begin{eqnarray}
\label{An5}
\left( \frac{\delta r_\mathrm{sh}}{r_\mathrm{sh}} \right)_\mathrm{ex}&=& \sum _{k=1}^n \mathcal{L}^{-1}[\mathcal{J}_k^*z_k^*],\\
&=& \sum _{k=1}^n (\mathcal{J}_k*z_k)(t), \\
\label{convolution}
&=& \sum _{k=1}^n  \int _0 ^t \mathcal{J}_k(t-\tau) z_k(\tau) \diff \tau .
\end{eqnarray}
We can hence obtain the evolution of the shock radius for any perturbation once we know $\mathcal{J}_k\ (k=1,\cdots ,n)$.

The function $\mathcal{I}^*[{\bf 0},{\bf y}_0](s)$ can be further divided into two parts:
\begin{eqnarray}
\mathcal{I}^*[{\bf y}_0]({\bf 0},s) &=& \mathcal{K}^*[{\bf y}_0](s) + \mathcal{S}^*(s),
\end{eqnarray}
where $\mathcal{K}^*$ and $\mathcal{S}^*$ describe the contributions from the initial perturbation in the shocked region and from the fluctuation at the inner boundary, respectively, and given as 
\begin{eqnarray}
\label{ini.frac}
\left( \frac{\delta r_\mathrm{sh}}{r_\mathrm{sh}} \right) _\mathrm{ini}^* &:=& \mathcal{K}^*[{\bf y}_0](s) := -\frac{{\bf a}^*(s)\cdot {\bf \tilde{h}}^*[{\bf y}_0](s) }{{\bf a}^*(s)\cdot \tilde{\Lambda}^*(s)(s {\bf c} +{\bf d})}, \\
\label{ib.frac}
\left( \frac{\delta r_\mathrm{sh}}{r_\mathrm{sh}} \right) _\mathrm{IB}^* &:=& \mathcal{S}^*(s) := \frac{b^*(s) }{{\bf a}^*(s)\cdot \tilde{\Lambda}^*(s)(s {\bf c}+{\bf d})},
\end{eqnarray}
Then the whole evolution of perturbed shock radius is given as the sum of these contributions:
\begin{equation}
\label{divi}
\frac{\delta r_\mathrm{sh}}{r_\mathrm{sh}} = \left(\frac{\delta r_\mathrm{sh}}{r_\mathrm{sh}} \right)_\mathrm{ini} +\left(\frac{\delta r_\mathrm{sh}}{r_\mathrm{sh}} \right)_\mathrm{ex} +\left(\frac{\delta r_\mathrm{sh}}{r_\mathrm{sh}} \right)_\mathrm{IB}.
\end{equation}

\section{E. The amplitude of each mode}\label{app.amplitude}
Let us write the evolution of  the perturbed shock radius as
\begin{equation}
\label{EqE1}
\frac{\delta r_\mathrm{sh}}{r_\mathrm{sh}}  = \sum _{j} a_je^{\Omega_j t}\sin(\omega _jt +\phi _j),
\end{equation}
and determine the amplitudes $a_j$ for a given upstream perturbation. In Eq.~(\ref{EqE1}) $\Omega _j$ and $\omega _j$ are the growth rate and oscillation frequency of the $j$-th mode ($j = 1, 2,3,\cdots $), respectively. To ensure the uniqueness of this expansion, we assume that $\omega _j \ge 0$ and $-\pi/2 \le \phi _j < \pi /2$ for each $j$. We also note that $\Omega _j$ and $a_j$ are real numbers. The amplitude $a_j$ is related with residue of $\delta r_\mathrm{sh}^*/r_\mathrm{sh}$ at the corresponding pole in the complex plane, which is directly calculated as
\begin{eqnarray}
\mathop{\Res}\limits_{s=\Omega_k +i\omega _k}\frac{\delta r_\mathrm{sh}^*}{r_\mathrm{sh}} 
&=& \mathop{\Res}\limits_{s=\Omega _k +i\omega _k}\sum _{j } \Laplace[a_j e^{\Omega _jt}\sin(\omega _jt +\phi _j)], \\
&=& \mathop{\Res}\limits_{s=\Omega _k+i\omega _k} \sum _j \frac{a_j}{2i}\left( \frac{e^{i\phi _j}}{s-\Omega _j-i\omega _j} - \frac{e^{-i\phi _j}}{s-\Omega _j+i\omega _j}\right), \\
\label{Res}
&=& \frac{a_k}{2i}e^{i\phi _k}.
\end{eqnarray}
On the other hand, the residue can be numerically obtained by means of the Cauchy's theorem:
\begin{eqnarray}
\label{Cauchy}
\mathop{\Res}\limits_{s=\Omega_k +i\omega _k}\frac{\delta r_\mathrm{sh}^*}{r_\mathrm{sh}} = \frac{1}{2\pi i}\oint _C \frac{\delta r_\mathrm{sh}^*}{r_\mathrm{sh}} \diff z,
\end{eqnarray}
where $C$ is any closed curve in the complex plane that includes only the $k$-th pole inside. The integral is conducted numerically by solving the initial-boundary-value problem given by Eqs.~(\ref{eq.L-linearized})-(\ref{IB}) for a set of $s$ on the contour $C$.

Once the residue for the $k$-th pole is obtained from Eq.~(\ref{Cauchy}), the amplitude $a_k$ is derived from Eq.~(\ref{Res}) as follows:
\begin{eqnarray}
a_k = -2(v_k\cos\phi_k -u_k\sin\phi_k) +2i(u_k\cos\phi_k +v_k\sin\phi_k),
\end{eqnarray}
where the $u_k$ and $v_k$ are the real and imaginary parts of the residue, respectively.
Since $a_k$ is real, the imaginary part must vanish. Therefore we find the result:
\begin{eqnarray}
\label{amplitude}
a_k = -2\mathop{\mathrm{sgn}}(v_k) \sqrt{u_k^2+v_k^2} = -2 \mathop{\mathrm{sgn}}\left[  \mathop{\mathrm{Im}}\left( \frac{\delta r_\mathrm{sh}^*}{r_\mathrm{sh}} \right) \right] \left| \mathop{\Res}\limits_{s=\Omega_k+i\omega_k} \frac{\delta r_\mathrm{sh}^*}{r_\mathrm{sh}} \right|,
\end{eqnarray}
and
\begin{eqnarray}
\tan \phi_k = -\frac{u_k}{v_k}. 
\end{eqnarray}

\section{F. Contribution from upstream perturbations}
Corresponding to the decomposition of $\delta r_\mathrm{sh}/r_\mathrm{sh}$, Eq.~ (\ref{divi}), its residue can be also divided into three parts:
\begin{eqnarray}
\label{EqAmp}
|a_k| = 2 \left| \sum _{j=1}^n\mathop{\Res}\limits_{s=\Omega _k +i\omega _k} [z^*_j(s)\mathcal{J}_j^*(s)] + \mathop{\Res}\limits_{s=\Omega _k +i\omega _k} [\mathcal{K}^*[{\bf y}_0](s) + \mathcal{S}^*(s)] \right| .
\end{eqnarray}
If we assume that ${\bf z}^*(s)$ does not have a pole at $s = \Omega _k +i\omega _k$, which is normally the case, the first term on the right hand side of the above equation is further calculated as
\begin{eqnarray}
\label{EqAmpP}
\sum _{j=1}^n\mathop{\Res}\limits_{s=\Omega _k +i\omega _k} [z^*_j(s)\mathcal{J}_j^*(s) ]= \sum _{j=1}^n z^*_j(\Omega _k +i\omega _k)\mathop{\Res}\limits_{s=\Omega _k +i\omega _k} \mathcal{J}_j^*(s) =: \sum_{j=1}^n W_j^*(s_k) .
\end{eqnarray}
Hence the contribution from the upstream perturbations can be easily obtained for any {\bf z}(t) once we derive the residue of $\mathcal{J}_j^*(s)$ $(j = 1,\cdots,n)$. If there is no perturbation initially in the shocked region and the inner boundary does not produce fluctuations, the amplitude is solely determined by the upstream perturbation as
\begin{eqnarray}
\label{EqAmpS}
|a_k| &=& 2 \left| \sum _{j=1}^n W_j^*(s_k) \right| , \\
\label{EqAmpS2}
&=&2 \sqrt{ \sum _{j=1}^n \left| W_j^*(s_k) \right| ^2  +\sum_{i\ne j} \mathrm{Re}\left[ W_i^*(s_k)\overline{W_j^*(s_k)} \right] },
\end{eqnarray}
where the horizontal line over functions means their complex conjugates.

\section{G. General upstream perturbations}\label{G}\textcolor{black}{
We here derive the excited amplitude for the generic upstream perturbation expressed in the following form:
\begin{equation}
z_j(t) = \sum _{\sigma} b_{j\sigma} \sin (\omega _{j\sigma} t + \phi _{j\sigma}).
\end{equation}
The corresponding Laplace transformed perturbation is given by
\begin{equation}
z_j^*(t) = \sum _{\sigma} b_{j\sigma} \frac{s\sin\phi_{j\sigma} + \omega _{j\sigma} \cos\phi_{j\sigma}}{s^2+\omega_{j\sigma}^2}.
\end{equation}
Inserting the above expression into Eqs.~(\ref{EqAmpS}) and (\ref{EqAmpS2}), we obtain
\begin{eqnarray}
|a_k| &=&  2 \left| \sum _\sigma \sum _{j=1}^n W_{j\sigma}^*(s_k) \right| ,\\
\label{Akup}
&=&2\sqrt{\sum _\sigma \left|\sum _{j=1}^n W_{j\sigma}^*(s_k)\right|^2+\sum_{\sigma\ne \rho} \mathrm{Re}\left[\sum _{i,j=1}^n W_{i\sigma}^*(s_k)  W_{j\rho}^*(s_k)\right]}, 
\end{eqnarray}
where $W^*_{j\sigma}$ is given by
\begin{equation}
W^*_{j\sigma}(s_k) := b_{j\sigma}\frac{(\Omega _k +i\omega _k)\sin \phi _{j\sigma}+\omega _{j\sigma} \cos \phi _{j\sigma}}{(\Omega _k +i\omega _k)^2+\omega _{j\sigma}^2}\mathop{\mathrm{Res}}\limits_{s=\Omega _k +i\omega _k}\mathcal{J}_j^*(s).
\end{equation}
We define $a_{k\sigma}$ as the amplitude of the $k$-th mode that would be excited by a single harmonic perturbation with a frequency $\omega _{j\sigma}$.
Recalling a relation $|a_{k\sigma}| = 2 |\sum_jW_{j\sigma}(s_k)|$, we rewrite Eq.~(\ref{Akup}) as
\begin{eqnarray}
|a_k| &=&\sqrt{\sum _\sigma |a_{k\sigma}|^2+ 4\sum_{\sigma\ne \rho} \mathrm{Re}\left[\sum _{i,j=1}^n W_{i\sigma}^*(s_k) W_{j\rho}^*(s_k)\right]} , \\
\label{ApprL}
&\sim& \sqrt{\sum _\sigma |a_{k\sigma}|^2}, 
\end{eqnarray}
where the last approximation holds unless $\sum_j W_{j\sigma}(s_k)$ $(\sigma =1,2,\cdots)$ are correlated to each other. The excited amplitude is then expressed as a square root of the sum of individual amplitudes squared and will be approximately given by the largest $a_{k\sigma}$.}


\begin{thebibliography}{}
\bibitem[Abdikamalov et al.(2015)]{Abd} Abdikamalov, E., Ott, C. D., Radice, D., Roberts, L. F., Haas, R., Reisswig, C., M\"{o}sta, P., Klion, H. \& Schnetter, E. 2015, \apj, 808, 70 
\bibitem[Abdikamalov et al.(2016)]{Abd2} \textcolor{black}{Abdikamalov, E., Zhaksylykov, A., Radice, D. \& Berdibek, S. 2016, arXiv:1605.09015v1}
\bibitem[Arnett \& Meakin(2011)]{Arnett} Arnett, W. D. \& Meakin, C. 2011, \apj, 733, 78
\bibitem[Asida \& Arnett(2000)]{Asida} Asida, S. M. \& Arnett, D. 2000, \apj, 545, 435
\bibitem[Bazan \& Arnett(1998)]{Bazan} Bazan, G. \& Arnett, D. 1998, \apj, 496, 316
\bibitem[Blondin et al.(2003)]{Blondin} Blondin, J. M., Mezzacappa, A. \& DeMarino, C. 2003, \apj, 584, 971
\bibitem[Bruenn(1985)]{Bruenn} Bruenn, S. W. 1985, ApJS, 58, 771
\bibitem[Bruenn et al.(2013)]{Bruenn13} Bruenn, S. W., Mezzacappa, A., Hix, W. P., Lentz, E. J., Messer, O. E. B., Lingerfelt, E. J., Blondin, J. M., Endeve, E., Marronetti, P. \& Yakunin, K. N. 2013, \apjl, 767, L6
\bibitem[Burrows et al.(2012)]{Burrows12} Burrows, A., Dolence, J. \& Murphy, J. W. 2012, \apj, 759, 5
\bibitem[Chatzopoulos et al.(2014)]{Chat} Chatzopoulos, E., Graziani, C. \& Couch, S. M. 2014, \apj, submitted (arXiv:1405.4873v1)
\bibitem[Couch \& Ott(2013)]{Couch} Couch, S. M. \& Ott, C. D. 2013, \apjl, 778, L7
\bibitem[Couch \& Ott(2015)]{Couch2} Couch, S. M. \& Ott, C. D. 2015, \apj, 799, 5
\bibitem[Couch et al.(2015)]{Couch3} Couch, S. M. \& Chatzopoulos, E., Arnett, W. D. \& Timm\textcolor{black}{e}s, F. X. 2015, \apjl, 808, L21
\bibitem[Fern\'{a}ndez et al.(2014)]{Fernandez14} Fern\'{a}ndez, R., M\"{u}ller, B., Foglizzo, T. \& Janka, H.-Th. 2014, MNRAS, 440, 2763
\bibitem[Fern\'{a}ndez(2015)]{Fernandez15} Fern\'{a}ndez, R. 2015, MNRAS, 452, 2071
\bibitem[Foglizzo(2009)]{Foglizzo09} Foglizzo, T. 2009, \apj, 694, 820
\bibitem[Foglizzo et al.(2006)]{Foglizzo06} Foglizzo, T.,  Scheck, L. \& Janka, H.-Th. 2006, \apj, 652, 1436
\bibitem[Foglizzo et al.(2007)]{Foglizzo07} Foglizzo, T., Galletti, P., Scheck, L. \& Janka, H.-Th. 2007, \apj, 654, 1006
\bibitem[Foglizzo et al.(2015)]{Foglizzo15} Foglizzo, T., Kazeroni, R., Guilet, J. et al. 2015, PASA, 32(e009), 17 
\bibitem[Guilet \& Foglizzo(2010)]{Guilet10} Guilet, J. \& Foglizzo, T. 2010, \apj, 711, 99
\bibitem[Guilet \& Foglizzo(2012)]{Guilet12} Guilet, J. \& Foglizzo, T. 2012, \mnras, 421, 546
\bibitem[Hanke et al.(2013)]{Hanke13} Hanke, F., M\"{u}ller, B., Wongwathanarat, A., Marek, A. \& Janka, H.-Th. 2013, \apj, 770, 66
\bibitem[Iwakami et al.(2014a)]{Iwakami} Iwakami, W., Nagakura, H. \& Yamada, S. 2014, \apj, 786, 118
\bibitem[Iwakami et al.(2014b)]{IwakamiB} Iwakami, W., Nagakura, H. \& Yamada, S. 2014, \apj, 793, 5
\bibitem[Lai \& Goldreich(2000)]{Lai} Lai, D. \& Goldreich, P. 2000, \apj, 535, 402
\bibitem[Meakin \& Arnett(2006)]{Meakin06} Meakin, C. A. \& Arnett, D. 2006, \apjl, 637, L53
\bibitem[Meakin \& Arnett(2007)]{Meakin07} Meakin, C. A. \& Arnett, D. 2007, \apj, 667, 448
\bibitem[\textcolor{black}{M\"{o}sta et al.}(2015)]{Moesta} \textcolor{black}{M\"{o}sta, P., Ott, C. D., Radice, D., Roberts, L. F., Schnetter, E. \& Haas, R. 2015, Nature, 528, 376}
\bibitem[M\"{u}ller \& Janka(2014)]{MJ14} M\"{u}ller, B. \& Janka, H.-Th. 2014, arXiv:1409.4783v1
\bibitem[M\"{u}ller et al.(2016)]{Muller16} M\"{u}ller, B., Viallet, M., Heger, A. \& Janka, H.-Th. 2016, arXiv:1605.01393v1
\bibitem[Ohnishi et al.(2006)]{Ohnishi} Ohnishi, N., Kotake, K. \& Yamada, S. 2006, \apj, 641, 1018
\bibitem[Peskin \& Schroeder(1995)]{Peskin} Peskin, M. E. \& Schroeder, D. V. 1995, An Introduction to Quantum Field Theory (Westview Press)
\bibitem[Sato et al.(2009)]{Sato} Sato, J., Foglizzo, T. \& Fromang, S. 2009, \apj, 694, 833
\bibitem[Sawai \& Yamada(2014)]{Sawai} Sawai, H. \& Yamada, S. 2014, \apjl, 784, L10
\bibitem[Schiff(1999)]{Text} Schiff, J. L. 1999, The Laplace Transform: Theory and Applications (Springer)
\bibitem[Scheck et al.(2006)]{Scheck} Scheck, L., Kifonidis, K., Janka, H.-Th. \& M\"{u}ller, E. 2006, A\&A, 457, 963
\bibitem[Shen et al.(1998)]{Shen} Shen, H., Toki, H., Oyamatsu, K. \& Sumiyoshi, K. 1998, NuPhA, 637, 435
\bibitem[Takahashi \& Yamada(2014)]{KT} Takahashi, K. \& Yamada, S. 2014, \apj, 794, 162
\bibitem[Takiwaki et al.(2014)]{Takiwaki14} Takiwaki, T., Kotake, K. \& Suwa, Y. 2014, \apj, 786, 83
\bibitem[Takiwaki et al.(2016)]{Takiwaki} Takiwaki, T., Kotake, K. \& Suwa, Y. 2016, arXiv:1602.06759v1
\bibitem[Yamasaki \& Foglizzo(2008)]{Yamasaki08} Yamasaki, T. \& Foglizzo, T. 2008, \apj, 679, 607
\bibitem[Yamasaki \& Yamada(2007)]{Yamasaki07} Yamasaki, T. \& Yamada, S. 2007, \apj, 656, 1019

\end{thebibliography}
\end{document}